\documentclass[12pt]{iopart}

\expandafter\let\csname equation*\endcsname\relax

\expandafter\let\csname endequation*\endcsname\relax

\usepackage{chemformula} % Formula subscripts using \ch{}
\usepackage[T1]{fontenc} % Use modern font encodings
\usepackage{amsmath}
\usepackage{amssymb}
\usepackage{bm}
\usepackage{algorithm}
\usepackage{algpseudocode}
\usepackage{booktabs}
\usepackage{graphicx}
\usepackage{subcaption}
\usepackage[hyperfootnotes=false]{hyperref}
\usepackage{cite}
\usepackage[symbol]{footmisc}

\newcommand*{\rvx}{\mathbf{x}}
\newcommand*{\mZ}{\mathbf{Z}}

\newcommand*{\bfzero}{\boldsymbol{0}}
\newcommand*{\mI}{\mathbf{I}}

\newcommand*{\rvepsilon}{\bm{\varepsilon}}
\newcommand*{\rvmu}{\bm{\mu}}
\newcommand*{\E}{\mathbb{E}}

\def\NoNumber#1{{\def\alglinenumber##1{}\State #1}\addtocounter{ALG@line}{-1}}
\makeatletter
\def\@makefnmark{\hbox{\@textsuperscript{\normalfont\@thefnmark}}}
\makeatother

\begin{document}

\title[Molecular relaxation by reverse diffusion]{Molecular relaxation by reverse diffusion with time step prediction}

\author{
Khaled Kahouli$^{1,2,*}$, 
Stefaan Simon Pierre Hessmann$^{1,2}$,
Klaus-Robert M\"uller$^{1,2,3,4}$,
Shinichi Nakajima$^{1,2,5}$,
Stefan Gugler$^{1,2}$\footnote[2]{Equal contribution}
and Niklas Wolf Andreas Gebauer$^{1,2,6}$\footnotemark[2]
}

\address{$^1$ Machine Learning Group, Technische Universit\"at Berlin, Berlin, Germany}
\address{$^2$ BIFOLD – Berlin Institute for the Foundations of Learning and Data, Berlin, Germany}
\address{$^3$ Department of Artificial Intelligence, Korea University, Seoul, Republic of Korea}
\address{$^4$ Max-Planck Institute for Informatics, Saarbrücken, Germany}
\address{$^5$ RIKEN Center for Advanced Intelligence Project, Tokyo, Japan}
\address{$^6$ BASLEARN -- TU Berlin/BASF Joint Lab for Machine Learning, Technische Universit\"at Berlin, Berlin, Germany}
\address{$^*$ Author to whom any correspondence should be addressed.}

\ead{khaled.kahouli@tu-berlin.de, stefan.gugler@tu-berlin.de and n.gebauer@tu-berlin.de}

\begin{abstract}
    Molecular relaxation, finding the equilibrium state of a non-equilibrium structure, is an essential component of computational chemistry to understand reactivity. Classical force field (FF) methods often rely on insufficient local energy minimization, while neural network FF models require large labeled datasets encompassing both equilibrium and non-equilibrium structures. As a remedy, we propose MoreRed, molecular relaxation by reverse diffusion, a conceptually novel and purely statistical approach where non-equilibrium structures are treated as noisy instances of their corresponding equilibrium states. To enable the denoising of arbitrarily noisy inputs via a generative diffusion model, we further introduce a novel diffusion time step predictor. Notably, MoreRed learns a simpler pseudo potential energy surface (PES) instead of the complex physical PES. It is trained on a significantly smaller, and thus computationally cheaper, dataset consisting of solely unlabeled equilibrium structures, avoiding the computation of non-equilibrium structures altogether. We compare MoreRed to classical FFs, equivariant neural network FFs trained on a large dataset of equilibrium and non-equilibrium data, as well as a semi-empirical tight-binding model. To assess this quantitatively, we evaluate the root-mean-square deviation between the found equilibrium structures and the reference equilibrium structures as well as their energies.
\end{abstract}

\vspace{2pc}

\section{Introduction}

    Geometry optimization is crucial for understanding reactivity in computational chemistry~\cite{schlegel2011},
    as it allows for the study of chemical reaction networks~\cite{broadbelt1994,broadbelt1996,broadbelt2005,fialkowski2005,gothard2012,kowalik2012,sameera2016,dewyer2018,maeda2011,feinberg2018,simm2019,unsleber2020,baiardi2022},
    which are fundamental in catalysis~\cite{deutschmann1998,zhu2005,gossler2019,ulissi2017,steiner2022},
    combustion~\cite{sankaran2007,harper2011},
    polymerization~\cite{vinu2012}, or
    atmospheric chemistry~\cite{vereecken2015}.
    Reactivity is governed by activation energy barriers connecting two equilibrium structures via a transition state, and are necessary for microkinetic modeling~\cite{proppe2017,proppe2019b,suleimanov2015,gao2016a,susnow1997,han2017}.
    Moreover, for generative~\cite{arspous2019}, or enumerative~\cite{gugler2020, reymond2015} explorations, e.g. in drug, battery, or catalyst design~\cite{hajduk2007decade_drug_fragments, hautier2011novel_batteries_Throughput, Bhowmik2019inverse_batteries, Freeze2019inverse_catalysts, Gantzer2020inverse_drugs, von2020exploring},
    a common approach is to use a computationally cheap method to generate a dataset,
    followed by a geometry optimization to obtain equilibrium structures for which most physical properties are defined.
    Equilibrium structures represent local minima on the Born–Oppenheimer potential energy surface (PES)~\cite{born1927,sutcliffe1992}
    and are identified by molecular relaxation, that is solving the electronic Schr\"{o}dinger equation while varying nuclear coordinates by iteratively following the negative gradients of the energy, i.e. the forces, until they converge to zero~\cite{jensen2017b,cramer2004,schlegel2011}. 
    
    Because iterative ab initio electronic structure calculations are computationally expensive and not feasible in high-throughput settings,
    methods for finding equilibrium structures need to be efficient~\cite{jensen2017b,cramer2004}.
    To address this limitation, 
    numerous approaches have been developed
    that speed up the computation of forces
    but also suffer from a loss in accuracy compared to ab initio methods. 
    These include
    classical force field (FF) methods like 
    MMFF94~\cite{halgren1996merck}, 
    the universal FF~\cite{rappe1992uff}, 
    or CHARMM~\cite{vanommeslaeghe2010charmm}, 
    on the one hand, and 
    semiempirical methods such as 
    GFN2-xTB~\cite{bannwarth2019gfn2}, PM6-7~\cite{pm6,pm7}, or OM2~\cite{om2}  on the other hand.
    Furthermore, 
    machine learning FF (MLFF) models~\cite{coulombmatrix, de2016comparing, behler2007generalized, faber2018fchl, dtnn, smith2017ani1_potential, dimenet, nequip, satorras2021n, so3krates, batatia2022mace, UnkeGems, allegro, tensorfieldnet, noe2020review, unke2021machine_review} have emerged as promising alternatives to physical models. When trained on a sufficient amount of ab initio calculations, MLFF models, using kernel methods such as sGDML~\cite{gdml,sgdml,Chmiela2023Accurate}
    or neural networks such as SchNet~\cite{schnet,Schuett2018SchNet}, learn the physical PES and very efficiently predict the forces. Moreover, they have been shown to produce promising results in relaxation tasks~\cite{physnet,spookynet}, while reducing computational cost by several magnitudes.
    Although MLFF models significantly accelerate gradient computations for structural relaxation, the training dataset must cover a wide range of the chemical space including equilibrium and non-equilibrium structures with accurately computed physical labels, introducing a large computational cost for generating a training dataset.

    An emerging machine learning-based approach to exploring chemical space is training generative models on a dataset of equilibrium structures in order to learn to generate new molecular structures.
    For instance, diffusion models have been used recently in molecule generation~\cite{edm, diffbridges, mdm, geoldm, moldiff}, conformer search~\cite{geodiff} and molecular graph generation~\cite{vignac2023digress, autoregressivediff}. They generate samples via iterative denoising, starting from a simple prior distribution like isotropic Gaussian noise.
    Several other generative models exist for 3D molecular structures, but they are generally not designed for denoising or generation from arbitrary states.
    Typically, they generate equilibrium structures from scratch by iteratively adding new atoms~\cite{gschnet, cgschnet, simm2020reinforcement_3d, simm2021symmetryaware_3d, meldgaard2021generating_3d} or by transforming samples from a prior distribution to a target distribution in one shot~\cite{noe2019boltzmann, kohler2020equivariant, garcia2021en_flow, klein2023timewarp}.
    Furthermore, generative models have been used to sample conformations given molecular graphs as input~\cite{mansimov2019molecular_graph_Translation, simm2020generative_graph_Translation, gogineni2020torsionnet_graph_Translation, ganea2021geomol, xu2021end_graph_Translation, lemm2021machine_graph_Translation, torsional_diff}.
    A common drawback of conventional generative methods is that, unlike relaxation-based methods, equilibrium structures are generated from scratch, which makes it difficult to steer the generation towards desired structures. 
    We define denoising as generating samples from the data manifold given arbitrarily noisy inputs. This task is fundamentally different from previous work \cite{Wang_denoising, pmlr-v202-feng23c, zaidi2023pretraining, liu2023molecular, godwin2022simple} based on the idea of denoising autoencoders \cite{DAE}, where different denoising, yet not generative, techniques are used on molecular structures as an auxiliary or pre-training task for the original regression task of predicting forces and other molecular properties, aiming to improve data efficiency, generalizability and robustness. Relative to our approach, Hsu \etal~\cite{hsu2022score} use a score-based model to eliminate thermal noise or perturbations in the atomic positions of condensed materials, dealing with relatively small noise magnitudes.
    
\begin{figure}
        \includegraphics[width=1.\linewidth]{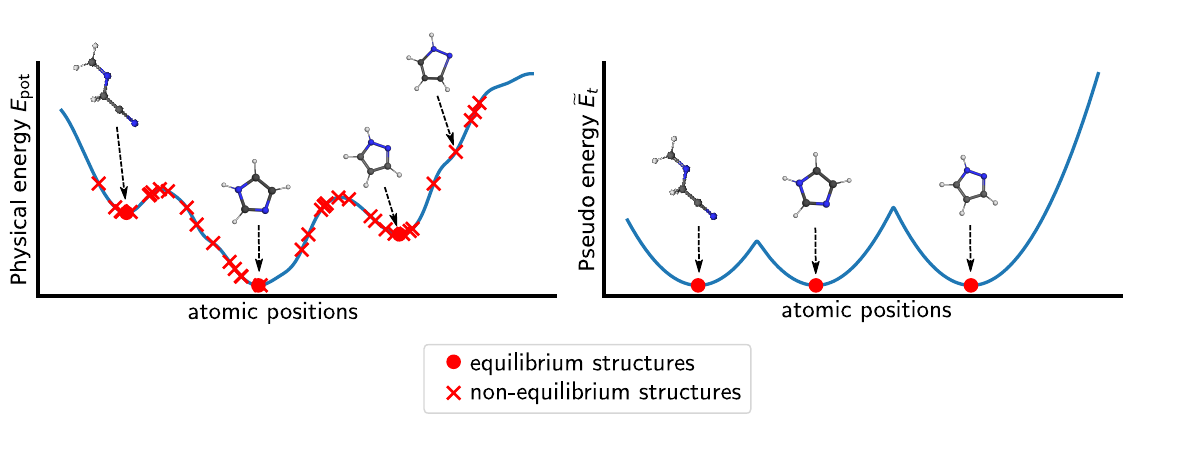}
        \caption{
        Schematic depictions of the physical PES (left) and the pseudo PES (right), which emerge from removing Gaussian noise from distorted structures.
        Learning the physical energy requires equilibrium (circles) and many non-equilibrium (crosses) training structures, while learning the pseudo PES only requires equilibrium training structures.
        }
        \label{fig:illustration}
    \end{figure}

    In this work, we propose a conceptually novel statistical approach to molecular relaxation through reverse diffusion, which we will call MoreRed.
    The distortion in a non-equilibrium input structure is interpreted as a noise level,
    as the structure has ``diffused away'' from its equilibrium state. In this setting,
    the molecular relaxation can be modeled as a denoising process which can be achieved by reverse diffusion.
    In contrast to MLFFs, MoreRed does not learn the physical PES, but a simple pseudo PES that emerges from removing Gaussian noise from distorted structures (see Figure~\ref{fig:illustration}).
    This offers a significant advantage over the MLFF models:
    Training MoreRed requires \emph{only} equilibrium structures \emph{without} labels for physical properties such as energy and forces, which considerably reduces the computational costs of generating training datasets.
    Therefore, it potentially expands the applicability of ML-based relaxation to domains where MLFFs cannot be trained because only equilibrium structures are reported.
    
    A key technical novelty of MoreRed is the \emph{diffusion time step predictor}.
    Existing diffusion models require the time step as an input
    that indicates how noisy the input is.
    However, in molecular relaxation the noise level, i.e. how far away from the equilibrium a structure lies, is unknown.
    Therefore, in MoreRed we predict the appropriate time step,
    enabling us to denoise distorted molecular structures at arbitrary noise levels.
    To this end, we provide a theoretical argument for why accurate time step prediction via a neural network is possible.
    In contrast to existing diffusion models, which follow a fixed pre-defined time step schedule, this allows furthermore for an adaptive schedule, where the time step is dynamically increased or decreased during denoising depending on the detected noise level, potentially correcting errors in the denoising process.
    We demonstrate the advantage of our adaptive schedule over classical, fixed sampling.
    
    We test the performance of our method on QM7-X \cite{qm7x}, a dataset containing 42 000 equilibrium structures found with third-order self-consistent charge density functional tight binding \cite{Mortazavi2018} (DFTB3) \cite{Seifert1996, Elstner1998, Gaus2011} and many-body dispersion (MBD) \cite{Tkatchenko2012, Ambrosetti2014} corrections.
    They cover all molecular graphs in GDB13~\cite{blum2009970} with up to 7 heavy atoms and include the elements H, C, N, O, S, and Cl.
    For each equilibrium structure, 100 non-equilibrium structures generated via normal-mode displacements of the equilibrium geometry are reported, including DFT calculations for energies and forces at the PBE0+MBD level\cite{Tkatchenko2012,Adamo1999,Perdew1996} with \texttt{FHI-aims}\cite{Blum2009,Ren2012}.
    In our experiments, we employ several baselines for comparison, including the classical FF MMFF94, the semiempirical method GFN2-xTB, as well as a MLFF model with an equivalent neural network backbone architecture, and show that MoreRed performs favourably.
    Specifically, MoreRed accurately maps non-equilibrium structures back to the data manifold of equilibrium structures that it has been trained on.
    This is despite being trained on two orders of magnitude fewer structures than the MLFF.
    Additionally, while MLFFs can only relax structures that are covered by the distribution of training data, the inherent augmentation of the training data through the diffusion process enhances the robustness of MoreRed against variations in the noise distribution of the non-equilibrium test structures.
    Consequently, MoreRed successfully identifies the correct equilibrium structures for non-equilibrium inputs where MLFFs fail.
    We also show that the difference in DFT energies between the reference equilibrium structures in QM7-X and the structures obtained by MoreRed through molecular relaxation falls below the threshold of chemical accuracy.

\section{Theory and methods}

    \subsection{Diffusion models}
        \label{sec:back_diff}

        \begin{figure}[t]
            \centering
            \includegraphics[width=1.\linewidth]{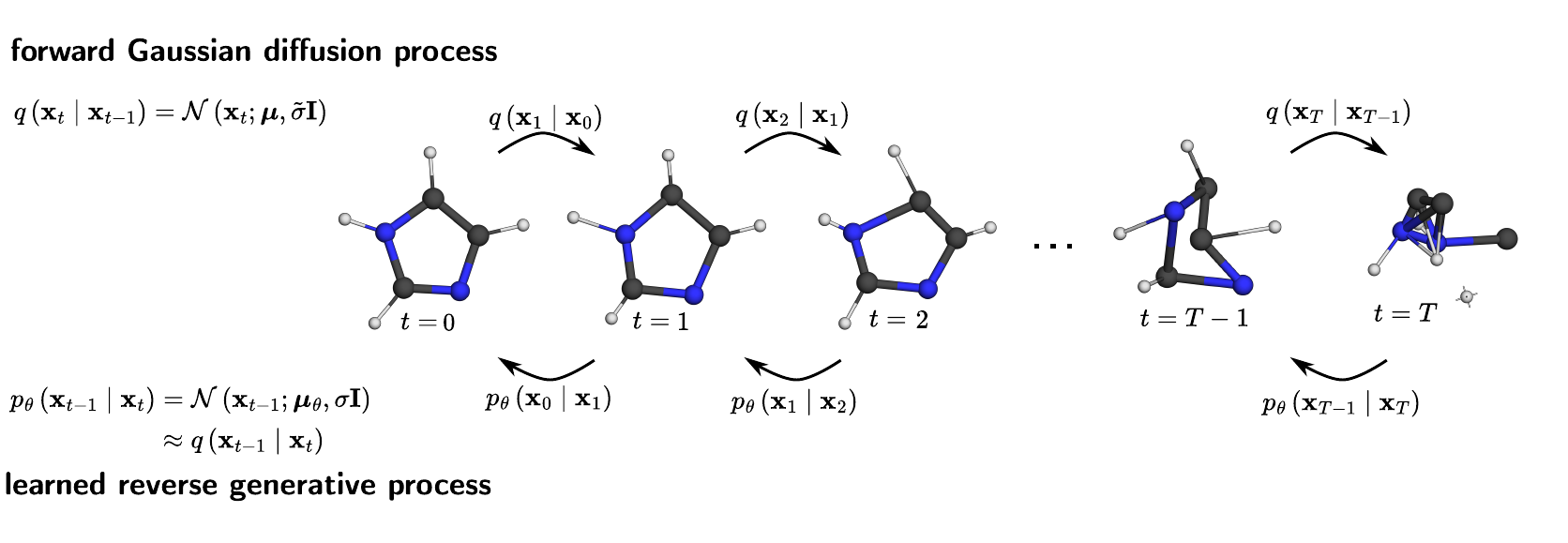}
            \caption{The diffusion model applied to the molecular structure of imidazole.
            It incorporates two stochastic processes:
            a forward process and a reverse process.
            The forward process involves a fixed diffusion kernel $q(\rvx_t|\rvx_{t-1})$
            to transform the original sample $\rvx_0 \sim q_\mathrm{data}(\rvx_0)$
            into complete noise sample, $\rvx_T \sim q_T(\rvx_T)$, usually isotropic Gaussian noise $\mathcal{N}(\bfzero, \mI)$.
            The backward process is a learned model with parameters $\theta$,
            which reverses the forward process, 
            i.e. $p_{\theta}(\rvx_{t-1}|\rvx_{t}) \approx q(\rvx_{t-1}|\rvx_{t})$.
            It maps a noise sample $\rvx_T$ from the tractable prior distribution $p_T(\rvx_T)=q_T(\rvx_T)$ to the complex target distribution of equilibrium structures, $q_\mathrm{data}(\rvx_0)$.}
            \label{fig:EDM}
        \end{figure}

        Introduced by Sohl-Dickstein \etal~\cite{diff_mod_sohl}, diffusion models are latent-variable generative models that can efficiently generate samples of a complex data distribution $q_\mathrm{data}(\rvx_0)$, where direct sampling is intractable, such as the distribution of equilibrium molecular structures.
        Instead of directly sampling from the target distribution, the idea is to obtain an initial sample $\rvx_T$ from a simple prior distribution $q_T(\rvx_T)$, often an isotropic Gaussian, and then use a learned mapping $h(\cdot)$ to transform $\rvx_T$ into a sample $\rvx_0 = h(\rvx_T)$ within $q_\mathrm{data}(\rvx_0)$.

        While defining and learning $h(\cdot)$ is challenging, the opposite task of transforming the complex distribution $q_\mathrm{data}(\rvx_0)$ into a simple distribution $q_T(\rvx_T)$ is manageable because it only involves simplifying the data by diminishing its signal, for instance by iteratively adding noise.
        Diffusion models leverage this concept by using two opposite processes (see Figure~\ref{fig:EDM}).
        A fixed, usually non-learned, forward diffusion process iteratively encodes $q_\mathrm{data}(\rvx_0)$ into a tractable latent distribution $q_T(\rvx_T)$ over $T$ steps. A backward or reverse process parametrized by a machine learning model then learns to reverse the forward diffusion to effectively map from $q_T(\rvx_T)$ back to $q_\mathrm{data}(\rvx_0)$, akin to the objective of the mapping $h(\cdot)$.

        Summarized in Figure~\ref{fig:EDM}, we focus on Denoising Diffusion Probabilistic Models (DDPM) \cite{DDPM_Ho} as a special class of diffusion models that defines the forward diffusion process as the fixed Markov process,
        \begin{equation}
                q(\rvx_{0:T}) = q_\mathrm{data}(\rvx_0) \prod_{t=1}^T q(\rvx_t|\rvx_{t-1}), \quad q(\rvx_t|\rvx_{t-1}) = \mathcal{N}(\rvx_t; \sqrt{1-\beta_t} \rvx_{t-1}, \beta_t \mI) \ ,
                \label{eq:forward_Traj}
        \end{equation}
        where $q_0(\rvx_0) = q_\mathrm{data}(\rvx_0)$
        and $\mathcal{N}(\rvx_t; \sqrt{1-\beta_t} \rvx_{t-1}, \beta_t \mI)$ denotes that $\rvx_t$ follows an isotropic Gaussian with mean $\sqrt{1-\beta_t} \rvx_{t-1}$ and variance $\beta_t \mI$,
        with the identity matrix $\mI$. 
        The diffusion process from Eq.~\eqref{eq:forward_Traj} can be simulated by sampling $\rvx_0 \sim q_\mathrm{data}(\rvx_0)$ from the training dataset representing the equilibrium positions of atoms and then iteratively applying 
        $\rvx_t = \sqrt{1-\beta_t} \rvx_{t-1} + \sqrt{\beta_t} \rvepsilon_t$ for $t = 1,2, \dots, T$, with Gaussian noise $\rvepsilon_t \sim \mathcal{N}(\bfzero, \mI)$.
        The variance $\beta_t$ follows a fixed monotonically increasing noise schedule $\{\beta_t \in (0,1)\}_{t=1}^T$, thus progressively injecting Gaussian noise with variance $\beta_t$ into the atom positions while diminishing the signal with the factor $\sqrt{1-\beta_t}$.
        This results in the generation of increasingly noisier molecular structures,
        $\rvx_1, \rvx_2, \dots, \rvx_T$,
        from the original sample $\rvx_0$
        with increasing diffusion time step $t$ (see Figure~\ref{fig:EDM}, forward Gaussian diffusion process, from left to right).
        At the endpoint $t=T$, the process destroys all the signal in the sample, converging to pure Gaussian noise, i.e. $ \rvx_T \sim \mathcal{N}(\bfzero, \mI)$.

        When generating progressively noisier samples $\{\rvx_t\}_{t=1}^T$, the diffusion process creates latent distributions $q_t(\rvx_t)$ that are increasingly smoother versions of the original data distribution $q_\mathrm{data}(\rvx_0)$, such that $\rvx_t \sim q_t(\rvx_t)$. These latent distributions can be derived from Eq.~\eqref{eq:forward_Traj} as $q_t(\rvx_t) = \int q(\rvx_t|\rvx_0)q_\mathrm{data}(\rvx_0) \mathrm{d}\rvx_0$ for $t \in [1,T]$.
        The perturbation kernel, $q(\rvx_t|\rvx_0)$, has the closed-form solution
        $\mathcal{N}(\rvx_t; \sqrt{\bar{\alpha}_t}\rvx_0, (1-\bar{\alpha}_t) \mI)$,
        where $\bar{\alpha}_t = \prod_{s=1}^t \alpha_s $ and $\alpha_s = 1 - \beta_s$ \cite{DDPM_Ho}.
        With this definition, we can directly sample from the forward diffusion process at any time step $t$ using an equilibrium structure $\rvx_0$ from the training data:
        \begin{equation}
            \label{eq:x_t}
            \rvx_t = \sqrt{\bar{\alpha}_t}\rvx_0 + \sqrt{1-\bar{\alpha}_t} \rvepsilon_t, \quad \rvepsilon_t \sim \mathcal{N}(\bfzero, \mI)
        \end{equation}
        avoiding the iterative simulation through all intermediate steps $\rvx_0,\dots,\rvx_t$.

        Reversing the forward diffusion process enables the generation of new samples by mapping back from $q_T(\rvx_T)$ to $q_\mathrm{data}(\rvx_0)$ using the reverse transition $q(\rvx_{t-1}|\rvx_t)$.
        Given that $T$ is large enough, the reverse transition is also Gaussian~\cite{diff_mod_sohl}.
        However, unlike the forward process, it is not tractable.
        Therefore, we need to approximate the reverse process, e.g. by learning a parametrized model $p_\theta(\rvx_{t-1}|\rvx_t) \approx q(\rvx_{t-1}|\rvx_t)$ such that \cite{DDPM_Ho}:
        \begin{equation}
            \label{eq:backward_traj}
            p_{\theta}(\rvx_{0:T}) = p_T(\rvx_T) \prod_{t=1}^T p_{\theta}(\rvx_{t-1}|\rvx_{t}), \quad p_{\theta}(\rvx_{t-1}|\rvx_{t}) = \mathcal{N}(\rvx_{t-1}; \rvmu_{\theta}(\rvx_t,t), \sigma_t^2 \mI)
        \end{equation}
        where $p_T(\rvx_T) = q_T(\rvx_T) = \mathcal{N}(\bfzero, \mI)$ is the endpoint of the forward process and the starting point (or latent prior) of the reverse process.
        The variance is $\sigma_t^2 = \frac{1-\bar{\alpha}_{t-1}}{1-\bar{\alpha}_t} \beta_t$ and the mean, $\rvmu_\theta(\rvx_t,t) = \frac{1}{\sqrt{\alpha_t}} \bigl( \rvx_t - \frac{\beta_t}{\sqrt{1-\bar{\alpha}_t}} \rvepsilon_{\theta}(\rvx_t,t) \bigr)$, is the only unknown quantity, 
        where $\rvepsilon_{\theta}(\rvx_t,t)$ is an estimate of the noise that was added to $\rvx_0$ to obtain $\rvx_{t}$. 
        This noise is predicted by a neural network that gets the noisy structure $\rvx_t$ and the current time step $t$ as an input.
        To train this network, we uniformly sample a diffusion time step $t \sim \mathcal U(1, T)$, and perform forward diffusion to generate the noisy sample $\rvx_t$ from a data point $\rvx_0$ using the sampled noise direction $\rvepsilon_t$ as described in Eq.~\eqref{eq:x_t}.
        Then, we minimize the mean squared error between the predicted and true noise direction, resulting in the following loss:
        \begin{equation}
            L_{\mathrm{DDPM}} =  \E_{t \sim \mathcal{U}(1,T),\rvx_0 \sim q_\mathrm{data}, \rvepsilon_t \sim \mathcal{N}(\boldsymbol{0} , \mI)}  \bigl[ || \rvepsilon_t - \rvepsilon_{\theta}(\rvx_t, t)||^2  \bigr], 
            \label{eq:DDPMLoss}
        \end{equation}

        Once the noise predictor, $\rvepsilon_{\theta}(\rvx_t,t)$, is trained, we can generate new samples $\rvx_0 \sim q_\mathrm{data}(\rvx_0)$ by simulating the reverse process in Eq.~\eqref{eq:backward_traj}.
        We first draw a starting sample $\rvx_T \sim p_T(\rvx_T)$ from the Gaussian noise distribution and then iteratively apply
        \begin{equation}
            \rvx_{t-1} =  \frac{1}{\sqrt{\alpha_t}} \bigl( \rvx_t - \frac{\beta_t}{\sqrt{1-\bar{\alpha}_t}} \rvepsilon_{\theta}(\rvx_t,t) \bigr) +  \sigma_t \bar{\rvepsilon}, \quad \bar{\rvepsilon} \sim \mathcal{N}(\boldsymbol{0} , \mI),
            \label{eq:DDPMReverse}
        \end{equation}
        for $t=T,T-1,\dots,1$, which progressively removes the noise from the sample to denoise it, in the optimal case ending with a sample $\rvx_0$ from the target $q_\mathrm{data}(\rvx_0)$ after $T$ reverse steps (see Figure~\ref{fig:EDM}, learned reverse generative process, right to left).
        This mimics molecular structure optimization by following atomistic forces.
        Here, the noise prediction $\rvepsilon_{\theta}(\rvx_t,t)$ defines the opposite of the force directions that minimize the energy and the scaling terms that depend on $\alpha_t$ and $\beta_t$ determine the magnitudes and the step sizes used at each optimization step.
        The noise $\bar{\rvepsilon}$ added at each denoising step results in a stochastic optimizer instead of a deterministic one, which can be helpful in the case of the presence of many shallow local minima.
        However, because noise prediction $\rvepsilon_{\theta}(\rvx_t,t)$ requires the diffusion time step $t$ as an input, diffusion models can only be used on structures with known noise level.
        For data generation, they start with samples from the known noise distribution at $t=T$, i.e. input structures that are pure noise.
        For molecular relaxation, in contrast, the non-equilibrium input structure can have an arbitrary level of perturbation such that the suitable initial time step is unknown, making the application of the standard plain diffusion models infeasible.

    \subsection{MoreRed: Molecular relaxation by reverse diffusion}
        We therefore introduce Molecular Relaxation by Reverse Diffusion (MoreRed) as a diffusion-based approach to finding minima on a PES.
        MoreRed reframes molecular relaxation as a denoising problem solved using a learned reverse diffusion process, where non-equilibrium molecular structures are considered as diffused noisy versions of their equilibrium counterparts.
        While diffusion models were initially designed to generate novel samples from complete noise $\rvx_T \sim p_T(\rvx_T)$, we adapt them to be applicable in this denoising framework, where we 
        initiate the reverse process from a noisy sample $\rvx_t \sim q_t(\rvx_t)$ at an arbitrary diffusion time step $t < T$ to reconstruct the nearest $\rvx_0$.
        Taking Figure~\ref{fig:EDM} as an illustration, the objective is to initiate the reverse process from any step $t$ within the trajectory where the structure of the yet noisy sample $\rvx_t$ remains identifiable, such as the third noisy structure from the left, and perfectly reconstruct the initial noiseless structure $\rvx_0$ on the far left.
        In contrast, starting from the complete noise sample $\rvx_T$ on the far right would yield different relaxed structures in repeated denoising attempts because no structure is apparent in the input. 
    
       Using the setup explained in section~\ref{sec:back_diff}, MoreRed learns the distribution of \emph{equilibrium} molecular structures as the target data distribution $q_\mathrm{data}(\rvx_0)$, and smoothed versions of it as the latent distributions,
        \begin{align}
            q_t(\rvx_t)  =   \int q(\rvx_t|\rvx_0)q_\mathrm{data}(\rvx_0) d\rvx_0,
            \qquad 
            \mbox{ for }
            t \in [1, T].
            \notag
        \end{align}
        This amounts to learning a pseudo PES, $\widetilde{E}_t = -\log q_t(\rvx)$, that depends on $t$ when diffusion models are trained to predict the noise direction \cite{song2021score}. This is conceptually similar to MLFF models, which implicitly learn the PES when trained on forces, where the diffusion noise could intuitively be seen as the opposite of the forces. 
        Yet, as depicted in Figure~\ref{fig:illustration}, $\widetilde{E}_t$ is much simpler than the physical potential energy $E_\mathrm{pot}(\rvx)$ that the existing MLFF models need to learn.
        Furthermore, MoreRed exhibits superior data efficiency compared to MLFF models, because \emph{it requires only equilibrium structures}.
        The whole input space of non-equilibrium structures, including physically non-plausible structures, is simply covered by the efficient forward diffusion process that adds Gaussian noise, as explained in section~\ref{sec:back_diff}.
        In contrast, for MLFF models to be reliable for any possible input structure, a large training dataset with many non-equilibrium structures derived from the physical PES is required, which can be infeasible to generate.
        However, if trained on extensive labeled data, MLFFs become applicable for molecular dynamics simulations.
        We note that MoreRed, on the contrary, \emph{only} targets molecular relaxation and is, at this stage, not intended for molecular dynamics simulations in its current form.
        
        We design our diffusion model such that the distribution is invariant with respect to rotations $\mathcal{R}(\cdot)$ and translations $\mathcal{T}(\cdot)$, i.e. $p_{\theta}(\mathcal{T} (\mathcal{R} (\rvx_t)) = p_{\theta}(\rvx_t)$.
        To guarantee translational invariance, we center the atomic positions after each forward or reverse step.
        As proven by Xu \etal~\cite{geodiff}, rotational invariance of the marginal distributions $p_\theta(\rvx_t)$ is achieved by using an invariant prior $p_T(\rvx_T)=\mathcal{N}(\bfzero, \mI)$, and an equivariant transition probability $p_\theta(\rvx_{t-1}| \rvx_t)$, which amounts to using an equivariant noise model  $\rvepsilon_{\theta}(\rvx_{t}, t)$.
        Therefore, we adopt the equivariant message passing architecture PaiNN~\cite{painn}, which allows us to directly predict equivariant tensor properties, such as the noise $\rvepsilon_{\theta}(\rvx_{t}, t)$, as well as invariant scalar properties.
        
        To perform reverse diffusion starting from a non-equilibrium molecular structure $\tilde{\rvx}$ at an arbitrary noise level, i.e. not sampled from the prior noise distribution $p_T(\rvx_T)$, it is necessary to set the initial diffusion time step $t<T$ for the reverse process appropriately.
        A starting time step that does not match the deviation of the noisy input structure from the data manifold of equilibrium structures would lead to inaccurate predictions of the noise direction and an incorrect number of denoising steps.
        Consequently, successful relaxation would not be possible.
        To address this issue, we introduce a time step predictor as a novel extension for diffusion models in the subsequent subsection.
        
    \subsection{Diffusion time step prediction}
    \label{sec:DiffusionTimePrediction}
        
        To identify the noise level of non-equilibrium molecular structures that we want to relax, we train a neural network $\tau_{\Theta}(\rvx_t)$ parametrized by $\Theta$ to predict the diffusion time step by minimizing the following loss:
        \begin{equation}
        \label{eq:L_dtp}
                L_{\mathrm{DTP}} = \E_{t \sim \mathcal{U}(1,T),\rvx_0 \sim q_\mathrm{data}, \rvepsilon_t \sim \mathcal{N}(\boldsymbol{0} , \mI)}  \bigl[ ( \tau_{\Theta}(\rvx_t) - a(t) )^2 \bigr],
                \mbox{ where }
                \rvx_t = \sqrt{\bar{\alpha}_t}\rvx_0 + \sqrt{1-\bar{\alpha}_t} \rvepsilon_t, 
        \end{equation}
        $a(t)$ is a monotonic function to scale the output, e.g. $a(t) = t/T$, and $t$ is sampled uniformally between $1$ and $T$.
        Analogous to the noise estimator $\rvepsilon_{\theta}(\rvx_t,t)$, we again adopt the neural network architecture PaiNN \cite{painn} for the model $\tau_{\Theta}(\rvx_t)$. 
        However, in this case, we use the scalar features to predict the time step, considering it as an invariant quantity similar to energy.
        In the following, we provide a theoretical argument on why an accurate prediction of the time step is feasible.
        Empirical evaluations of the time step prediction performance are found in the results section~\ref{sec:Exp.DiffusionTimePrediction}.

        In Eq.~\eqref{eq:x_t}, the latent distribution $q_t$ at time step $t$ is derived by applying isotropic Gaussian noise to the training data points representing equilibrium structures in the input space, denoted as $\rvx_0^{(i)} \sim q_\mathrm{data}(\rvx_0)$, where $i$ represents the index of the training data points. This process transforms each $\rvx_0^{(i)}$, which is a Dirac delta function, into a Gaussian distribution, $\mathcal{N}(\rvx_t^{(i)}; \sqrt{\bar{\alpha}_t}\rvx_0^{(i)}, (1-\bar{\alpha}_t)\mI)$. Therefore, considering that the equilibrium structures, $\rvx_0 \sim q_\mathrm{data}(\rvx_0)$, are isolated from each other in the input space, up to the symmetry operations, $q_t$ essentially forms a mixture of Gaussians with each Gaussian component centered around one of the training equilibrium structures, $\rvx_0^{(i)}$.
        Moreover, the variance term $(1-\bar{\alpha}_t)$ of these Gaussian components, which is defined by the diffusion noise schedule, increases monotonically with the diffusion time step $t$. Consequently, predicting the time step $t$ from a sample $\rvx_t^{(i)} \sim q_t(\rvx_t)$ amounts to estimating its noise level, $(1-\bar{\alpha}_t)$, or the distance from $\rvx_0^{(i)}$.
        This estimation is feasible when the dimension $D$ of the input space is large and the different mixture components do not overlap, due to the following reasons.
        
        Let us transform the Gaussian distribution with variance $(1-\bar{\alpha}_t)$ from the Euclidean to the polar coordinate system. By marginalizing out the polar directions, we can compute the marginal distribution over the (scaled) radius $\tilde{r} = \frac{r}{\sqrt{D (1-\bar{\alpha}_t)}}$ as:
        \begin{equation}
            p(\tilde{r}) =  \frac{ D^{D/2}\tilde{r}^{D-1}}{ {{2^{D/2-1}}}  \Gamma(D/2)  } \exp\left(-\frac{D  \tilde{r}^2}{2 } \right) , 
        \label{eq:GaussianPolar}
        \end{equation}
        where $\Gamma(\cdot)$ denotes the Gamma function. 
        As discussed in Bishop \etal~\cite{bishopPRML}, for large $D$, $p(\tilde{r})$ has a sharp peak at 
        $\hat{r} \approx 1$, as illustrated in the left plot in Figure~\ref{fig:DiffusionTimePredictionPerformance} in the supplementary information.
        This implies that each Gaussian component of $q_t$ at time step $t$ represents a sphere centered at a training sample $\rvx_0^{(i)}$, i.e. its density, represented by the set of diffused samples $\rvx_t^{(i)}$, is concentrated in a thin shell at radius $\hat{r} \approx 1$. Therefore, most of the samples $\rvx_t^{(i)}$ have similar distance from $\rvx_0^{(i)}$. Accordingly, assuming that the model can learn the data manifold, the distance, which corresponds to the noise level $(1-\bar{\alpha}_t)$, is easy to identify from a single sample $\rvx_t^{(i)}$, as long as the noise level is small such that the mixture components (spheres) do not overlap.
        When the noise level increases, the diffused samples $\rvx_t^{(i)}$ from different training data points $\rvx_0^{(i)}$ overlap with each other. This overlap makes the estimation of the diffusion time step difficult, as indicated by the right plot in Figure~\ref{fig:DiffusionTimePredictionPerformance}. Further explanation and discussion are provided in~\ref{sec:time_step_Theory} in the supplementary information, where we also show empirical evidence together with the derivation of Eq.~\eqref{eq:GaussianPolar}.

    \subsection{Variants of reverse diffusion}
    \label{sec:Variants}
    
        We compare three variants of MoreRed that differ in how they handle the diffusion time step prediction (section \ref{sec:Exp.DiffusionTimePrediction}).
        In the first variant, called \textit{MoreRed initial time prediction} (MoreRed-ITP), 
        only the initial diffusion time step, defining the start of the denoising process, is predicted.
        Given a non-equilibrium structure $\tilde{\rvx}$, MoreRed-ITP estimates an appropriate starting time step, $\hat{t} = \tau_{\Theta}(\tilde{\rvx})$, sets $\rvx_{\hat{t}} =  \tilde{\rvx}$, and performs the iterative update described in Eq.~\eqref{eq:DDPMReverse}
        for $t = \hat{t}, \hat{t}-1, \ldots, 0$, instead of starting from $t=T$.

        As a second variant, we use a more flexible process where the time step prediction is performed before \emph{every} denoising step instead of only at the start.
        We call this approach \textit{MoreRed adaptive scheduling} (MoreRed-AS).
        It iterates through a time-adaptive version of  Eq.~\eqref{eq:DDPMReverse}:
        \begin{align}
            \hat{t} = \tau_{\Theta}(\tilde{\rvx}), \quad\quad
            &
            x_{\hat{t}-1} = \frac{1}{\sqrt{\alpha_{\hat{t}}}} \bigl( \rvx_{\hat{t}} - \frac{\beta_{\hat{t}}}{\sqrt{1-\bar{\alpha}_{\hat{t}}}} \rvepsilon_{\theta}(\rvx_{\hat{t}},{\hat{t}}) \bigr) +  \sigma_{\hat{t}} \bar{\rvepsilon}, \quad \bar{\rvepsilon} \sim \mathcal{N}(\boldsymbol{0} , \mI)
        \label{eq:MoreRed-AS}
        \end{align}
        In contrast to the fixed schedule $t,t-1,\dots,0$, which always decreases towards $0$ by a one-step decrement, this adaptive approach allows the denoising process to move back and forth in the trajectory.
        In this way, errors in the noise prediction $\rvepsilon_\theta(\rvx_t, t)$, which lead to unexpected noise levels in the subsequent structure, can be compensated.
        For instance, if after one denoising step the resulting sample has less noise and converges faster than expected, the prediction $\hat{t}$ will be smaller than $t-1$ to jump more than 1 step towards $0$.
        If, on the other hand, the resulting sample has a higher noise level than expected, the prediction $\hat{t}$ will be higher than in the previous step.
        Similar to classical molecular relaxation methods, we define a convergence criterion for stopping the adaptive denoising process, i.e. we require time step predictions smaller than a threshold $\hat{t} \leq \underline{t}$.
        
        A third variant,
        \textit{MoreRed joint training} (MoreRed-JT), uses the same adaptive reverse diffusion process as described in Eq.~\eqref{eq:MoreRed-AS} for MoreRed-AS but differs in the model definition and training.
        For MoreRed-AS we employ two separate neural networks with separate backbone representations, where one is used to predict the noise $\rvepsilon_{\theta}$ and the other one to predict the time step $\tau_{\Theta}$.
        For MoreRed-JT, we use one neural network as a shared backbone representation, and we add two prediction heads on top, one for the noise and one for the time step.
        This forces the noise and the time step heads to learn a joint molecular representation. 
        We train this joint network by minimizing the joint loss $L_{\mathrm{joint}} = \eta L_{\mathrm{DDPM}} + (1 - \eta) L_{\mathrm{DTP}}$, for $\eta \in [0, 1]$ defining a trade-off between the two losses and combining Eq.~\ref{eq:DDPMLoss} and Eq.~\ref{eq:L_dtp}.
        In the supplementary information, we provide details for the training in Algorithm~\eqref{alg:training} and for the sampling in Algorithm~\eqref{alg:sampling} in~\ref{app:morered}, and for the models in~\ref{app:Architecture}.

\section{Results and discussion}

    An integral part of relaxation with MoreRed is the time step predictor.
    It estimates how far the non-equilibrium input is away from the learned data manifold of equilibrium structures, which determines the appropriate time step for the reverse diffusion process and the number of denoising steps.
    Therefore, we first evaluate the time step predictor.
    Subsequently, we validate the relaxation performance of MoreRed with respect to the root-mean-square deviation (RMSD) and DFT energy. We compare the results against those obtained from a baseline MLFF, as well as from the FF model MMFF94~\cite{halgren1996merck} and the semi-empirical model GFN2-xTB~\cite{bannwarth2019gfn2} in several experiments of molecular relaxation.

    \subsection{Diffusion time step prediction performance}
    \label{sec:Exp.DiffusionTimePrediction}
        
        \begin{figure}[t]
            \begin{center}
            \includegraphics[width=1.\linewidth]{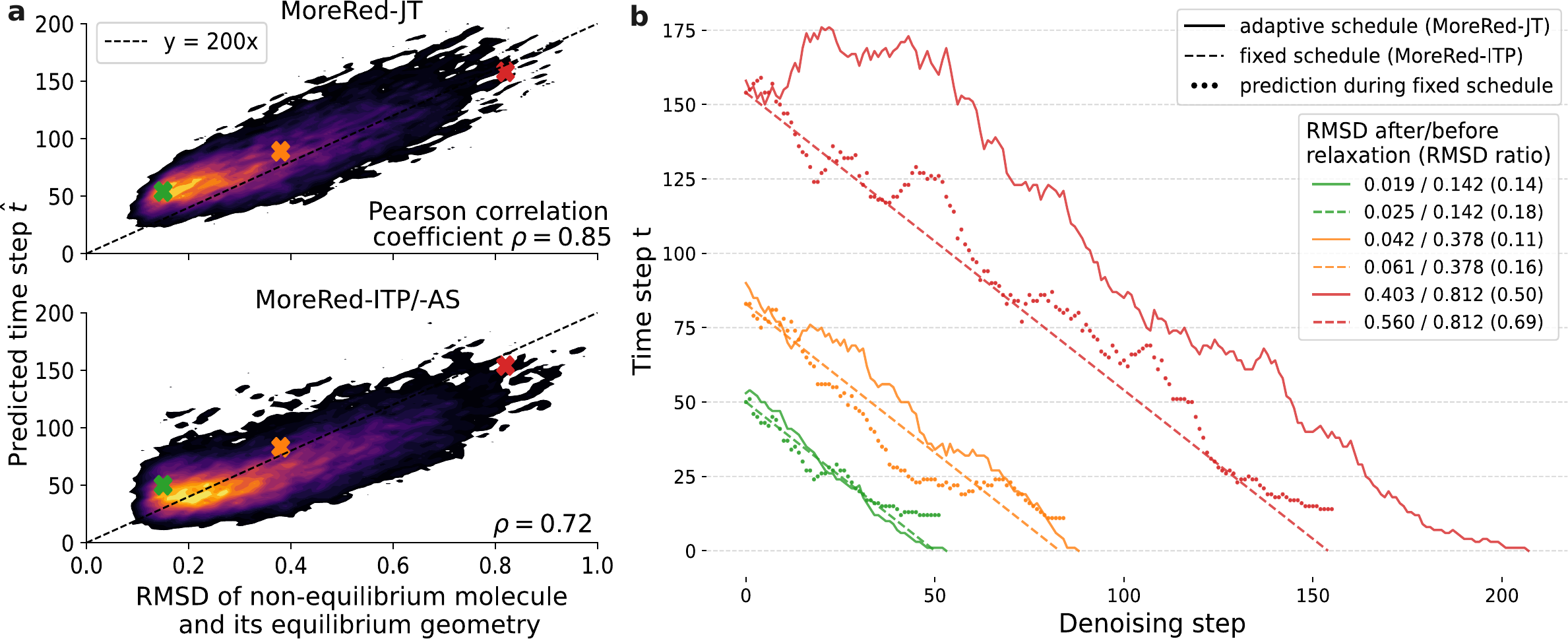}
            \end{center}
            \caption{
            \textbf{a}: 
            Density plots of the 
            RMSD of 10 000 non-equilibrium test structures from QM7-X 
            and their equilibrium structures 
            vs. the initial diffusion time step, $\hat{t}$, 
            predicted by MoreRed-JT (top)
            and MoreRed-ITP/-AS (bottom).
            A brighter color (yellow) indicates a higher density of scatter points,
            a darker color (purple) indicates a lower density.
            For both, we show the Pearson correlation coefficient, $\rho$.
            The three crosses (red, orange, green)
            mark the structures for which a time step trajectory is shown in plot \textbf{b}
            in matching colors.          
            \textbf{b}: Comparison of time step trajectories for relaxation of three non-equilibrium test structures from QM7-X in red, orange, and green (matching the crosses in plot \textbf{a}).
            MoreRed-JT follows an adaptive schedule (the time step and noise are predicted at each denoising step; solid lines) and MoreRed-ITP follows a fixed schedule (the time step is only predicted at the initial denoising step; dashed lines). The dots show the predicted (but not utilized) time steps for MoreRed-ITP during denoising with the fixed schedule.
            All two methods start from a comparable initial time step for each non-equilibrium structure.
            The inset box shows the RMSDs between the reference equilibrium geometry and the non-equilibrium structure after and before relaxation, respectively.}
        \label{fig:rmsd_corr_Traj}
        \end{figure}

        The diffusion time step predictor determines the starting step $\hat{t}$ of the reverse diffusion for molecular relaxation.
        The further a non-equilibrium structure deviates from its equilibrium structure, 
        the more denoising steps are required,
        which means that a higher starting step $\hat{t}$ should be predicted.
        As can be seen in Figure~\ref{fig:rmsd_corr_Traj}a, the predicted starting steps correlate well with the RMSDs between the non-equilibrium and equilibrium test structures from QM7-X.
        This is a notable insight, as the non-equilibrium structures in QM7-X stem from DFTB normal-mode displacements of equilibrium geometries.
        All non-equilibrium examples used for training the time predictors, on the other hand, stem from diffusing equilibrium structures with Gaussian noise.
        Nevertheless, the time step predictors reliably predict $\hat{t}>0$ for all the 10 000 test structures, highlighting the robustness in identifying non-equilibrium structures even if they do not contain Gaussian noise.
        Moreover, we observe that a joint model for predicting both time step and noise, as in MoreRed-JT, leads to fewer outliers in the predictions of $\hat{t}$ and, consequently, a higher Pearson correlation coefficient ($\rho=0.85$; top) than the separately trained time step predictor in MoreRed-ITP/AS ($\rho=0.72$; bottom).
        
        Usually, diffusion models follow a fixed schedule where the time step $t$ is reduced by one after each denoising step until it reaches $t=0$.
        In MoreRed-ITP, we follow such a fixed schedule and only use the time step predictor once to obtain a suitable starting step $\hat{t}$.
        As described in Eq.~\eqref{eq:MoreRed-AS} before, the other two variants, MoreRed-AS/-JT, utilize the time step predictor at \emph{every} denoising step to obtain a new time step estimate $\hat{t}$.
        This results in an adaptive schedule, where the relaxation ends after a variable number of steps, as soon as $\hat{t}=0$ is predicted.
        Figure~\ref{fig:rmsd_corr_Traj}b shows the merit of this approach, where the time step trajectories during geometry relaxations of three different test structures are plotted in red, orange, and green.
        When following the fixed schedule (dashed lines),
        errors can occur and accumulate~\cite{song2021score}.
        If not corrected, they lead to a mismatch between the true noise level in the structure and the time step $t$.
        Therefore, the relaxation may end before the sample reaches the equilibrium geometry.
        This can be observed for all three examples in the plot:
        The predicted but not enacted time step values associated with the fixed schedule (dotted lines) show a high value when the denoising with the fixed schedule ends, as
        the dashed line reaches $t=0$.
        In contrast, the adaptive schedule (solid lines), MoreRed-JT in this case, can account for such errors by adapting $\hat{t}$ at each denoising step.
        After converging to $\hat{t}=0$, the relaxed structures are significantly closer to the ground truth equilibrium geometry than those obtained with the fixed schedule, despite starting from comparable initial time steps (see the RMSD values after relaxation in the right-hand side box of Figure~\ref{fig:rmsd_corr_Traj}b).
        In Figure~\ref{appfig:rmsd_corr_Traj_different} in the \ref{sec:details_improved_gen} we present similar trajectories
        to Figure~\ref{fig:rmsd_corr_Traj}b but for three cases where the predicted initial time steps are very different for both methods. It highlights that even when MoreRed-ITP starts from a higher initial step, it is still outperformed by MoreRed-JT.

        To conclude, we find that the time step predictor accurately identifies non-equilibrium structures, where larger time steps are predicted if the RMSD from the equilibrium structure is larger.
        Moreover, employing an adaptive schedule that utilizes the time step predictor to determine the time step at every denoising iteration proves beneficial compared to constantly decreasing the time step at a fixed rate.
        We provide further experiments in~\ref{app:data_generation} in the supplementary information, where we show that the time step predictor also significantly enhances the performance of diffusion models in the original task of novel structure generation from complete noise.

    \subsection{Molecular relaxation performance}
    \label{sec:Exp.MoleculeRelaxation}

        In the following, we compare the relaxation performance of MoreRed with a baseline MLFF, as well as the FF method MMFF94,~\cite{halgren1996merck} and the semi-empirical method GFN2-xTB~\cite{bannwarth2019gfn2}.
        As a molecular representation, all variants of MoreRed as well as the baseline MLFF use the same equivariant message passing neural network architecture PaiNN~\cite{painn} as implemented in the open-source software package SchNetPack~\cite{schuett2018schnetpack, schutt2023schnetpack}.
        Details on the models' architectures including the hyperparameters are shown in the supplementary information~\ref{app:arch_and_hyperparams}.
        While training MoreRed requires only different unlabeled equilibrium structures, MLFFs have to be trained on non-equilibrium structures as well and require the energies and forces as labels.
        For this reason, we use QM7-X~\cite{qm7x}, the only labeled dataset that provides both equilibrium (42 000) as well as the corresponding non-equilibrium structures (100 each) of different chemical compositions, enabling the training of both MoreRed and MLFFs.
        However, the dataset has a mismatch between the computational methods employed for finding the equilibrium structures (DFTB3+MBD) and those for computing their energy and force labels (PBE0+MBD), introducing a challenge in comparing the performance of MoreRed and the MLFF.
        On one hand, evaluating the geometric deviation, such as RMSD, between a structure after molecular relaxation and the corresponding equilibrium structure reported in the QM7-X is in favour of MoreRed.
        This is because it is trained to learn the data manifold of these reference structures, which were determined using DFTB3+MDB.
        On the other hand, comparing the DFT energies of the structures after relaxation and the equilibrium structures in QM7-X using PBE0+MBD reference calculations favours the MLFF because it is trained on the energies and forces resulting from PBE0+MBD calculations.
        Therefore, we must carefully integrate our findings on both metrics before drawing conclusions.

        For our evaluation, we have reserved a test set of 6504 reference equilibrium structures from QM7-X which are not utilized for training the neural networks (see supplementary information~\ref{app:qm7x} for details).
        To cover a wide range of test examples, 
        we sort the 100 non-equilibrium structures of each reference structure based on the RMSD to their equilibrium geometry and choose three of them: the closest, one from the middle, and the most distant.
        This results in almost 20 000 non-equilibrium test inputs, $\tilde{\rvx}$, for relaxation.
        Note that whenever we compute the RMSD, 
        the rotation and translation of structures are aligned. 
        For molecular relaxation, we employ Open Babel's~\cite{Openbabel} built-in routines and optimizer for MMFF94. Additionally, we utilize the L-BFGS optimization algorithm implemented in ASE \cite{ase-paper} for both the MLFF and the semi-empirical GFN2-xTB, where we set a convergence threshold of $f_{\text{max}} = 1.15 \cdot 10^{-3}~\text{kcal/mol/\AA}$ for the forces and use the tblite\footnote{https://github.com/tblite/tblite} Python interface for the official GFN2-xTB model implementation. Besides, we set a maximum number of relaxation steps $T=1000$ for all methods, including MoreRed, and a convergence criterion of $\hat{t} \leq 0$ for MoreRed-AS/-JT.
        
        \begin{figure}
            \centering
                \includegraphics[width=1.\linewidth]{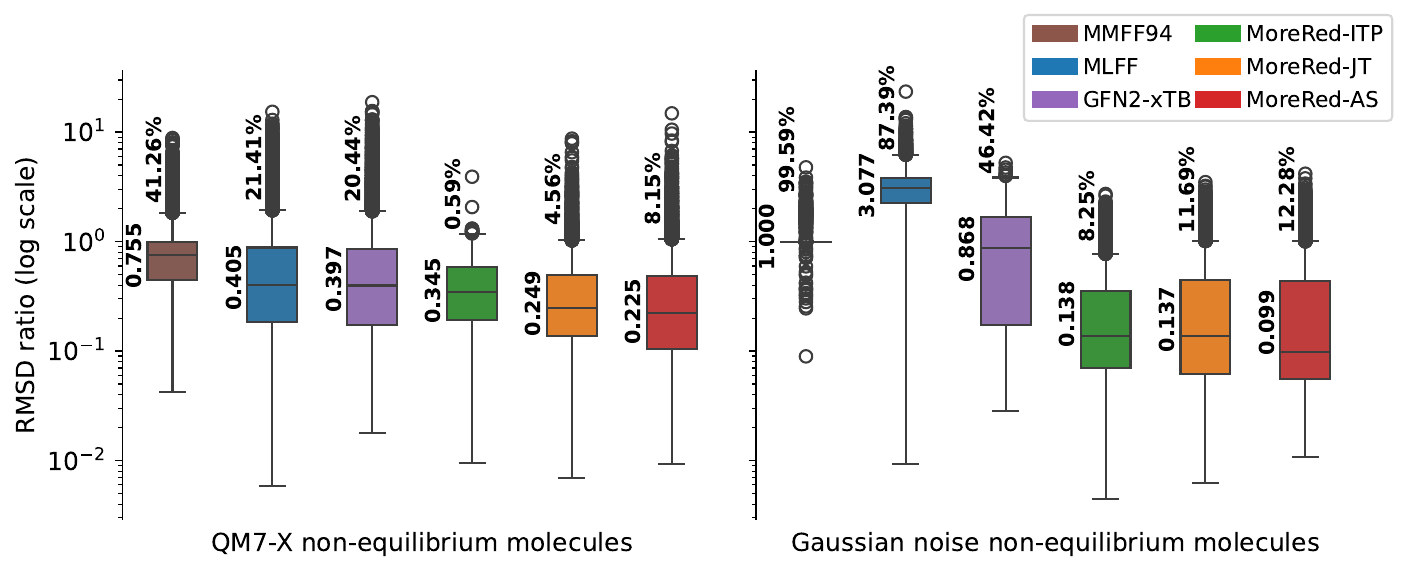}
            \caption{
                The RMSD ratios, i.e. the RMSD after relaxation divided by the RMSD of the non-equilibrium inputs, of structures relaxed with the baselines and the MoreRed variants, trained on the DFTB3+MBD equilibrium structures reported in QM7-X, for 20 000 non-equilibrium structures from the QM7-X test split (left) and 6500 non-equilibrium structures obtained by adding Gaussian noise over 250 forward diffusion steps to equilibrium structures from the QM7-X test split (right). 
                The median values and the percentage of failure cases, i.e. the cases where the RMSD ratio $\geq 1$, are shown to the left of each box plot in bold.
                In Figure~\ref{fig:confusion_matrix}, we provide examples of relaxed structures comparing both MoreRed and the MLFF model.
            }
        \label{fig:relaxation}
        \end{figure}

        We first evaluate the geometric deviation of structures relaxed with MoreRed and the different baselines from the reference equilibrium structures in QM7-X.
        To this end, we calculate the RMSD ratio, which is the RMSD of the reference structure from the test structure after relaxation divided by the RMSD of the reference structure from the test structure before relaxation.
        It captures to which extent the non-equilibrium test structure was brought closer to the reference equilibrium structure.
        We define \emph{failure cases} as cases where the RMSD ratio exceeds 1, which means that the RMSD increased during molecular relaxation.
        Those failures correspond to cases where the structure diverges or the relaxation converges to a different local minimum in the PES.
        Figure \ref{fig:relaxation}a (left) shows boxplots of the RMSD ratio for the MLFF model, the FF method MMFF94, the semi-empirical method GFN2-xTB, as well as MoreRed-ITP, MoreRed-JT, and MoreRed-AS.
        The lowest median RMSD ratios and the lowest percentages of failure cases are all achieved by MoreRed.
        There are almost no failure cases for the model with a fixed time step schedule, MoreRed-ITP, and the median RMSD ratio is particularly low for the two variants with an adaptive time step schedule, MoreRed-JT/-AS.
        These low ratios translate well to low absolute RMSDs between the relaxed structures and the reference structures, where the variants MoreRed-ITP/-JT/-AS achieve a median RMSD of 0.12~\AA, 0.06~\AA, and 0.05~\AA, respectively.
        Further details and results based on the absolute RMSD are provided in Figure~\ref{appfig:raw-rmsd} in the supplementary information~\ref{app:extended_relaxation_experiments}.
        Our findings show that the MoreRed variants, especially when using an adaptive schedule, excel in reliably bringing the test structures close to the reference equilibrium structures.
        The classic FF method MMFF94 shows the highest number of failure cases, which is above 40\%, and has the worst median RMSD ratio.
        Interestingly, the MLFF and GFN2-xTB show very similar performance to each other, with more than 20\% failure cases and a median RMSD ratio close to 0.4.
        However, we note that the baseline methods might capture minima from slightly different PES than the one described by the reference structures, for instance, due to the discussed structure-label mismatch in QM7-X.
        For a more comprehensive understanding, we first test the robustness of all approaches and then proceed to evaluate the DFT energy levels of the relaxed structures.

        We assess the robustness of the methods on synthetically generated inputs by diffusing equilibrium test structures from QM7-X with $250$ forward diffusion steps. 
        We ensure that the resulting median RMSD between the diffused configurations and the equilibrium test structures is within the range of the RMSD between the non-equilibrium structures and the equilibrium test structures from QM7-X.
        Figure~\ref{fig:relaxation}a (right) shows the RMSD ratios after molecular relaxation of the diffused structures.
        For all MoreRed variants, the median RMSD ratio further improves compared to relaxing the QM7-X test structures, which is expected as our method is trained to denoise diffused structures.
        However, there is an increase in failure cases, which we attribute to more often ending up in equilibrium states different from the reference geometry.
        We hypothesize that this is caused by the physically less plausible deviations in the diffused structures compared to the normal mode displaced structures in QM7-X.
        The existence of physically less plausible deviations is supported by the results of the baseline methods, where we observe a clear deterioration in performance.
        MMFF94 completely fails to handle input structures perturbed with Gaussian noise and, in nearly all cases, just returns the non-equilibrium input structure, leading to a median RMSD ratio of $1$ and resulting in $99.6$\% failure cases.
        The median RMSD ratio of GFN2-xTB as well as its percentage of failure cases are more than doubled.
        Most notably, the MLFF fails to get closer to the reference geometry in almost 90\% of the cases.
        For the MLFF, this is expected as the training data distribution of non-equilibrium structures from QM7-X does not cover all the chemical space, including the Gaussian diffused inputs.
        Therefore, relaxation often completely fails, leading to disconnected structures even if the input structure does not appear to be overly distorted.
        We show an example of this in Figure~\ref{fig:relaxation_examples}b, with a detailed discussion in~\ref{app:examples_relaxed_structures}, in the supplementary information.
        This means that, although the MLFF uses 100 times more training data than MoreRed, it cannot easily be transferred to relax the diffused structures.
        In contrast, MoreRed performs well on both the diffused samples and on the non-equilibrium structures from QM7-X albeit requiring only the unlabeled equilibrium structures for training.
        Accordingly, the diffusion training scheme leads to a more robust method for relaxation that can be used for input structures that are obtained from different sources, e.g. different datasets, various empirical FFs, or other generative models.
        This also means that MoreRed will oftentimes find a reasonable structure even if the input was physically not plausible, which should be considered by practitioners using the method.

        \begin{figure}[t]
            \centering
            \includegraphics[width=1\linewidth]{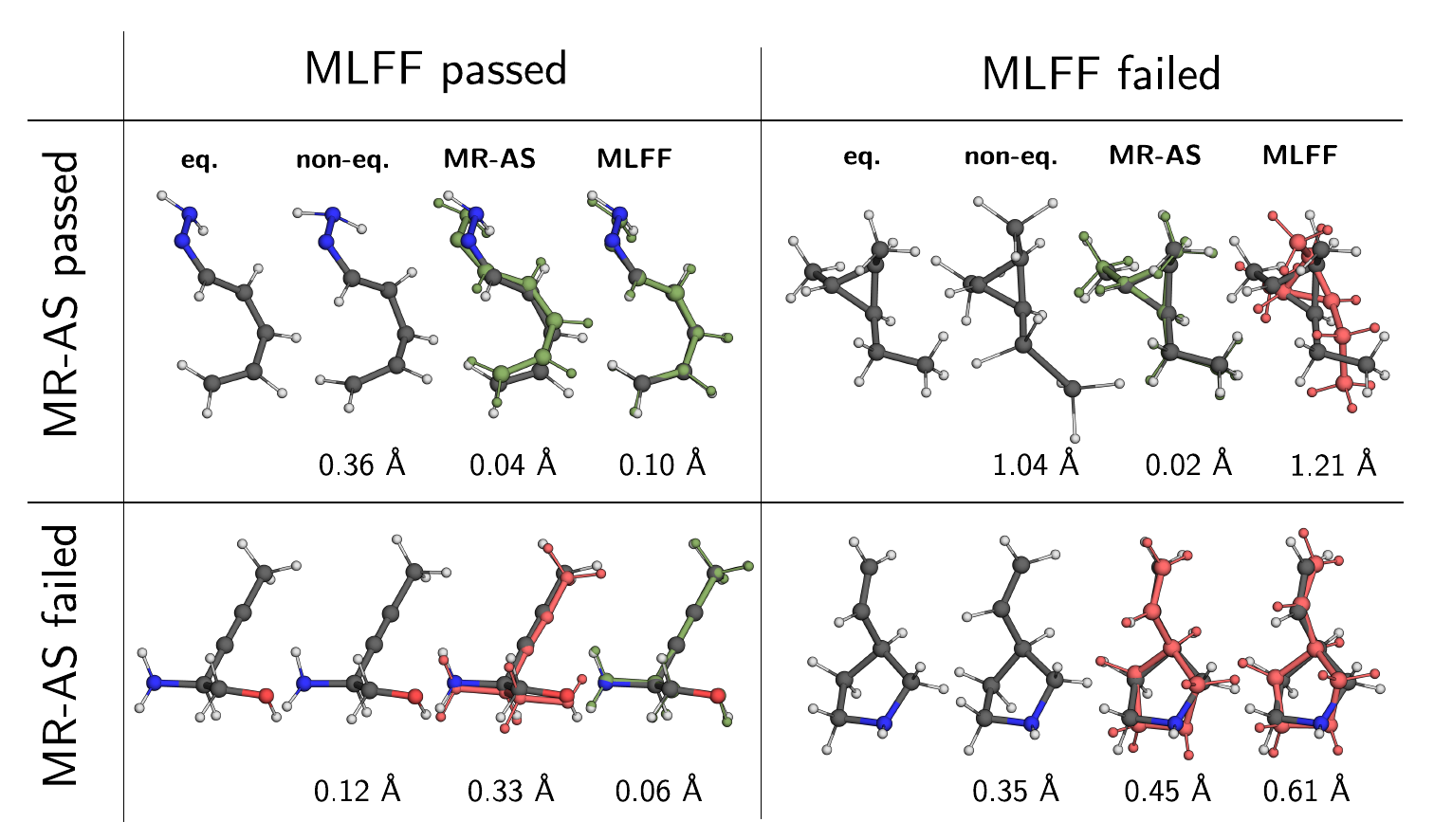}
            \caption{
            A pairwise comparison of molecular structures from QM7-X that were relaxed with 
            both MoreRed-AS (MR-AS) and MLFF. 
            In each block, 
            from left to right,
            the equilibrium structure is shown (eq.), 
            followed by the non-equilibrium structure (non-eq.)
            and the structures relaxed by MoreRed-AS and MLFF,
            all labelled with their respective RMSD to the equilibrium structure in angstrom (\AA).
            The first two structures follow the CPK colouring,
            i.e. white for hydrogen, gray for carbon, blue for nitrogen, and red for oxygen.
            In the subsequent two depictions, the equilibrium structure (CPK) is superimposed with MoreRed-AS and MLFF relaxed structure. Structures with an decreased RMSD value relative to the non-equilibrium structure are shown in green, while failure cases with a higher RMSD compared to the initial structure are highlighted in salmon}. 
            \label{fig:confusion_matrix}
        \end{figure}

        For illustration, we show a series of pairwise comparisons of molecular structures from QM7-X relaxed by both MoreRed-AS and MLFF methods in Figure~\ref{fig:confusion_matrix}. 
        Each panel contains a sequence of depictions, beginning with the equilibrium structure (eq.), followed by the corresponding non-equilibrium structure, and then structures relaxed using MoreRed-AS and MLFF respectively. 
        The RMSD values, relative to the equilibrium structure, are provided for each case.
        For both models, we show examples that were successfully relaxed as well as failure cases.
        Additionally, in Figure~\ref{fig:relaxation_examples} and \ref{app:examples_relaxed_structures} in the supplementary information, we provide further examples and discussion of relaxed structures using all MoreRed variants and baseline models, including MMFF94 and GFN2-xTB.
        
\begin{figure}
            \centering
            \includegraphics[width=0.6\linewidth, height=0.5\linewidth]
            {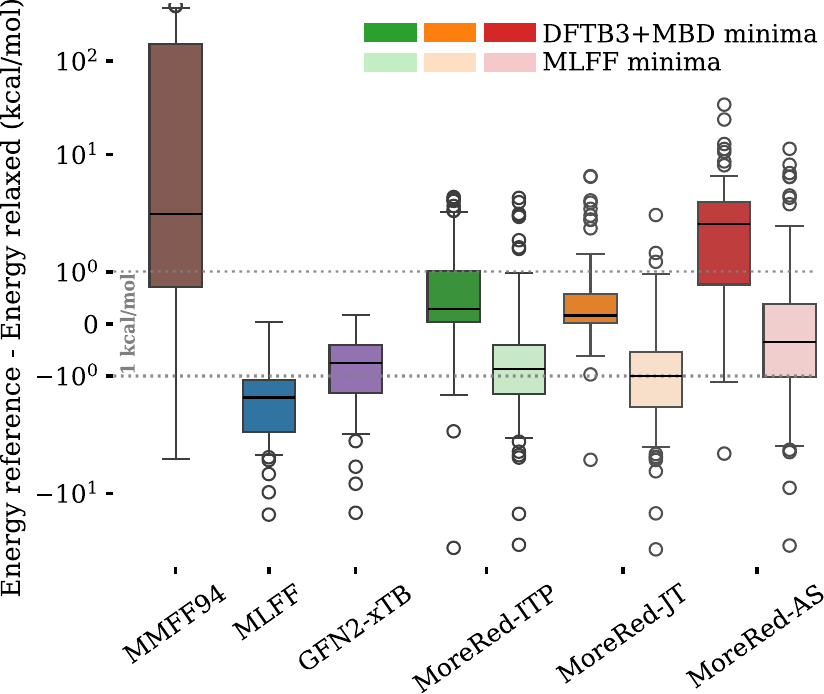}
            \caption{
            Deviation of the energy of structures resulting from relaxation with the different methods from the energy of the reference structures from QM7-X. Negative values indicate lower energy than the equilibrium reference structure.
            The DFT energy calculations were performed with PBE0+MBD and a def2-TZVP basis for 100 randomly sampled test structures from QM7-X.
            Each MoreRed variant has two versions: the darker color represents the model trained on the original DFTB3+MBD equilibrium structures from QM7-X, which have non-negligible PBE0+MBD residual forces, and uses the same relaxation results used for Figure~\ref{fig:relaxation}. On the other hand, the lighter color represents the model trained on equilibrium structures resulting from further relaxation of the QM7-X reference structures to minimize the residual forces, using the MLFF trained on the more accurate PBE0+MBD force labels.}
            \label{fig:dft_energies}
        \end{figure}

        Finally, we compare the energies of relaxed structures using DFT calculations.
        We randomly sample 100 non-equilibrium test inputs from QM7-X and compute the energy of the corresponding equilibrium reference structure as well as of the structures resulting from the relaxation of the test inputs with all methods.
        The energies are calculated with PBE0+MBD \cite{Tkatchenko2012, Ambrosetti2014} and a def2-TZVP \cite{weigend2005balanced} basis, using the PySCF \cite{Sun2015, Sun2017, Sun2020} implementation.
        The results are reported in Figure~\ref{fig:dft_energies}, where we compare the deviation of the energy of the structures resulting from relaxation with the different methods from the energy of the reference equilibrium structures reported in QM7-X.
        Positive energy differences occur when the relaxation method yields a structure with higher energy than the reference structure, and a negative difference indicates that the molecular relaxation yields a structure with lower energy than the reference.
        The structures relaxed with MMFF94 mostly have significantly larger energies than the reference structures.
        Hence, MMFF94 is clearly outperformed by MoreRed and the other baselines, as it results in the largest structural deviations in terms of the RMSD ratio and the worst energy levels in our DFT calculations.
        The MLFF, on the other hand, mostly finds structures with lower energy levels compared to the reported reference structures in QM7-X.
        According to the previously discussed structure-label mismatch in the dataset, this can be attributed to training the MLFF on energy and force labels calculated with the more accurate PBE0+MBD, whereas the reference equilibrium structures were found with DFTB+MBD. These reported equilibrium structures still have a mean and median force magnitude greater than $5$ kcal/mol/$\text{\AA}$ when calculated with PBE0+MBD instead of DFTB+MBD, i.e. they are no minima on the PES defined by PBE0+MBD. 
        The energy of structures relaxed with GFN2-xTB is higher than with the MLFF but still lower than the energy of the DFTB+MBD reference structures.
        This shows that the larger RMSD ratios found for the MLFF and GFN2-xTB stem from finding minima on a more accurate PES that is different from the one described by the QM7-X reference equilibrium structures.
        The energy levels, indicated by the darker colors in Figure~\ref{fig:dft_energies}, of structures relaxed with MoreRed-ITP/-JT, trained on the reference DFTB3+MBD minima from QM7-X, match the energy levels of these reference structures the closest. For a majority of relaxed structures, the energy is less than 1 kcal/mol higher, i.e. within chemical accuracy. 
        Only for MoreRed-AS, which had the smallest median RMSD ratio, we observe higher energy levels than expected. It also performs more reverse denoising steps than the other two variants most of the time, further reducing the structural deviation while underestimating the interatomic distances, which are more strongly penalized in energy calculations than in the RMSD metric. Overall, we observe a mismatch between the RMSD results and the DFT energies of the MLFF and all MoreRed variants that were trained on equilibrium structures from QM7-X (darker colors in Figure~\ref{fig:dft_energies}), where methods having lower RMSD ratios result in higher energies and vice versa.
        Therefore, to further investigate this mismatch and ensure a fair comparison, we further relaxed the DFTB+MBD equilibrium structures of QM7-X using the MLFF to minimize the forces and consequently the energies. We then retrained all MoreRed variants on this MLFF-relaxed dataset and used them to relax the non-equilibrium test structures from QM7-X. The resulting DFT energies, represented with light-colored boxes in Figure~\ref{fig:dft_energies}, show that MoreRed models, trained on the MLFF-relaxed structures, achieve much lower PBE0+MBD energies compared to those trained on the original DFTB+MBD minima. However, they exhibit higher RMSD ratios, as demonstrated in Figure~\ref{fig:rmsd_ratios_mlff_relaxed} in the supplementary information. This outcome aligns with the results achieved by the MLFF and supports the notion that the RMSD-energy mismatch is related to the structure-label mismatch. Specifically, the MLFF finds lower PBE0+MBD energy minima that differ from the reference DFTB+MBD minima reported in QM7-X, resulting in higher RMSD ratios. Further details on this experiment are provided in \ref{app:training_on_mlff_relaxations} of the supplementary information.

        In summary, we find that MoreRed accurately captures the data manifold of equilibrium structures as it outputs relaxed structures that are close to the reference structures in both structural deviation and in terms of energy difference.
        Moreover, using equilibrium structures with lower energy during training further improves MoreRed's energy performance.
        MoreRed outperforms the classical MMFF94 method in all of our experiments and metrics.
        While the MLFF and GFN2-xTB find structures with larger structural deviations, the obtained structures have lower energies according to reference calculations with PBE0+MBD.
        This was explained by the mismatch of computational methods employed in QM7-X to obtain equilibrium structures (DFTB+MBD) and to compute the energy and force labels (PBE0+MBD).
        Besides, MoreRed shows improved robustness to distorted inputs compared to all baseline methods, allowing it to relax structures from different sources albeit requiring only unlabeled equilibrium structures for training.
        When considering all metrics together, MoreRed-JT, which uses an adaptive schedule and predicts both the time step as well as the noise for reverse diffusion with the same neural network, performs better than MoreRed-ITP/-AS and is therefore recommended.

        Efficiency-wise, we note that training MoreRed takes $1.45$ days on average, while the MLFF model needs more than $7$ days on NVIDIA P100.
        The relaxation of one single structure with MMFF94 takes $0.022$s. A single relaxation step per structure with the MLFF, GFN2-xTB, and MoreRed takes $0.02$s, $1.5$s, and $0.03$s, respectively. In contrast to the other methods that relax one structure at a time, MoreRed relaxes the structures batchwise, which could yield an even higher speed-up on tensor units. In our experiments, using batches of 128 structures results in $0.05$s per relaxation step per batch, i.e. $0.05s/128 \approx 0.0004$s per relaxation step per structure. A more detailed analysis of the computation times can be found in the supplementary information~\ref{app:computation_Times}. Moreover, the effect of the size of the molecule on the performance and efficiency of MoreRed is studied in \ref{app:molecule_size}.

\section{Conclusion}

In this study, we introduced MoreRed, a conceptually novel and data-efficient approach for molecular relaxation employing reverse diffusion with a time step prediction component.
MoreRed learns the data manifold of equilibrium structures and accurately maps non-equilibrium structures to equilibrium structures, without the need for forces, energies, or non-equilibrium training data. 
Compared to the other tested methods, its performance in relaxing non-equilibrium structures distorted by either normal modes or Gaussian noise remains robust.

A key technical novelty of our diffusion model lies in the integration of a time step predictor, which estimates the distortion level within a molecular structure. This enables the denoising of input structures with arbitrary noise levels, extending the applicability of diffusion models. Additionally, it allows for a novel adaptive schedule, enhancing MoreRed's capability to rectify accumulated errors in the reverse denoising process. To this end, we provided both theoretical arguments and empirical evidence supporting the feasibility of time step prediction in high-dimensional spaces.
Three variants of MoreRed were introduced:
i) MoreRed-ITP (Initial Time Prediction), which estimates the distortion level for only the initial input structure;
ii) MoreRed-AS (Adaptive Schedule), which predicts a new time step for each denoising step of the reverse diffusion process, providing enhanced flexibility to move back and forth in time; and
iii) MoreRed-JT (Joint Training), which retains the adaptive schedule of MoreRed-AS but estimates both the diffusion noise and time step using one joint neural network instead of two separate ones.
Our adaptive approach not only proves beneficial for molecular relaxation but also for molecular generation tasks.

While not directly comparable due to the mismatch in computational methods used to create the dataset, MoreRed exhibits accurate structure relaxation performance with significantly fewer training points and reduced training time compared to machine learning FFs. 
The data efficiency might prove beneficial for larger systems or extensive databases where
generating a sufficient amount of accurately labeled training data, especially non-equilibrium structures, is challenging.
The results also revealed an inherent issue when using mismatched methods for structure relaxation and properties calculation, as observed in the QM7-X dataset.
While machine learning FFs produce more accurate minima by following the PBE0+MBD-based forces, MoreRed accurately learns the data manifold of the provided DFTB+MBD-based equilibrium structures.
Utilizing more accurate minima during training can further improve MoreRed's energy performance while maintaining its data efficiency advantage.

\section*{Code and data availability}
The code and its associated data are made public in Zenodo \cite{kahouli_2024_10927872} and on Github at \url{https://github.com/khaledkah/MoreRed}. The datasets utilized for training the models, namely QM7-X \cite{qm7x} and QM9 \cite{ruddigkeit2012enumeration, qm9}, are also publicly accessible.

\ack

This work was partly funded by the German Ministry for Education and Research (BMBF) as BIFOLD – Berlin Institute for the Foundations of Learning and Data (BIFOLD24B) (under refs 01IS14013A-E, 01GQ1115, 01GQ0850, 01IS18056A, 01IS18025A and 01IS18037A) and BBDC/BZML. Furthermore, Klaus-Robert M\"uller was partly supported by the Institute of Information \& Communications Technology Planning \& Evaluation (IITP) grants funded by the Korean Government (MSIT) (No. 2019-0-00079, Artificial Intelligence Graduate School Program, Korea University and No. 2022-0-00984, Development of
Artificial Intelligence Technology for Personalized Plug-and-Play Explanation and Verification of
Explanation). We thank Stefan Chmiela, Jonas Lederer, and Elron Pens for insightful discussions and feedback.

\newpage

\appendix

\section{MoreRed: details}
\label{app:morered}
    
    \subsection{Diffusion time step prediction}
    \label{sec:time_step_Theory}
        
        \paragraph{Derivation of Eq.~\eqref{eq:GaussianPolar}:}
        \label{sec:Derivation}
            
            In the polar coordinate system $(r, \bm{\varphi})$, 
            the marginal distribution of the radius $r$ of the centered isotropic Gaussian over the direction $\bm{\varphi}$ is given by
            \begin{align}
                p(r)
                &= \int p(r, \bm{\varphi}) d\bm{\varphi} \\
                &= \frac{1}{(2 \pi \sigma^2)^{D/2}} \exp\left(-\frac{r^2}{2 \sigma^2} \right) \cdot r^{D-1} S_D 
                \notag\\
                &= \frac{ r^{D-1}}{2^{D/2-1}   \Gamma(D/2) \sigma^D } \exp\left(-\frac{r^2}{2 \sigma^2} \right),
                \notag
            \end{align}
            where $S_D = \frac{2 \pi^{D/2}}{\Gamma(D/2)}$ is the surface area of the $(D-1)$-dimensional unit sphere embedded in the $D$-dimensional space, $\Gamma(\cdot)$ denotes the Gamma function \cite{bishopPRML} and $\sigma^2$ is the variance, which is equal to $(1-\bar{\alpha}_t)$ in Eq.~\eqref{eq:GaussianPolar}.
            By changing the radius variable $r$ to the scaled version, $\tilde{r} = \frac{r}{\sqrt{D} \sigma}$, we get
            \begin{align}
                p(\tilde{r}) 
                &= p(r) \frac{dr}{d\tilde{r}} \\
                &= \frac{ ( \sqrt{D}\sigma \tilde{r})^{D-1}}{2^{D/2-1}  \Gamma(D/2) \sigma^D } \exp\left(-\frac{( \sqrt{D}\sigma \tilde{r})^2}{2 \sigma^2} \right) \cdot \sqrt{D}\sigma
                \notag\\
                &= \frac{ D^{D/2}\tilde{r}^{D-1}}{2^{D/2-1}  \Gamma(D/2)  } \exp\left(-\frac{D  \tilde{r}^2}{2 } \right) ,
                \notag
            \end{align}
            which gives Eq.~\eqref{eq:GaussianPolar}.

        \begin{figure}
            \begin{minipage}{0.49\linewidth}
                \centering
                \includegraphics[width=.85\linewidth]{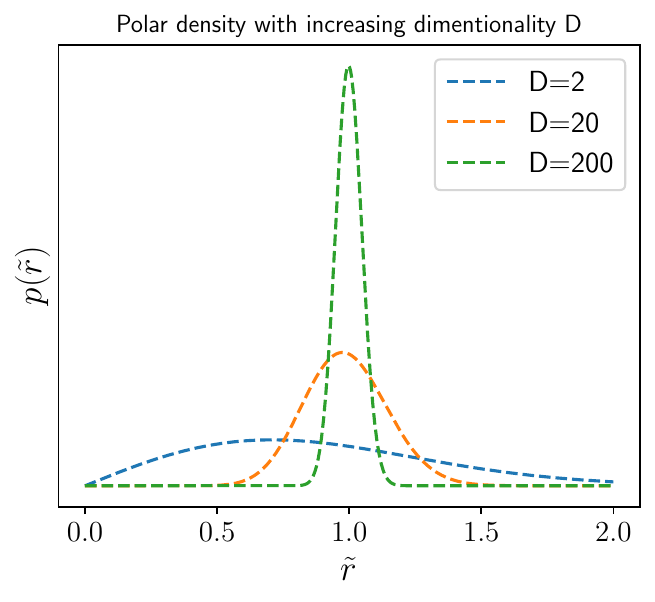}
            \end{minipage}
            \hfill
            \begin{minipage}{0.49\linewidth}
                \centering
                \includegraphics[width=1.\linewidth]{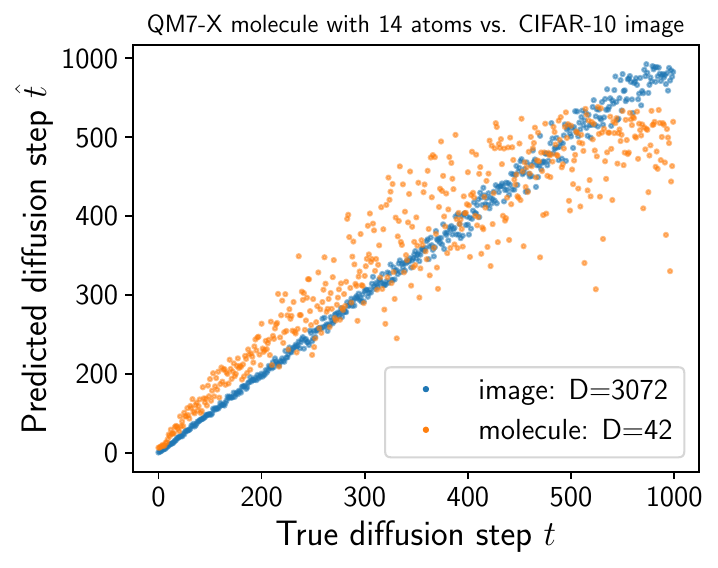}
            \end{minipage}
            \hfill
            \caption{ \textbf{Left:} Distribution of the (scaled) radius of one Gaussian component in the polar coordinate system (Eq.~\eqref{eq:GaussianPolar}), illustrating sharper density with increasing dimensionality $D$. \textbf{Right:} Diffusion time prediction performance using a molecule (orange) with 14 atoms from QM7-X, resulting in an input sample with $14 \cdot 3 =42$ dimensions, and using an image (blue) from CIFAR-10 with $32 \cdot 32 \cdot 3 = 3072$. The time prediction becomes more accurate with increasing dimensions and/or decreasing noise variance, meaning low time step $t$.}
            \label{fig:DiffusionTimePredictionPerformance}
        \end{figure}
        
        \paragraph{Discussion:} 
        \label{sec:time_pred_discussion}
        
            Our diffusion time step predictor essentially predicts the noise level of the input sample, which reduces to predicting the variance of the Gaussian component, from which the noisy input sample is drawn, using solely this \textit{single} perturbed sample. Intuitively, this is too challenging in a low-dimensional space because the distances between the mean and different samples from the same Gaussian are broadly distributed. This intuition does not apply to a high-dimensional space.
    
            As discussed in section~\ref{sec:DiffusionTimePrediction}, the marginal distribution of the radius in the polar coordinate system, provided in Eq.~\eqref{eq:GaussianPolar} and depicted in the left plot in Figure~\ref{fig:DiffusionTimePredictionPerformance},
            implies that most of the Gaussian perturbed samples lie in a thin shell with an equal distance to the center of the Gaussian. This implies that a neural network, which can learn the data manifold, can predict the variance of the perturbation noise, and consequently, the diffusion time step.
            
            One might still worry that training samples drawn from two overlapping Gaussian components will deteriorate the performance of the diffusion time prediction. Indeed, such overlapping makes the prediction harder, as can be empirically seen in the right plot in Figure \ref{fig:DiffusionTimePredictionPerformance}.
            However, with high dimensional input space, such overlapping does not significantly affect the time prediction performance for samples perturbed with small noise variance, meaning when the diffusion time step $t$ is low.
            Assume that there are two training molecules $\rvx_a, \rvx_b$ with the Euclidean distance $r = \|\rvx_a-\rvx_b\|$, and consider the Gaussian component centered at $\rvx_a$, which represents its noisy versions with standard deviation $\sigma = r$.
            Although, in this situation, $\rvx_b$ lies in the high-density shell of this Gaussian component (the bump in Figure~\ref{fig:DiffusionTimePredictionPerformance} left), the noisy samples of $\rvx_a$ are uniformly distributed all over the high $(D-1)$-dimensional shell. Therefore the probability that the Gaussian noise produces a sample close to $\rvx_b$ is extremely low.
            On the other hand, many training samples from the neighbourhood of $\rvx_b$ are fed to the diffusion time predictor as slightly noisy versions of $\rvx_b$, because of its high density.
            Accordingly, the network is trained to recognize the molecules close to $\rvx_b$ as low noise samples resulting from $\rvx_b$ without being disturbed by high noise samples from $\rvx_a$.
            This intuition can be mathematically confirmed by computing the density ratio between two Gaussian components around $\rvx_b$, i.e. $\mathcal{N}(\rvx_b + \rvepsilon; \rvx_b, \delta^2 \mI) / \mathcal{N}(\rvx_b+ \rvepsilon; \rvx_a, r^2 \mI)$ for $\| \rvepsilon\| \sim \delta \ll r$, which is extremely high unless $D$ is very small.            
            
            To conclude, the high dimensionality of the data space enables accurate diffusion time step prediction, especially for the samples close to one of the training equilibrium molecules.
            This can also be observed empirically.
            Figure~\ref{fig:DiffusionTimePredictionPerformance} (right) with orange dots shows a scatter plot of the true diffusion time vs. its prediction by our diffusion time predictor after training on the equilibrium molecules from the QM7-X dataset.  
            We also show the performance of the diffusion time predictor trained on CIFAR10~\cite{cifar10} --- a common image benchmark dataset --- where the images have a higher dimensionality than the molecule data, as blue dots.
            As discussed above, the diffusion time prediction is easier when the dimension $D$ is large, and the true diffusion time, i.e. the noise level, is small.
    
    \subsection{Algorithms}

        Algorithm~\ref{alg:training} shows the training procedure for MoreRed-JT. 
        For the other two variants, we instead train two separate architectures and use only the first part of the loss in line 7 to train the denoising model and the second part to train the time step prediction model separately.
        Algorithm~\ref{alg:sampling} describes the sampling with the adaptive MoreRed variants (AS and JT).
        MoreRed-ITP uses a fixed schedule $i=\hat{t}, \hat{t}-1,...,1$ but starts from a predicted initial time step $i=\hat{t}$ instead of a fixed value.
        
        \noindent%
        \begin{minipage}{0.49\textwidth}
        \begin{algorithm}[H]
            \caption{Training}
            \label{alg:training}
            \raggedright
            \textbf{Input:} \text{ $q_\mathrm{data}(\rvx_0)$}, $a(t)$, $\eta$, $\theta$, $\Theta$\\
            \textbf{Output:} \text{$\rvepsilon_{\theta}$, $\tau_{\Theta}$} 
            \begin{algorithmic}[1]
                \Repeat
                \State $\rvx_0 \sim q(\rvx_0)$
                \State $t \sim \mathcal{U}(1,T)$
                \State $\rvepsilon \sim \mathcal{N}(\boldsymbol{0} , \mI)$
                \State \text{subtract center of geometry from} $\rvepsilon$
                \State $\rvx_t = \sqrt{\bar{\alpha}_t}\rvx_0 + \sqrt{1-\bar{\alpha}_t} \rvepsilon$
                \State \text{Take SGD step with the gradient}
                \NoNumber{ $\nabla_{(\theta, \Theta)} \bigg[ \eta || \rvepsilon - \rvepsilon_{\theta}(\rvx_t, \tau_\Theta(\rvx_t)) ||^2$}
                \NoNumber{ $\quad + (1-\eta) || \tau_\Theta(\rvx_t) - a(t) ||^2\bigg]$}
                \Until {convergence}
            \end{algorithmic}
        \end{algorithm}
        \end{minipage}
        \hfill
        \begin{minipage}{0.49\textwidth}
        \begin{algorithm}[H]
            \caption{Sampling}
            \label{alg:sampling}
            \raggedright
            \textbf{Input:} \text{$\rvepsilon_{\theta}$, $\tau_{\Theta}$} \\
            \textbf{Output:} \text{new sample $\rvx_i$, \#iterations $i$} 
            \begin{algorithmic}[1]
                \State $i=0$
                \State $\rvx_i \sim \mathcal{N}(\boldsymbol{0} , \mI)$
                \While{\text{$\tau_{\Theta}(\rvx_i) \neq 0$}}
                    \State $\hat{t} = \tau_{\Theta}(\rvx_i)$
                    \State $\bar{\rvepsilon} \sim \mathcal{N}(\boldsymbol{0} , \mI)$ 
                    %\text{if $\hat{t} > 1$, else $\bar{\rvepsilon}=0$}
                    \State \text{subtract center of geometry from} $\bar{\rvepsilon}$
                    \State $\rvepsilon_{\theta}=\rvepsilon_{\theta}(\rvx_i,\hat{t})$
                    \State \text{subtract center of geometry from} $\rvepsilon_{\theta}$
                    \State { $\rvx_{i+1} = \frac{1}{\sqrt{\alpha_{\hat{t}}}}
                    \left( \rvx_i - \frac{\beta_{\hat{t}}}{\sqrt{1-\bar{\alpha}_{\hat{t}}}} \rvepsilon_{\theta} \right) + \sigma_{\hat{t}} \bar{\rvepsilon} $}
                    \State $i = i + 1$
                \EndWhile
                \State \Return {$\rvx_i$, $i$}
            \end{algorithmic}
        \end{algorithm}
        \end{minipage}

\section{Further experiments and details}

    \subsection{Datasets}
    \label{app:qm7x}
    
        \paragraph{QM7-X:}
            QM7-X ~\cite{qm7x} is a comprehensive dataset that was derived from 7000 molecular graphs sampled from the GDB13 chemical space with up to 7 heavy atoms, including types C, N, O, S, and Cl. 
            For each SMILES string, structural and constitutional isomers were obtained using the MMFF94 FF and subsequently optimized with DFTB3+MBD computations, leading to 42 000 equilibrium structures. 
            To capture the PES close to the equilibrium molecules, non-equilibrium molecules were generated by displacing each equilibrium molecule along a linear combination of normal mode coordinates computed with DFTB3+MBD, such that the energy difference between the non-equilibrium and equilibrium structures follow a Boltzmann distribution.
            For each equilibrium structure, 100 non-equilibrium configurations were generated, leading to 4 200 000 non-equilibrium structures in total, where forces and energies for each structure were computed with DFT calculations at the PBE0+MBD level with FHI-aims.
            
            For our experiments, we split the dataset into individual sets for training, validation, and testing of all methods.
            This is done at the molecular graph level to prevent bias leakage between different sets due to related isomers and conformations originating from the same graph.
            Specifically, we use the molecules resulting from 4500 graphs for training, 1250 for validation and the rest for testing.
            Note that  MoreRed does not utilize the non-equilibrium configurations for training, effectively decreasing the training set size by a factor of 100 compared to the training set of the MLFF model.

        \paragraph{QM9:}
            We evaluate the molecular structure generation performance, discussed in details in ~\ref{app:data_generation}, on the QM9 dataset \cite{ruddigkeit2012enumeration,qm9}, a widely used benchmark for molecular generation tasks \cite{cgschnet, gschnet, edm}. It comprises approximately 130k equilibrium organic molecules, each containing up to 9 heavy atoms of types C, O, N, and F. We use 55k molecules for training, 10k for validation, e.g. for scheduling the learning rate, and define the rest as the test split.

    \subsection{Extended analysis of relaxation with MoreRed}
    \label{app:extended_relaxation_experiments}
    
        \begin{figure}
            \centering
                \includegraphics[width=1.0\linewidth]{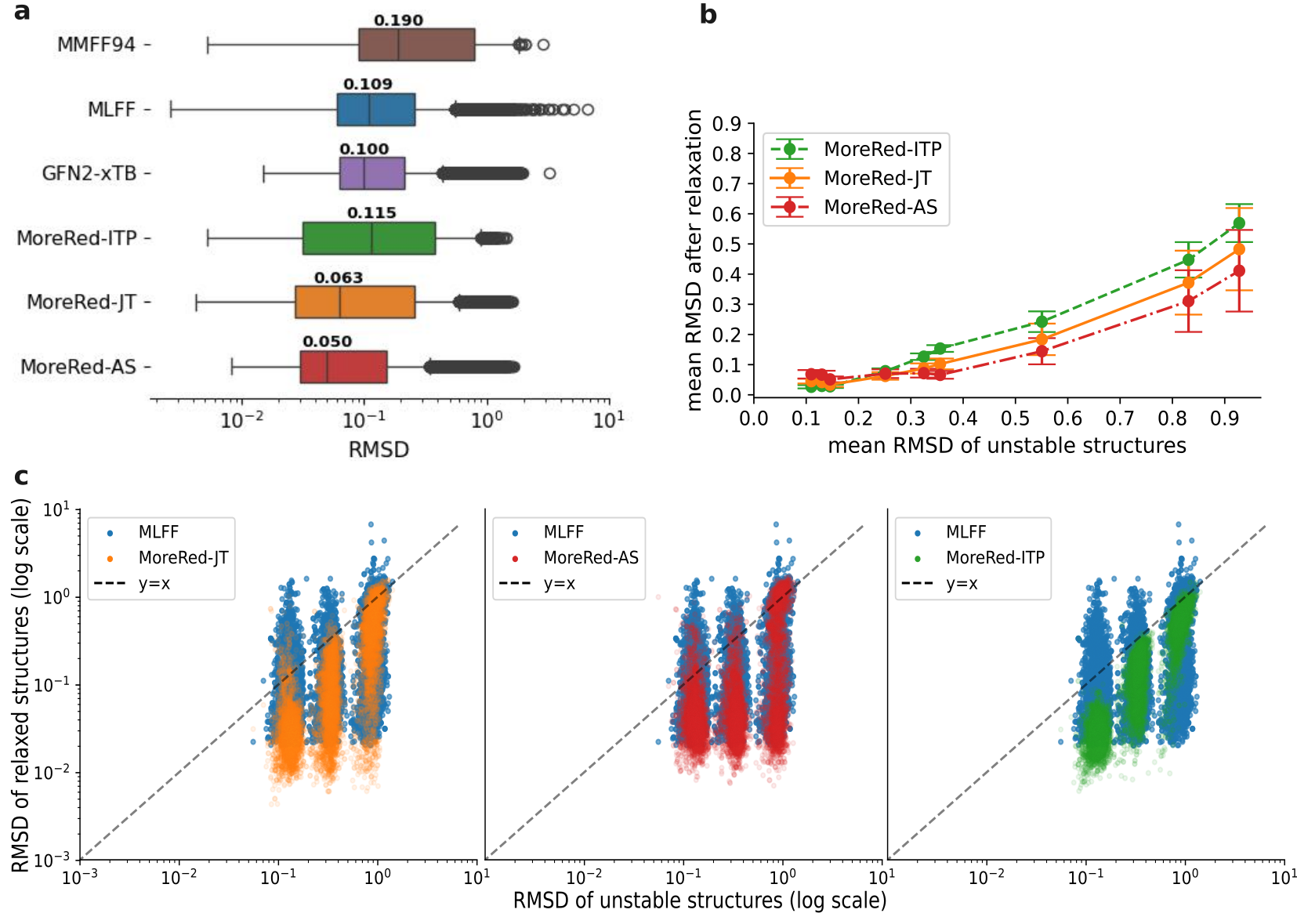}
            \caption{\textbf{a}:  The RMSD of molecules relaxed with the baselines and the MoreRed variants for 20 000 non-equilibrium structures from the QM7-X test split. \textbf{b}: The mean RMSD of 20k non-equilibrium structures from the QM7-X test split vs. their mean RMSD after relaxation. The mean is taken over bins of 2k non-equilibrium molecules with increasing RMSD. The bars show the standard deviation of the RMSD after relaxation over five runs with the respective MoreRed variant. \textbf{c}: RMSD of 10k non-equilibrium structures from the QM7-X test split before (x-axis) and after (y-axis) relaxation for all three variants of MoreRed and the baseline MLFF model. The three partitions occur because the test structures are sampled to cover low/intermediate/high RMSD values, where we sample 3 different non-equilibrium structures per equilibrium structure.}
            \label{appfig:raw-rmsd}
        \end{figure}
    
        Here we further analyze the three different variants of MoreRed by discussing extended results from our experiments in section~\ref{sec:Exp.MoleculeRelaxation} on relaxing non-equilibrium structures from the QM7-X test set.
        In Figure~\ref{appfig:raw-rmsd}, we analyze the RMSD values of the optimized molecules in comparison to their equilibrium structure, instead of the RMSD ratio as it is reported in the main text.
        First of all, in Figure~\ref{appfig:raw-rmsd}a we compare the RMSD values of the three MoreRed variants to the baseline models, including MMFF94, MLFF and GFN2-xTB.
        
        Notably, the two variants with adaptive scheduling have lower median RMSDs than all baselines, while the median RMSD of MoreRed-ITP is slightly worse than that of MLFF and GFN2-xTB.
        The reason for this can be seen in Figure~\ref{appfig:raw-rmsd}c, where the RMSD values after relaxation are compared to the RMSD values of the initial non-equilibrium structures for all three MoreRed variants.
        While the performance of MoreRed-ITP (green) is particularly good for structures that are already close to the equilibrium state, its performance is impaired for structures that initially have a high RMSD.
        The adaptive variants, MoreRed-JT/-AS (orange, red), show a more balanced performance, successfully relaxing structures over the whole spectrum of non-equilibrium test molecules.
        This suggests that the adaptive scheduling with the time step prediction improves the relaxation of molecules that are further away from the data manifold, which is in line with our findings in section~\ref{sec:Exp.DiffusionTimePrediction}.
        This comes at the cost of a higher number of relaxation steps for the adaptive variants and more failure cases (see section~\ref{sec:Exp.MoleculeRelaxation} and Figure~\ref{fig:relaxation}).
        
        Furthermore, to investigate the stochasticity of our method, we analyze the mean RMSD values and their standard deviation from the mean after optimization, subject to the RMSD of the initial structures.
        For this, we created 8 bins of initial structures based on their RMDS and measured the RMSD after optimization (see Figure \ref{appfig:raw-rmsd}b).
        It shows that not only does the mean RMSD increase based on the initial RMSD, but also the standard deviation of the RMSD after optimization increases, with MoreRed-AS having the lowest variance and MoreRed-ITP having the highest. However, the variance is still small in comparison to the mean RMSD.
        This is expected, because structures with large RMSD are assigned to high time steps, resulting in higher variance values for the diffusion reverse kernel.
        Besides, with high time steps, MoreRed needs more optimization steps until convergence and with every step, a small amount of stochasticity is added to the positions.
        Considering statistics over all test structures, the deviations across multiple relaxation runs with MoreRed are very low and therefore not reported in the boxplots.
    
        In conclusion, for initial structures with higher levels of perturbation, the RMSD of the optimized structures increases and the variants with adaptive scheduling, MoreRed-JT/-AS, via time step prediction provide the required flexibility to perform more accurate optimization.
        On the other hand, the fixed schedule variant, MoreRed-ITP, provides fast and very accurate equilibrium molecules for initial structures with low levels of noise. 

    \subsection{Examples of relaxed structures}
    \label{app:examples_relaxed_structures}

        \begin{figure}
            \centering
                \includegraphics[width=1.\linewidth]{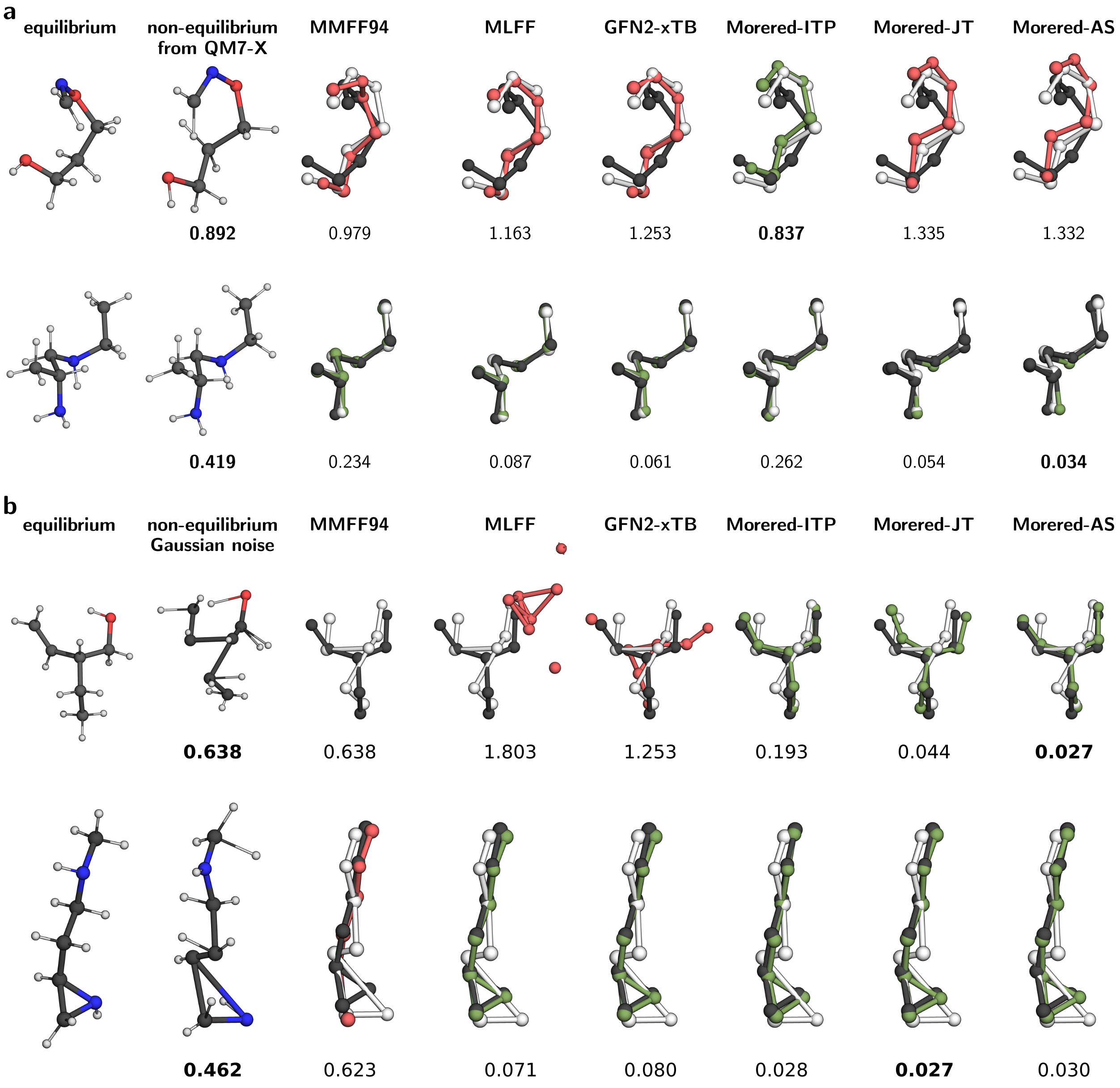}
            \caption{
            Exemplary relaxation results for all baselines and the MoreRed variants. Equilibrium geometries from the QM7-X test split and corresponding non-equilibrium structures are shown on the left-hand side with nitrogen in blue, oxygen in red, hydrogen in white, and carbon in grey. They are followed by results from the different relaxation methods, where the hydrogen atoms are suppressed for clarity and the equilibrium and the non-equilibrium structures are super-imposed in black and white respectively, together with the relaxed structure of the respective method. Failure cases are drawn in red and success cases in green. The RMSD from the equilibrium geometry is denoted below each structure. In \textbf{a}, the non-equilibrium molecule is taken from QM7-x. In \textbf{b}, the non-equilibrium molecule is obtained by applying the forward diffusion process, i.e. adding Gaussian noise, for 250 steps. In both panels, the top row shows a non-equilibrium structure with larger RMSD that was more difficult to converge for many methods than the example in the bottom row.
            }
        \label{fig:relaxation_examples}
    \end{figure}

    For illustration, in Figure \ref{fig:relaxation_examples} we visualize relaxation of non-equilibrium structures from different sources.
    Example structures are shown in equilibrium and non-equilibrium on the left-hand side.
    On the right-hand side, the different relaxation methods are shown where the hydrogen atoms are suppressed.
    In panel \textbf{b}, we use the non-equilibrium molecules obtained by perturbing the equilibrium geometries using $250$ steps of Gaussian diffusion.
    Although the diffused non-equilibrium examples are only distorted up to an extent where the reference structure is visually still recognizable and the median noisy RMSD matches that of the non-equilibrium structures from QM7-X, they become hard to relax for the baseline methods.
    They fail at relaxing the first example while the three MoreRed variants converge towards the ground-truth equilibrium geometry.
    Especially the results from the models with adaptive schedule, MoreRed-JT/-AS, match the reference structure better than MoreRed-ITP.
    The MLFF gives a physically implausible result.
    In the bottom example, where the non-equilibrium structure has a lower RMSD from the equilibrium geometry, the MLFF and the semi-empirical method also converge to the reference geometry.
    However, all MoreRed variants achieve a significantly lower RMSD and the simple force field baseline, MMFF94, fails at recovering the reference.
    In panel \textbf{a}, we use the QM7-X non-equilibrium structures. %, which were obtained with normal model displacements of the equilibrium geometry.
    In the bottom example, all methods manage to find the reference equilibrium, where it is matched most closely by the two MoreRed variants with adaptive schedule (JT/AS).
    While the low RMSD values are a good indicator that MoreRed has accurately captured the distribution of equilibrium reference structures reported in QM7-X, the top example shows why additional metrics should be considered in the analysis.
    In this case, where the non-equilibrium structure has a higher RMSD from the equilibrium geometry, none of the methods recovers the reference.
    Nevertheless, the obtained configurations might be local minima on the PES that structurally deviate even further from the equilibrium structure reported in QM7-X than the non-equilibrium starting point of the relaxation.
    Therefore, we evaluate DFT-computed energies of relaxed structures in section~\ref{sec:Exp.MoleculeRelaxation} in the main text to gain further insights into the performance of all methods.

    \subsection{Training on low energy minima from MLFF relaxations:}
    \label{app:training_on_mlff_relaxations}

        \begin{figure}
            \centering
            \includegraphics[width=0.7\linewidth]{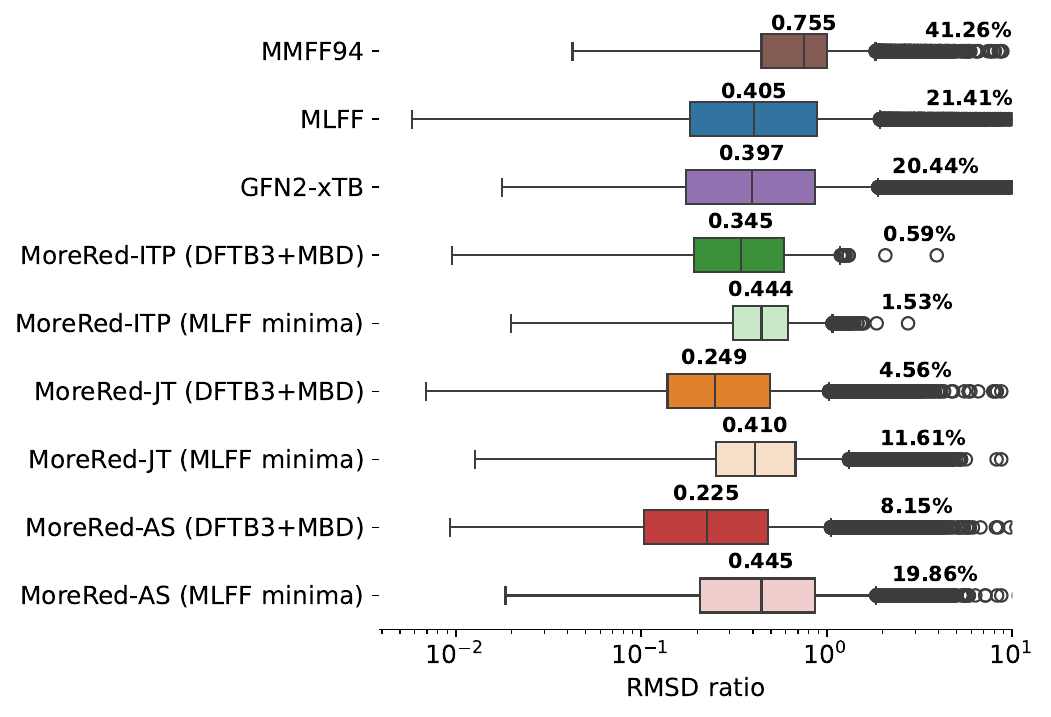}
            \caption{
                The RMSD ratios of structures relaxed with the baselines and the different MoreRed variants for 20,000 non-equilibrium structures from QM7-X, with each MoreRed variant having two versions. One version is trained on the original DFTB3+MBD equilibrium structures from QM7-X, while the other is trained on equilibrium structures resulting from further relaxation of the QM7-X reference structures using the MLFF, as in Figure~\ref{fig:dft_energies}. The median values and the percentage of failure cases, defined as those with an RMSD ratio $\geq 1$, are displayed on top of each box plot.
            }
            \label{fig:rmsd_ratios_mlff_relaxed}
        \end{figure}
    
        We investigate the structure-label mismatch in QM7-X, where force labels were computed using a more accurate and expensive method, PBE0+MBD, than for identifying the reference equilibrium structures, DFTB3+MBD. To this end, we first used the MLFF model, which was trained on the PBE0+MBD labels, to further relax the DFTB3+MBD equilibrium structures to lower energy levels. We then retrained all MoreRed variants on the relaxed data. Subsequently, we performed relaxation with the new MoreRed models on the non-equilibrium test structures from QM7-X, and computed the DFT energies of the resulting samples and their RMSD to the QM7-X reference minima, maintaining the same setup as in the main experiments.

        If the inconsistency between the RMSD and energy results observed when training on the QM7-X references is due to the structure-label mismatch in the dataset, then training MoreRed on the MLFF-relaxed structures should result in similar energy and RMSD levels to those of MLFF. This is because MoreRed, by definition, learns the data manifold described by the training data and outputs structures with similar characteristics.

        We summarize the results of the RMSD ratios in Figure~\ref{fig:rmsd_ratios_mlff_relaxed} and the DFT energies in Figure~\ref{fig:dft_energies}. The three MoreRed variants achieve higher RMSD ratios but lower DFT energies, comparable to MLFF, when trained on the relaxed structures from MLFF (the lighter colors in the Figures) rather than the original DFTB3-MBD reference structures from QM7-X (the darker colors). Thus, we can conclude that the structure-label mismatch in QM7-X causes the inconsistency between RMSD and energy results observed in the original experiments. Specifically, MLFF found minima with lower energy states but different structures than the reference minima in QM7-X, resulting in higher RMSD but lower DFT energy, as it was trained on more accurate labels. These accurate labels exhibit non-negligible forces, indicating that the equilibrium structures identified by DFTB3-MBD still have higher energies.
    
    \subsection{Generalization with different equivariant representations}
    \label{sec:Exp.so3net}
    
        \begin{figure}[t]
             \begin{minipage}{.49\linewidth}
                 \begin{center}
                 \includegraphics[width=1.\linewidth]{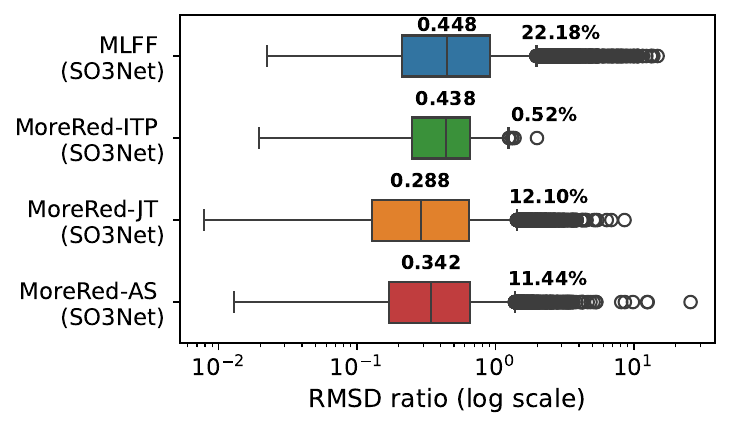}
                 \end{center}
             \end{minipage}
             \begin{minipage}{.49\linewidth}
                 \begin{center}
                 \includegraphics[width=1.\linewidth]{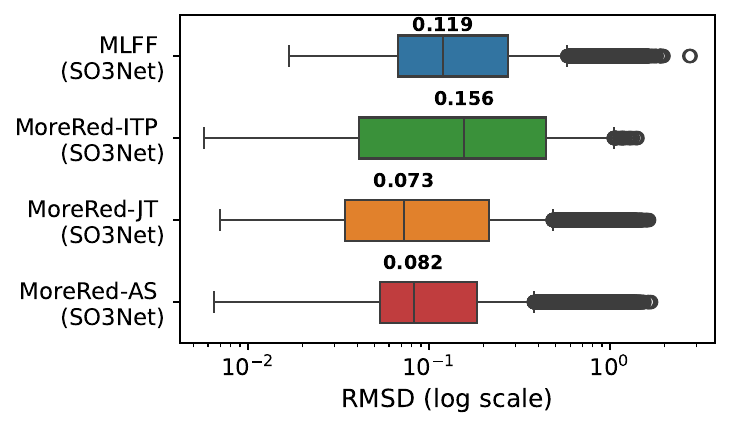}
                 \end{center}
             \end{minipage}
             \caption{The RMSD ratios (left) and RMSD (right) of molecules relaxed with the baseline MLFF model and with our MoreRed variants for $20$k non-equilibrium structures from the QM7-X test split using SO3Net implementation in SchNetPack \cite{schutt2023schnetpack} as an alternative equivariant backbone representation for PaiNN. 
             The median values and the percentage of relaxation failure cases, i.e. the cases where the RMSD ratio exceeds $1$, are shown above each box plot.}
            \label{fig:relaxation_so3net}
        \end{figure}
        
         To further assess the robustness of our method, we conducted a set of experiments employing an alternative equivariant molecular representation to PaiNN. Specifically, we trained the MLFF model and all three variants of MoreRed using SO3Net \cite{schutt2023schnetpack} as a backbone representation. This representation incorporates spherical harmonics in the spirit of Tensor Field Networks \cite{tensorfieldnet} and NequIP \cite{nequip} to handle SO(3)-equivariance, distinguishing it from the PaiNN architecture.
        
        Utilizing the same data splits as in the PaiNN experiments, we tested the models on the same set of 20 000 non-equilibrium structures from the test split of QM7-X. All other experimental details align with those outlined in section \ref{sec:Exp.MoleculeRelaxation} for PaiNN. Given the long training time of the MLFF model (7 days) in comparison to MoreRed, we opted to use half the number of parameters employed in PaiNN to expedite the experiments. However, to maintain fairness, we used the identical model hyperparameters for both MoreRed and MLFF. Additional hyperparameter details for SO3Net are provided in section \ref{sec:so3net_hyperparams}.
        
        Our findings, summarized in Figure \ref{fig:relaxation_so3net}, affirm that our approach performs comparably well with this alternative equivariant neural network backbone, consistently outperforming the MLFF model in terms of structure accuracy with MoreRed-AS and MoreRed-JT. Yet, the overall performance for all models, including the MLFF, is slightly worse than reported in Figure \ref{fig:relaxation} with PaiNN, and there is a subtle discrepancy in performance between MoreRed-AS and MoreRed-JT. We attribute these differences to the lack of hyperparameter tuning with SO3Net and the use of half the number of parameters employed in PaiNN.

    \subsection{Molecular generation with time step prediction}
    \label{app:data_generation}
    
        \begin{figure}
            \begin{minipage}{.59\linewidth}
                \begin{center}
                    \includegraphics[width=1.\linewidth]{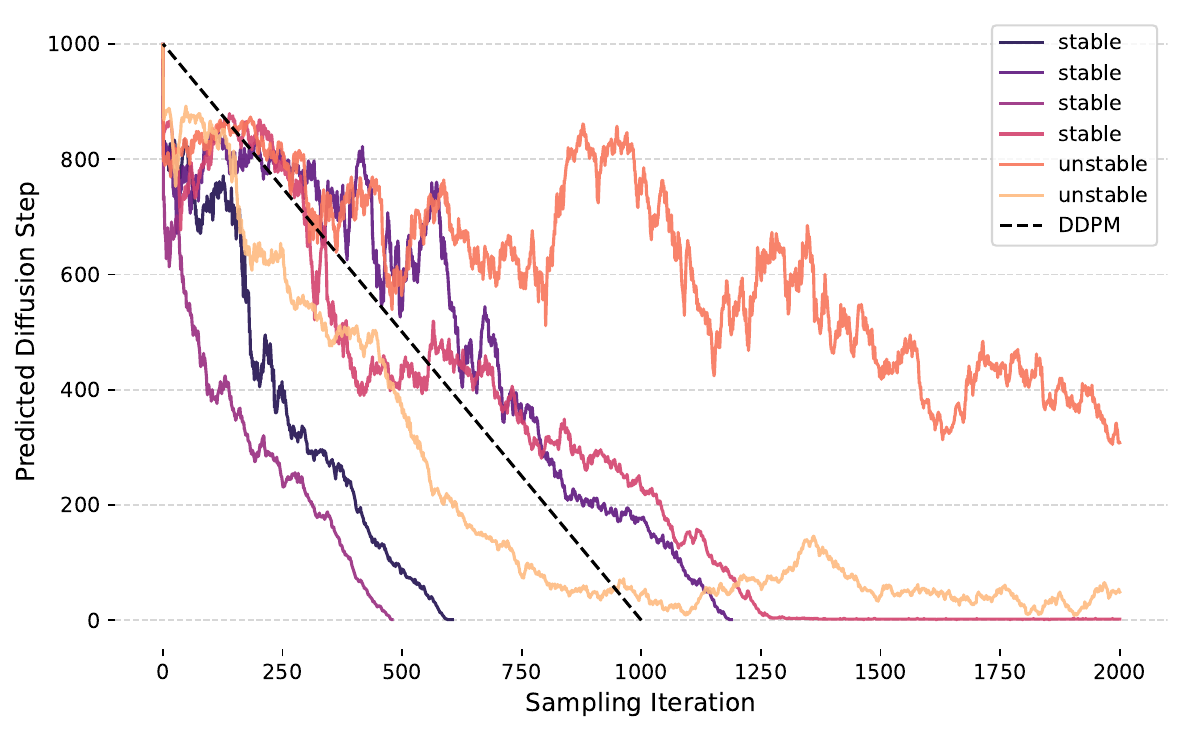}
                \end{center}
            \end{minipage}
            \begin{minipage}{.4\linewidth}
                \begin{center}
                    \includegraphics[width=1.\linewidth]{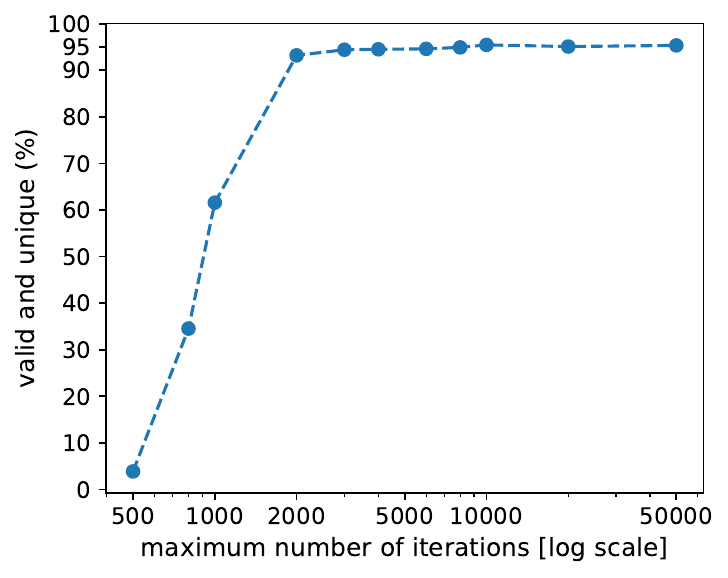}
                \end{center}
            \end{minipage}
            \caption{\textbf{Left:} Examples of time step trajectories from sampling molecular structures from complete noise using MoreRed-JT. The solid lines indicate the predicted time steps by MoreRed through the sampling iterations. For reference, the dashed line indicates the fixed decreasing time step schedule used in the usual reverse diffusion from plain DDPM. Each colour indicates a different sample. In the box to the top-right, we note whether the sampled molecule passes the validity check explained in section~\ref{sec:details_improved_gen} (stable) or not (unstable). \textbf{Right:} The evolution of validity and uniqueness when increasing the maximum number of iterations for MoreRed-large (short for MoreRed-JT-Large) in the molecule generation task on QM9.}
            \label{appfig:sampling-traj-stability-plot}
        \end{figure}
        
        \paragraph{Improved molecule generation with adaptive schedule:}
        \label{sec:details_improved_gen}

        \begin{figure}
            \centering
            \includegraphics[width=.7\linewidth]{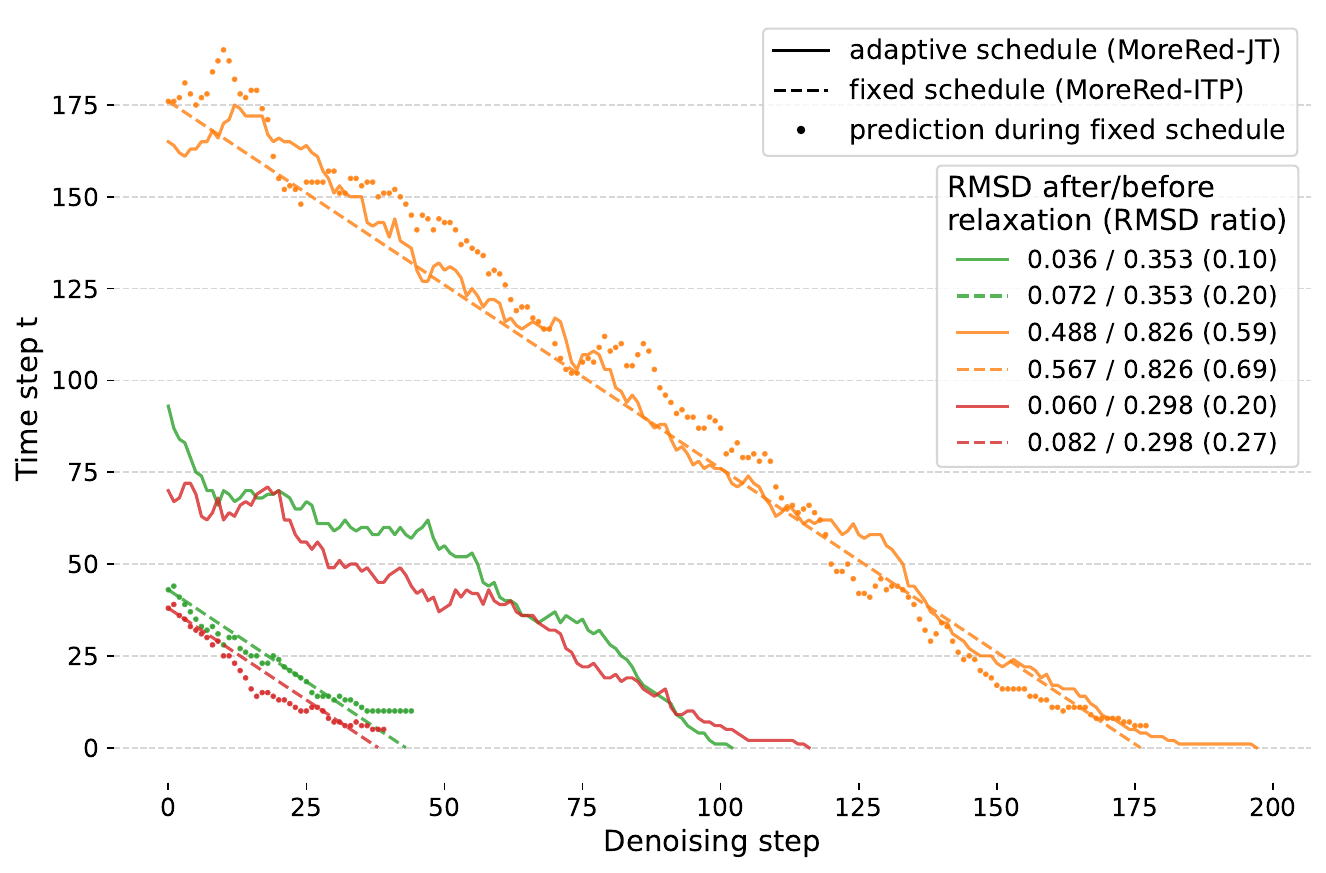}
            \caption{
            Comparison of time step trajectories for relaxation of three non-equilibrium test structures from QM7-X in red, orange, and green, similar to Figure~\ref{fig:rmsd_corr_Traj}b but with significantly different initial time step predictions.
            MoreRed-JT follows an adaptive schedule (the time step and noise are predicted at each denoising step; solid lines) and MoreRed-ITP follows a fixed schedule (the time step is only predicted at the initial denoising step; dashed lines). The dots show the predicted (but not utilized) time steps for MoreRed-ITP during denoising with the fixed schedule.
            The inset box shows the RMSDs between the reference equilibrium geometry and the non-equilibrium structure after and before relaxation, respectively. MoreRed-JT outperforms MoreRed-ITP in all three cases, including the orange trajectory, where MoreRed-ITP starts from a higher initial step.
            }
            \label{appfig:rmsd_corr_Traj_different}
        \end{figure}
        
\begin{figure}
            \centering
            \includegraphics[width=.9\linewidth]{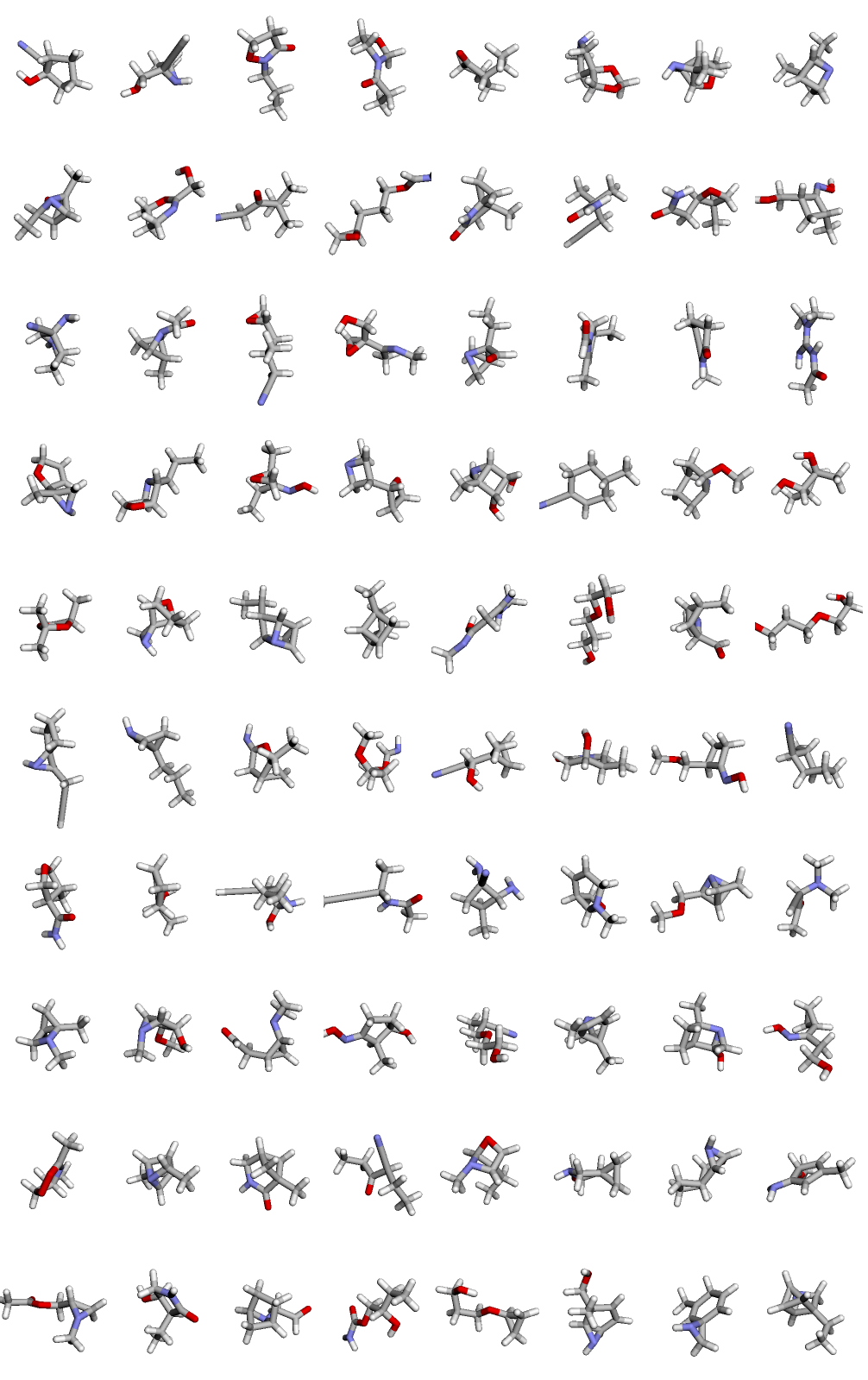}
            \caption{Batch of generated molecules with MoreRed-JT trained on QM9.}
            \label{fig:qm9_pred_samps}
        \end{figure}

\begin{figure}
            \centering
            \includegraphics[width=.9\linewidth]{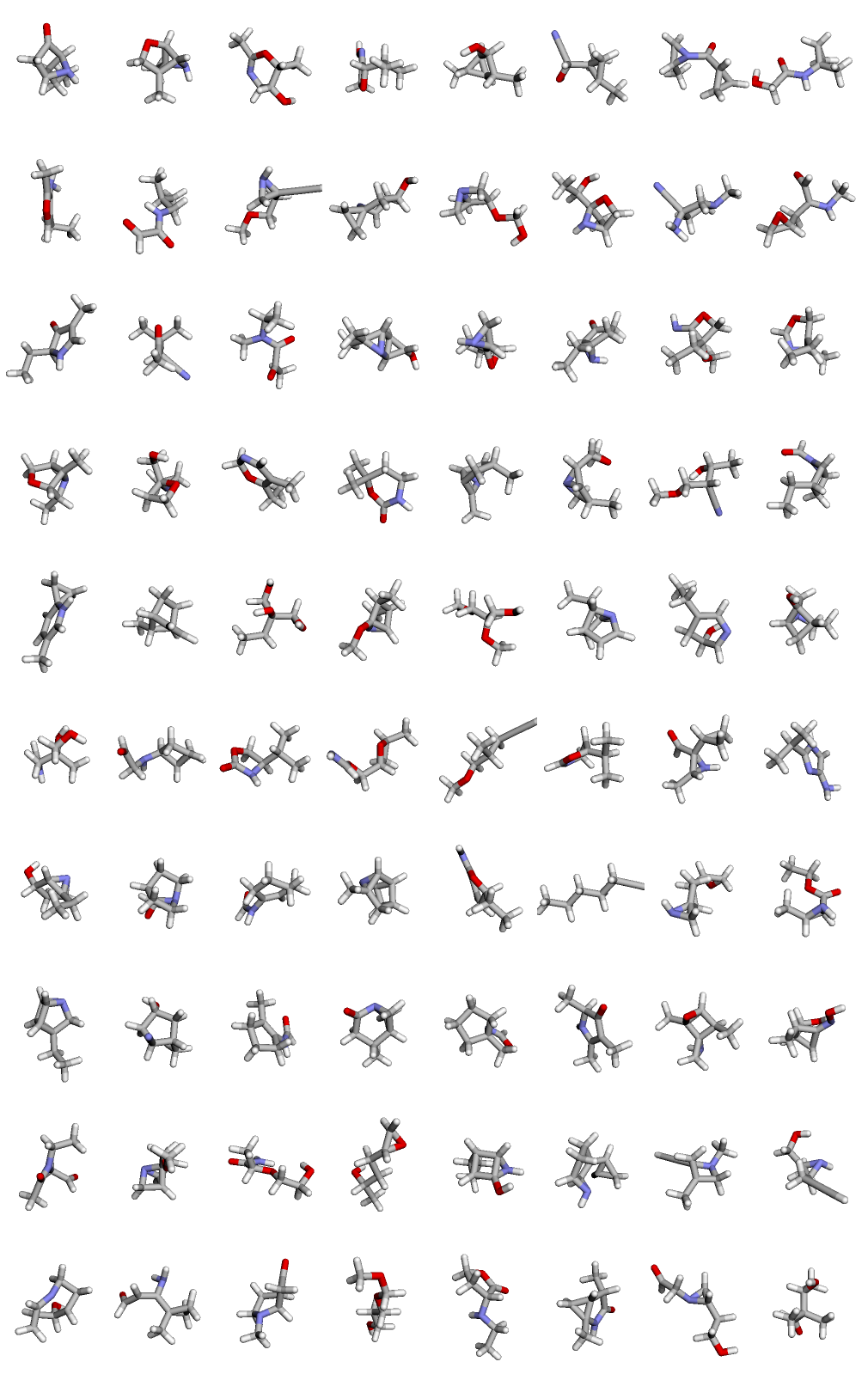}
            \caption{Batch of generated molecules with DDPM trained on QM9.}
            \label{fig:qm9_cond_samps}
        \end{figure}

            In contrast to existing diffusion models, e.g DDPM in our case, MoreRed-AS/-JT adaptively control the reverse diffusion process with the diffusion time step predictor, as observed in the relaxation task (see Figure \ref{fig:rmsd_corr_Traj}b in the main text for similar initial time step predictions and \ref{appfig:rmsd_corr_Traj_different} for different initial time step predictions).
            This means that we follow an adaptive time step schedule where the time step at each denoising/sampling iteration is estimated by a neural network $\hat{t} = \tau_{\Theta}(\rvx_i)$.
            Sampling in existing diffusion models follows a pre-defined fixed schedule where exactly $T$ denoising steps with decreasing time step values $t=T, T-1, \ldots, 1$ are done, as illustrated by the dashed line in the left plot in Figure~\ref{appfig:sampling-traj-stability-plot}.
            In the following, we compare the sampling performance with the adaptive schedule of our MoreRed to the standard, fixed diffusion model sampling process in the task of data generation from complete noise that diffusion models were originally designed for.
            To this end, given a molecular composition $\mZ$, complete noise samples from the isotropic Gaussian prior distribution, $\mathcal{N}(\boldsymbol{0}, \mI)$, are used as initial molecular structures and are then denoised by both models to sample valid molecular structures.
            For a fair comparison, we employ the same noise model for classical sampling with a fixed schedule, denoted as DDPM, and for adaptive sampling using the time step predictor, i.e. MoreRed.
            In the rest of this section, we refer to MoreRed-JT as MoreRed.
            Examples of generated structures can be found in Figure~\ref{fig:qm9_pred_samps} and Figure~\ref{fig:qm9_cond_samps} for MoreRed and DDPM, respectively.
            
            We evaluate the generation performance on the QM9 dataset~\cite{ruddigkeit2012enumeration,qm9}, which is a widely used benchmark for molecular generation tasks and described in \ref{app:qm7x}.
            For data generation with MoreRed, we set the convergence criteria to $\hat{t}\leq0$ or a maximum number of sampling iterations equal to $2000$.
            For the standard DDPM \cite{DDPM_Ho}, we use a fixed schedule with $T=1000$ sampling iterations (same as during training).
            For our evaluation, we generate 10 000 structures starting from the latent prior distribution where the atomic compositions $\mZ$ of molecules are randomly drawn from the QM9 test split.
            
            As metrics, we adopt validity, uniqueness and novelty as proposed by Gebauer \etal~\cite{gschnet}, using their publicly available analysis script for comparability.
            It translates the generated structures to canonical SMILES~\cite{weininger1988smiles} encodings, which is a string representation of molecular graphs.
            A molecule is considered valid if all its atoms are connected and possess the proper valency in that encoding.
            Furthermore, unique and novel molecules are identified by comparing the canonical SMILES strings of all generated structures to each other and those of all molecules in QM9, respectively.
            Table~\ref{tab:stability} summarizes the results.
            We observe that MoreRed, i.e. adaptive scheduling, performs better than DDPM, i.e. fixed scheduling, in all criteria.
            The same tendency can also be observed for architectures with more parameters, i.e. MoreRed-large.
            This confirms our hypothesis that our adaptive reverse diffusion procedure based on the time step prediction is beneficial for unconditional sampling from complete Gaussian noise and is not restricted to relaxation from noisy non-equilibirum structures.
            
            MoreRed can dynamically adapt the time step at each sampling iteration to match the current noise level in the sample, correcting for the errors caused by the noise predictor, $\rvepsilon_\theta$, as can be seen in some exemplary sampling trajectories in Figure~\ref{appfig:sampling-traj-stability-plot} left.
            This adaptive scheduling dynamically determines the number of reverse (sampling) iterations, providing a dynamic solution to the issue identified in Song \etal~\cite{song2021score} of one step noise prediction with fixed reverse diffusion leading to potential sample deviation from the optimal reverse trajectory.
            Namely, MoreRed mitigates this problem by automatically adjusting the time steps, offering a promising solution without manual hyperparameter tuning, which is necessary for previously proposed solutions that use correction steps after each reverse diffusion step, such as using second-order SDE/ODE solvers or running Langevin dynamics iterations~\cite{song2021score, diff_edm}.
    
            \begin{table}
                \centering
                \caption{Quality of 10 000 generated molecules after training on QM9. DDPM-large and MoreRed-large are larger neural networks that use 4 times more learnable parameters than DDPM and MoreRed, respectively. V=valid, U=unique, N=novel.}
                \label{tab:stability}
                \begin{tabular}{lccc}
                    \hline
                    Model & V (\%) & V+U (\%) & V+U+N (\%) \\
                    \hline
                    DDPM & 78.2 & 77.3 & 62.5 \\
                    MoreRed & \textbf{89.3} & \textbf{88.0} & \textbf{68.6} \\
                    \hline
                    DDPM-large & 86.6 & 85.3 & 63.8 \\
                    MoreRed-large & \textbf{94.7} & \textbf{92.4} & \textbf{66.7} \\
                    \hline
                \end{tabular}
            \end{table}

        \paragraph{Maximum number of sampling iterations:}
        \label{sec:max_iterations}
            The results presented in table~\ref{tab:stability} are obtained using a maximum number of sampling steps of $2000$.
            However, to gain a more comprehensive understanding of the model's behaviour, we conduct experiments with varying maximum numbers of sampling steps using the large version of MoreRed-JT, i.e. MoreRed-large, and summarize the validity and uniqueness results in the right plot in Figure \ref{appfig:sampling-traj-stability-plot}.
            
            We observe that the model can generate around $4\%$ valid and unique molecules with less or equal to $500$ sampling iterations and up to $35\%$ with no more than 800 steps, which is less than $1000$ steps during training.
            Interestingly, MoreRed-large yields around $25\%$ fewer valid and unique molecules compared to the standard diffusion model DDPM-large when using $1000$ steps.
            Yet, it outperforms it by approximately $7\%$ using $2000$ sampling steps at most.
            This suggests that the error introduced by the stochastic predictor of the diffusion step may slow down convergence in certain cases but ultimately lead to finer samples.
            Furthermore, by observing the evolution of the curve, we deduce that using $2000$ steps is sufficient to achieve results close to the best performance.
            Beyond $2000$ steps and up to 50k sampling steps, it exhibits only marginal improvements compared to the significant progress observed between $500$ and $2000$ steps.
            The latter aligns with the findings of Song \etal~\cite{song2021score}, who showed that using exactly $2000$ steps of a predictor-corrector sampler with manually tuned hyperparameters instead of only $1000$ steps of a predictor-only sampler, like DDPM, to sample from a diffusion model trained on $1000$ steps enhances performance in images.
            Nevertheless, in our method, we do not fix the exact number of iterations to $2000$ for all the samples and we do not need a manual tuning of hyperparameters but MoreRed dynamically sets the number of steps by iteratively predicting the time step, eventually ending sampling after less than $2000$ steps if the convergence criteria, $\hat{t} \leq 0$, is met.

    \section{Model architectures and hyperparameters}
\label{app:arch_and_hyperparams}

    \subsection{Architectures}
        \label{app:Architecture}
        \begin{figure}
            \centering
            \includegraphics[width=0.5\linewidth]{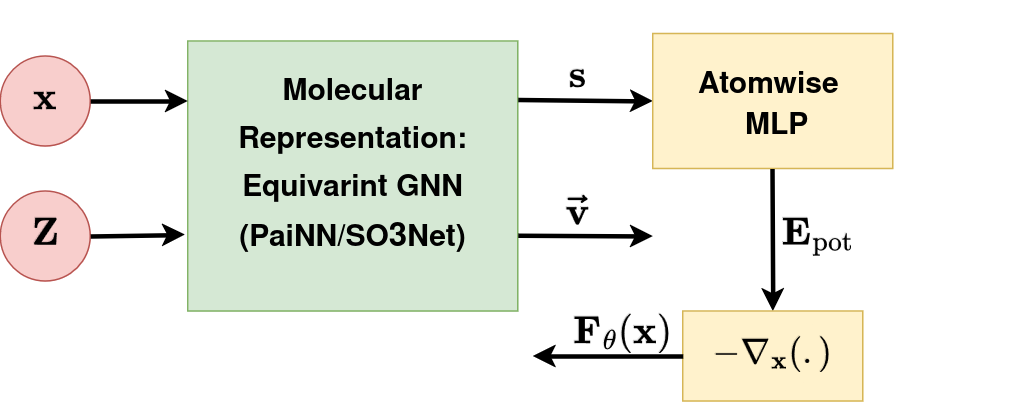}
            \caption{MLFF architecture: $\rvx$ are the atom positions, and $\mathbf{Z}$ are the atom types. 
            An equivariant graph neural network, e.g. PaiNN, is used to learn an equivariant molecular representation that extracts one vector of invariant features $\mathbf{s}$ and one vector of equivariant vectorial features $\vec{\mathbf{v}}$ per atom. The scalar features $s$ are used to estimate the energy $E_\mathrm{pot} = \sum_{i=1}^N{E_\mathrm{pot}^{(i)}}$ using an atomwise MLP $E^\mathrm{pot}_\theta = E_\mathrm{pot}^{(i)}$, where $
            \theta$ represents the neural network weights. The derivatives with respect to the positions $\rvx$ represents the forces prediction, i.e. ${F}_{\theta}(\rvx) = -\nabla_\rvx E_\mathrm{pot}$.}
            \label{fig:ff_model}
        \end{figure}
    
        \paragraph{MLFF:}
            The architecture of MLFF is illustrated in Figure \ref{fig:ff_model}. Using the atomwise invariant features $\mathbf{s}$ from PaiNN, an atomwise multi-layer perceptron (MLP) predicts the atom-wise energies $\{E_\mathrm{pot}^{(i)}\}_{i=1}^N$, which are then aggregated to form a permutation-invariant potential energy $E_\mathrm{pot} = \sum_{i=1}^N{E_\mathrm{pot}^{(i)}}$, where $N$ is the number of atoms in the molecule. The gradients, i.e. the interatomic forces, are computed as the derivative of the potential energy $E_\mathrm{pot}$ with respect to the atom positions $\rvx$, which ensures energy conservation and equivariant predictions of interatomic forces.

        \begin{figure}
            \begin{minipage}{.49\linewidth}
                \begin{center}
                    \includegraphics[width=1.\linewidth]{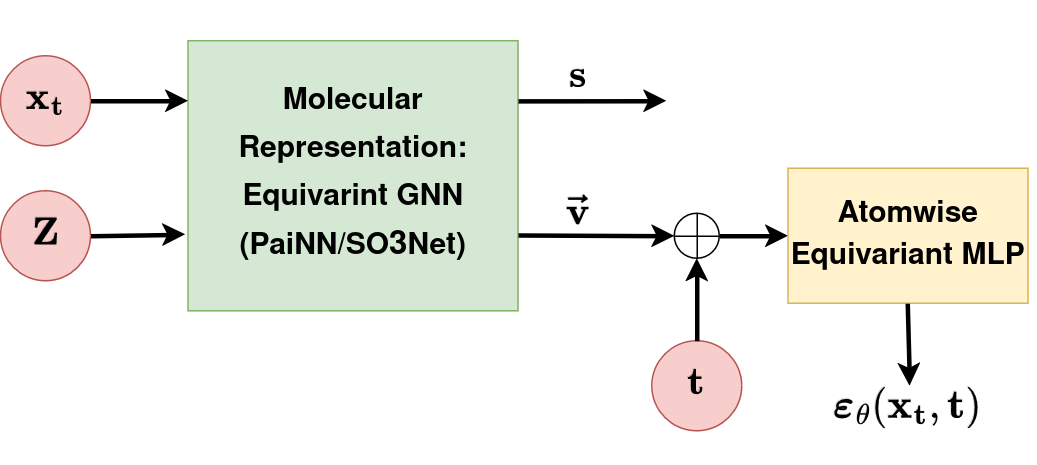}
                \end{center}
            \end{minipage}
            \hfill
            \begin{minipage}{.49\linewidth}
                \begin{center}
                    \includegraphics[width=1.\linewidth]{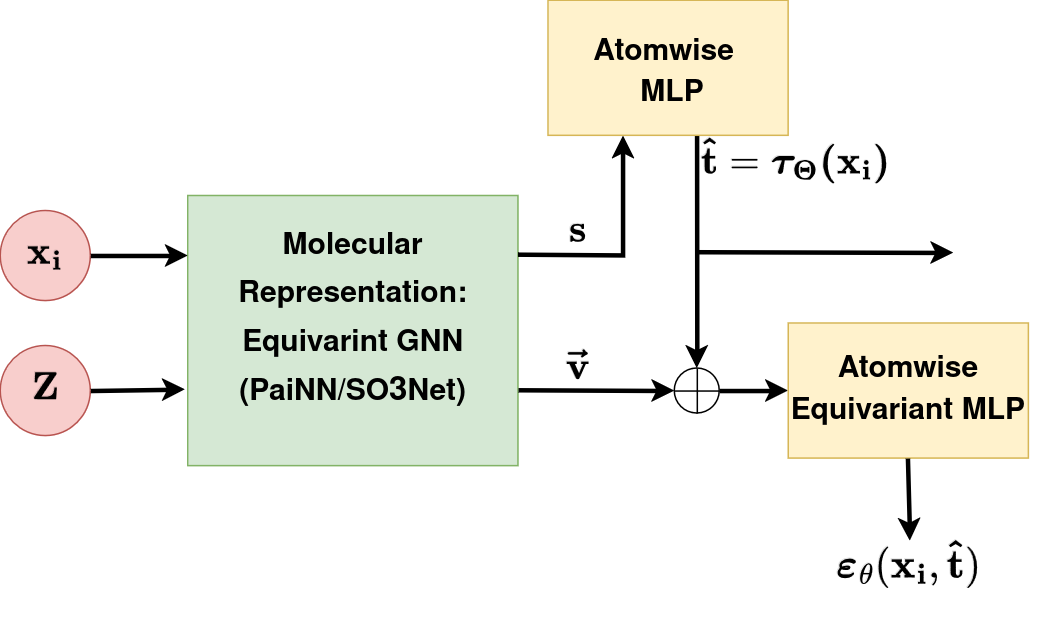}
                \end{center}
            \end{minipage}
            \caption{$\rvx_t$ and $\rvx_i$ are the (noisy) atom positions at true time step $t$ or unknown step $i$ respectively, and $\mZ$ is a vector of integers representing the atom types. $\oplus$ denotes the elementwise vector concatenation operation. Both architectures use an equivariant graph neural network, e.g. PaiNN, to learn an equivariant molecular representation that extracts one vector of invariant features $\mathbf{s}$ and one vector of equivariant vectorial features $\vec{\mathbf{v}}$ per atom. \textbf{Left:} The model architecture used to predict the diffusion noise in the plain DDPM (section \ref{sec:back_diff}). The diffusion step $t \in \{0,1,\dots,T\}$ is normalized by dividing by the maximum time step $T$, i.e. $t/T \in [0,1]$, and concatenated to the vectorial features $\vec{\mathbf{v}}$ of each atom to directly predict one equivariant noise vector $\rvepsilon_{\theta}(\rvx_t,t)$ per atom using an equivariant gated MLP. \textbf{Right:} The model architecture for MoreRed-JT. It is identical to the DDPM architecture except that the unknown time step of $\rvx_i$ is first predicted by an atomwise MLP $\hat{t} = \tau_\Theta(\rvx_i)$ on top of the invariant scalar features $\mathbf{s}$, similar to predicting the invariant potential energy in MLFF. MoreRed-ITP/-AS use the same architecture as MoreRed-JT except that $\tau_\Theta(\rvx_i)$ and $\rvepsilon_{\theta}(\rvx_i,\hat{t})$ use two separate GNN molecular representations, i.e. two separate green boxes in the diagram. In our experiments, we use two identical PaiNN architectures to compute the seperate features for the time and noise head when using MoreRed-ITP/-AS.}
            \label{fig:model_arch}
        \end{figure}
        
        \paragraph{MoreRed variants, including DDPM:}
            The architectures of the different variants are illustrated in Figure \ref{fig:model_arch}. On the right side is the architecture of MoreRed-JT, where the time and noise head share the same backbone molecular representation, e.g. PaiNN. Similar to the invariant energy in MLFF, the time $\hat{t} = \tau_\Theta(\rvx_i)$ is predicted by an atomwise MLP using the invariant features as input. In contrast to the energy-conservative forces in MLFF, the diffusion noise is directly predicted as an equivariant tensorial quantity using an equivariant gated MLP on top of the equivariant vectorial features $\Vec{\mathbf{v}}$. For MoreRed-ITP and MoreRed-AS, we use the same model architecture as for MoreRed-JT, with the only difference being that the time head $\tau_\Theta(\rvx_i)$ and the noise head $\rvepsilon_{\theta}(\rvx_i,\hat{t})$ use two separate, but identical, molecular representation networks that are trained separately, instead of using one joint network.
            On the left side is the architecture of the plain DDPM \cite{DDPM_Ho} (section \ref{sec:back_diff}) used as a baseline model in the molecular generation experiments in~\ref{app:data_generation}. 
            The overall architecture is similar to MoreRed, except that the diffusion time step $t$ is explicitly provided by the user as input to the noise head, rather than being dynamically predicted by a neural network.
    
    \subsection{Hyperparameters}
    \label{app:Hyperparameters}

        The models used in our experiments with molecules are implemented and trained using SchNetPack \cite{schutt2023schnetpack}. For all our experiments except the generalization experiments in section~\ref{sec:Exp.so3net}, we use PaiNN \cite{painn} with $3$ interaction blocks and $20$ Gaussian radial basis functions with a cosine cutoff of $5$Å as molecular representation for all the models. After computing the molecular representation, the number of atomic features is halved in each layer of the output heads, with a total of $3$ layers for each head for all the models. We use the AdamW~\cite{loshchilov2018decoupled} optimizer for all the models and train them until complete convergence. Moreover, we use the exponential moving average (EMA) of the model parameters with a decay of $0.999$ for all models across all training epochs during validation, testing and inference rather than using the most recent parameter updates. Additionally, we use a learning schedule that halves the learning rate during training if the validation loss stagnates for a predefined number of epochs, allowing for finer steps near the local minima and avoiding fluctuating around them. We use early stopping to stop the training process when the validation loss stops decreasing after some epochs instead of using a fixed number of epochs and we use the model checkpoint with the lowest validation loss for testing and inference. The specific details for the different models are listed below.
        
        \paragraph{MLFF:} 
            Overall, for MLFF training, we follow the hyperparameters and the training details reported in the original work \cite{painn}, but we further tuned the batch size on $\{ 10, 64, 128\}$, the learning rate on $\{ 10^{-3}, 10^{-4}\}$ and the atomic features on $\{ 64, 128, 256\}$. We found that a batch size of $10$, learning rate of $10^{-3}$ and $128$ atomic features achieve the lowest loss, which aligns with the results from previous work using PaiNN \cite{painn, spookynet}. We found that using more atomic features than 128, i.e. more parameters, for MLFF hurt the performance. 
            Additionally, we use a patience of 15 epochs for the learning rate schedule and 30 epochs for early stopping. The resulting baseline MLFF achieves a mean square error of 0.376 kcal/mol for energy and 0.519 kcal/mol/$\text{\AA}$ for forces after training, which is within chemical accuracy of 1 kcal/mol and on par with the benchmarks on QM7-X as reported in Unke \etal~\cite{spookynet}.

\begin{figure}
                \centering
                \begin{minipage}{.5\textwidth}
                  \includegraphics[width=1.\linewidth]{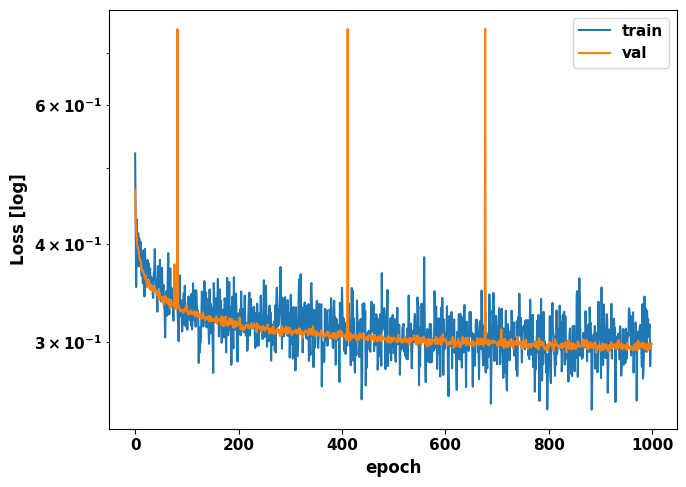}
                \end{minipage}%
                \begin{minipage}{.5\textwidth}
                    \includegraphics[width=1.\linewidth]{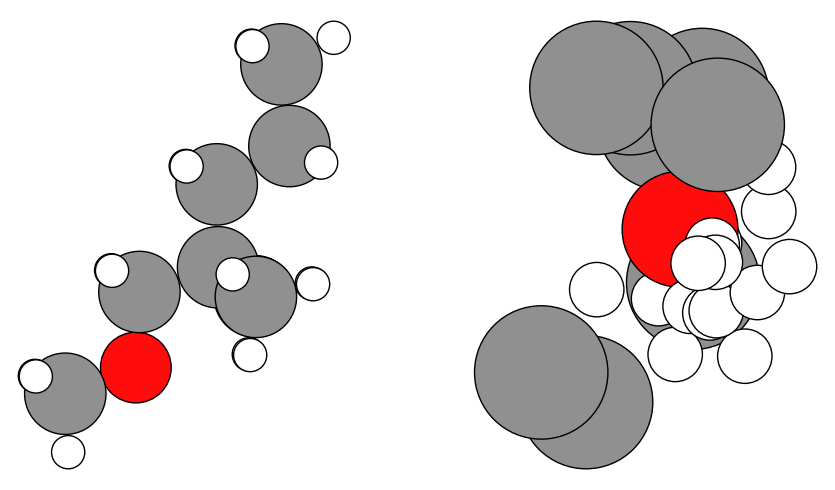}
                \end{minipage}
                \caption{ \textbf{Left:} Plot showing the evolution of the training loss in blue and validation loss in orange in the log scale. The validation loss has fewer fluctuations than the training loss, showcasing the effect of applying EMA on the model's parameters. The validation loss shows three high peaks due to exploding model outputs. Subsequent analysis revealed that this explosive behaviour occurs only for high diffusion steps above $t=850$, which was associated with highly noisy and unrealistically dense molecules where all atoms were tightly packed with tiny atomic distances, as can be seen in the rightmost figure. \textbf{Right:} Example of a degenerated molecule after $916$ diffusion steps. On the left is the original molecule. On the right is the diffused molecule, where all the hydrogens (white nodes) are tightly condensed, resulting in tiny atomic distances and explosive behaviour of the model. However, this issue does not occur frequently during training.}
                \label{fig:degen_case}
            \end{figure}

        \paragraph{MoreRed variants, including DDPM:} 
            For all diffusion models used in our work, we employ the polynomial approximation of the cosine noise schedule \cite{edm} with $T=1000$ discretization steps, and a precision parameter of $s=10^{-5}$ to prevent the atoms from undergoing large unrealistic movements during the initial sampling steps. 
            We use a large batch size of $128$ molecules to improve the accuracy of the loss estimation, as it involves uniformly sampling a single diffusion step $t$ per molecule per batch instead of using the whole trajectory for each molecule in the batch. We set the number of atomic features to $256$ for all variants of MoreRed, including the plain DDPM model used in molecular generation experiments in section \ref{app:data_generation}. In these molecular generation experiments, we also employ architectures with more parameters, namely MoreRed-large and DDPM-large. Here we use $512$ atomic features and $5$ layers for the noise head instead of $3$, resulting in circa $10M$ parameters instead of $2.5M$. For MoreRed-JT, we found that setting $\eta$ to $0.9$ works well because the noise prediction provides more signals ($3N$  per molecule with $N$ atoms) compared to diffusion step prediction (one step per molecule). While we use a separate time predictor with a separate representation for MoreRed-AS and MoreRed-ITP, all hyperparameters are kept consistent across all MoreRed variants for all experiments and datasets. As depicted in Figure \ref{fig:degen_case}, uniformly sampling one time step $t$ per molecule per batch results in a noisy training loss since the model uses different diffusion steps at each training iteration instead of the entire diffusion trajectory. To mitigate this issue, we used EMA of the parameters with a decay of $0.999$ across all training epochs during validation, testing and inference rather than using the most recent parameter updates. This approach yielded smoother learning evolution, as reflected in the less noisy validation loss in Figure~\ref{fig:degen_case} because the EMA of the parameters better maintains the previously learned signal from the different diffusion steps seen per molecule. We use a patience of 300 epochs for early stopping and 150 epochs instead of 15 for the learning rate schedule because MoreRed uses only non-equilibrium molecules resulting in 100 times fewer data and fewer iterations per epoch. 

            In initial experiments with MoreRed, we observed that the model outputs occasionally explode, resulting in exploding gradients and divergence of the training, as illustrated by the high peaks in the validation loss in Figure \ref{fig:degen_case} after $400$ epochs. Subsequent analysis revealed that this phenomenon occurs only for high diffusion steps above $t=850$, which was associated with highly noisy and unrealistically dense molecules where all atoms were tightly packed with tiny atomic distances, as shown in Figure \ref{fig:degen_case}. Hence, we added gradient clipping with a global norm of $0.5$ to mitigate this explosive behaviour.
        
        \paragraph{Details for SO3Net:}
        \label{sec:so3net_hyperparams}
        
            In the generalization experiments, in section \ref{sec:Exp.so3net}, using SO3Net instead of PaiNN, we maintain a consistent configuration with 128 atomic features across the entire architecture and utilize 2 hidden layers for all property prediction heads for all models, including the MLFF. The latter results in identical model sizes for both MLFF and MoreRed, with MoreRed using fewer parameters than in the main experiments. Additionally, we set $l_{\text{max}}=1$ for the maximum degree of the spherical harmonics features. To speed up training, a large batch size of 512 and a learning rate of $2 \cdot 10^{-3}$ is utilized for all models. All other experimental details remain unchanged from those employed in the main experiments with PaiNN.

    \section{Computation time}
    \label{app:computation_Times}
    
        Here we give a more detailed analysis of the computation time for the three MoreRed variants and the MLFF model that have been used for the experiments in section \ref{sec:Exp.MoleculeRelaxation}.
        The median number of steps until convergence as well as the accuracy of the equilibrium structures strongly depend on the MoreRed variant, ranging from 53 steps for the fastest MoreRed-ITP to 1000 steps for the most accurate MoreRed-AS.
        For MoreRed-AS, we observe many trajectories where the model predicts many consecutive low-time steps until reaching the maximum number of allowed steps, but the optimization does not converge due to the strict convergence criterion of $\hat{t}\leq0$.
        Applying less strict convergence criteria might decrease the number of steps per structure optimization significantly, which is a direction for future work.
        Furthermore, the computation time per structure during inference dramatically improves if many structures are evaluated in batches instead of sequentially.
        While batch-wise optimization can be done straightforwardly with all MoreRed variants, batch-wise structure optimization utilizing the MLFF model is not trivial, due to the dependency on the L-BFGS optimizer.
        In the following, we give a rough estimate of the inference time by either using sequential optimization or batch-wise optimization.
        For the analysis, we neglect any computational cost that is not directly related to model inference.
        
        \paragraph{Sequential optimization} 
            Comparing MoreRed variants to the MLFF model, we observed an average inference time for a single structure, not a batch, of 0.03s for MoreRed and 0.02s for the MLFF model. To compute the mean inference time per structure optimization performed sequentially, we need to use not the median, as reported above, but the mean. The mean number of optimization steps until convergence was measured as 64 steps for MoreRed-ITP, 489 steps for MoreRed-JT,  992 for MoreRed-AS, and 122 for the MLFF model, which results in an average total inference time per structure optimization of $0.03s \cdot 64 = 1.92s$ for MoreRed-ITP, 14.67s for MoreRed-JT, 29.76s for MoreRed-As and 2.44s for the force field.
        
        \paragraph{Parallelized optimization} 
            For efficient optimization of a large number of structures, as is usually the case in many applications, model inference is preferably done in batches. Assuming batch-wise relaxation with the MLFF model is possible, we observed an inference time for evaluating a batch of 128 molecules of 0.03 s for the force field and 0.05 s  for the MoreRed variants. Since the batch-wise relaxation is done until all the structures in the batch have converged, in a worst-case scenario, which is more likely to happen the larger the batch is, both methods need the maximum number of allowed steps, which is 1000. This would result in an average inference time per structure optimization of $\frac{0.03s \cdot 1000}{128} = 0.23s$ for the force field and $\frac{0.05s \cdot 1000}{128}=0.39 s$ for the MoreRed variants.
    
    \section{Effect of molecular size on performance and efficiency}
    \label{app:molecule_size}
\begin{figure}
                \centering
                \includegraphics[width=1.\linewidth]{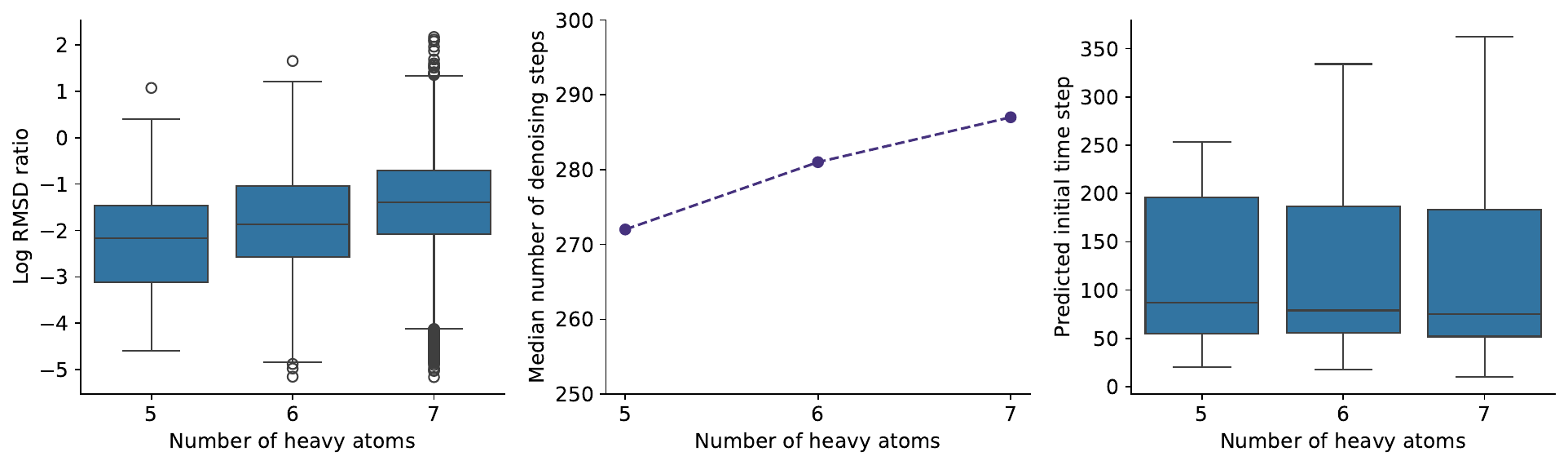}
            \caption{
            Analysis of molecular relaxation outcomes categorized by the count of heavy atoms, spanning from 5 to 7. The graph illustrates the RMSD ratio (\textbf{left}), the count of denoising steps required for convergence (\textbf{mid}), and the estimated initial timestep (\textbf{right}). The initial structures are derived from the non-equilibrium configurations of QM7-X and we use MoreRed-JT for the denoising.}
            \label{fig:num_atoms}
        \end{figure}

        For many chemistry applications, it is required to perform molecular relaxation on large systems with many atoms.
        To evaluate how MoreRed performs on molecules with different numbers of atoms, we use the results of Section \ref{sec:Exp.MoleculeRelaxation}, where molecular relaxation with MoreRed-JT is performed on 20 000 non-equilibrium structures of QM7-X, and categorize the results by the number of heavy atoms per molecule.
        In Figure \ref{fig:num_atoms} we report the RMSD ratio, the number of denoising steps per relaxation trajectory and the predicted initial timestep for all structures with 5 to 7 heavy atoms.
        We find that, excluding some outliers, the predicted initial time step remains largely unaffected by the number of heavy atoms.
        Conversely, as the number of atoms grows, the median number of denoising steps needed for convergence increases by up to 10 steps, and the RMSD ratios also rise.
        However, using the RMSD ratios as a metric is in favor of small structures because local inaccuracies tend to affect the whole molecule, significantly impacting the atom positions at the molecule's periphery.
        Therefore, similar inaccuracies at one point can lead to significantly larger RMSD of positions if the molecule is larger.
        Furthermore, the number of possible local minima increases with the number of heavy atoms.

\newpage

%\References
\section*{References}
\bibliographystyle{unsrt}
\bibliography{morered}

\begin{thebibliography}{100}

\bibitem{schlegel2011}
Schlegel, H.~B.
\newblock Geometry optimization.
\newblock {\em WIREs Computational Molecular Science}, 1(5):790–809, May 2011.

\bibitem{broadbelt1994}
Broadbelt, L.~J., Stark, S.~M., and Klein, M.~T.
\newblock Computer generated pyrolysis modeling: On-the-fly generation of species, reactions, and rates.
\newblock {\em Industrial \& Engineering Chemistry Research}, 33(4):790–799, April 1994.

\bibitem{broadbelt1996}
Broadbelt, L.~J., Stark, S.~M., and Klein, M.~T.
\newblock Computer generated reaction modelling: Decomposition and encoding algorithms for determining species uniqueness.
\newblock {\em Computers \& Chemical Engineering}, 20(2):113–129, February 1996.

\bibitem{broadbelt2005}
Broadbelt, L.~J. and Pfaendtner, J.
\newblock Lexicography of kinetic modeling of complex reaction networks.
\newblock {\em AIChE Journal}, 51(8):2112–2121, June 2005.

\bibitem{fialkowski2005}
Fialkowski, M., Bishop, K. J.~M., Chubukov, V.~A., Campbell, C.~J., and Grzybowski, B.~A.
\newblock Architecture and evolution of organic chemistry.
\newblock {\em Angewandte Chemie International Edition}, 44(44):7263–7269, November 2005.

\bibitem{gothard2012}
Gothard, C.~M., Soh, S., Gothard, N.~A., Kowalczyk, B., Wei, Y., Baytekin, B., and Grzybowski, B.~A.
\newblock Rewiring chemistry: Algorithmic discovery and experimental validation of one‐pot reactions in the network of organic chemistry.
\newblock {\em Angewandte Chemie International Edition}, 51(32):7922–7927, July 2012.

\bibitem{kowalik2012}
Kowalik, M., Gothard, C.~M., Drews, A.~M., Gothard, N.~A., Weckiewicz, A., Fuller, P.~E., Grzybowski, B.~A., and Bishop, K. J.~M.
\newblock Parallel optimization of synthetic pathways within the network of organic chemistry.
\newblock {\em Angewandte Chemie International Edition}, 51(32):7928–7932, July 2012.

\bibitem{sameera2016}
Sameera, W. M.~C., Maeda, S., and Morokuma, K.
\newblock Computational catalysis using the artificial force induced reaction method.
\newblock {\em Accounts of Chemical Research}, 49(4):763–773, March 2016.

\bibitem{dewyer2018}
Dewyer, A.~L., Arg\"{u}elles, A.~J., and Zimmerman, P.~M.
\newblock Methods for exploring reaction space in molecular systems.
\newblock {\em WIREs Computational Molecular Science}, 8(2), November 2017.

\bibitem{maeda2011}
Maeda, S., Komagawa, S., Uchiyama, M., and Morokuma, K.
\newblock Finding reaction pathways for multicomponent reactions: The passerini reaction is a four‐component reaction.
\newblock {\em Angewandte Chemie International Edition}, 50(3):644–649, December 2010.

\bibitem{feinberg2018}
Feinberg, E.~N., Sur, D., Wu, Z., Husic, B.~E., Mai, H., Li, Y., Sun, S., Yang, J., Ramsundar, B., and Pande, V.~S.
\newblock Potentialnet for molecular property prediction.
\newblock {\em ACS Central Science}, 4(11):1520–1530, November 2018.

\bibitem{simm2019}
Simm, G.~N., Vaucher, A.~C., and Reiher, M.
\newblock Exploration of reaction pathways and chemical transformation networks.
\newblock {\em The Journal of Physical Chemistry A}, 123(2):385–399, November 2018.

\bibitem{unsleber2020}
Unsleber, J.~P. and Reiher, M.
\newblock The exploration of chemical reaction networks.
\newblock {\em Annual Review of Physical Chemistry}, 71(1):121–142, April 2020.

\bibitem{baiardi2022}
Baiardi, A., Bosia, F., Brunken, C., Csizi, K.-S., Grimmel, S.~A., Gugler, S., Haag, M.~P., Heuer, M.~A., M\"{u}ller, C.~H., Polonius, S., Simm, G.~N., Sobez, J.-G., Steiner, M., T\"{u}rtscher, P.~L., Unsleber, J.~P., Vaucher, A.~C., Weymuth, T., and Reiher, M.
\newblock qcscine/utilities: Release 5.0.0, 2022.

\bibitem{deutschmann1998}
Deutschmann, O. and Schmidt, L.~D.
\newblock Modeling the partial oxidation of methane in a short‐contact‐time reactor.
\newblock {\em AIChE Journal}, 44(11):2465–2477, November 1998.

\bibitem{zhu2005}
Zhu, H., Kee, R.~J., Janardhanan, V.~M., Deutschmann, O., and Goodwin, D.~G.
\newblock Modeling elementary heterogeneous chemistry and electrochemistry in solid-oxide fuel cells.
\newblock {\em Journal of The Electrochemical Society}, 152(12):A2427, 2005.

\bibitem{gossler2019}
Gossler, H., Maier, L., Angeli, S., Tischer, S., and Deutschmann, O.
\newblock Carmen: An improved computer-aided method for developing catalytic reaction mechanisms.
\newblock {\em Catalysts}, 9(3):227, March 2019.

\bibitem{ulissi2017}
Ulissi, Z.~W., Medford, A.~J., Bligaard, T., and Nørskov, J.~K.
\newblock To address surface reaction network complexity using scaling relations machine learning and dft calculations.
\newblock {\em Nature Communications}, 8(1), March 2017.

\bibitem{steiner2022}
Steiner, M. and Reiher, M.
\newblock Autonomous reaction network exploration in homogeneous and heterogeneous catalysis.
\newblock {\em Topics in Catalysis}, 65(1–4):6–39, January 2022.

\bibitem{sankaran2007}
Sankaran, R., Hawkes, E.~R., Chen, J.~H., Lu, T., and Law, C.~K.
\newblock Structure of a spatially developing turbulent lean methane–air bunsen flame.
\newblock {\em Proceedings of the Combustion Institute}, 31(1):1291–1298, January 2007.

\bibitem{harper2011}
Harper, M.~R., Van~Geem, K.~M., Pyl, S.~P., Marin, G.~B., and Green, W.~H.
\newblock Comprehensive reaction mechanism for n-butanol pyrolysis and combustion.
\newblock {\em Combustion and Flame}, 158(1):16–41, January 2011.

\bibitem{vinu2012}
Vinu, R. and Broadbelt, L.~J.
\newblock Unraveling reaction pathways and specifying reaction kinetics for complex systems.
\newblock {\em Annual Review of Chemical and Biomolecular Engineering}, 3(1):29–54, July 2012.

\bibitem{vereecken2015}
Vereecken, L., Glowacki, D.~R., and Pilling, M.~J.
\newblock Theoretical chemical kinetics in tropospheric chemistry: Methodologies and applications.
\newblock {\em Chemical Reviews}, 115(10):4063–4114, April 2015.

\bibitem{proppe2017}
Proppe, J. and Reiher, M.
\newblock Reliable estimation of prediction uncertainty for physicochemical property models.
\newblock {\em Journal of Chemical Theory and Computation}, 13(7):3297–3317, June 2017.

\bibitem{proppe2019b}
Proppe, J. and Reiher, M.
\newblock Mechanism deduction from noisy chemical reaction networks.
\newblock {\em Journal of Chemical Theory and Computation}, 15(1):357–370, December 2018.

\bibitem{suleimanov2015}
Suleimanov, Y.~V. and Green, W.~H.
\newblock Automated discovery of elementary chemical reaction steps using freezing string and berny optimization methods.
\newblock {\em Journal of Chemical Theory and Computation}, 11(9):4248–4259, August 2015.

\bibitem{gao2016a}
Gao, C.~W., Allen, J.~W., Green, W.~H., and West, R.~H.
\newblock Reaction mechanism generator: Automatic construction of chemical kinetic mechanisms.
\newblock {\em Computer Physics Communications}, 203:212–225, June 2016.

\bibitem{susnow1997}
Susnow, R.~G., Dean, A.~M., Green, W.~H., Peczak, P., and Broadbelt, L.~J.
\newblock Rate-based construction of kinetic models for complex systems.
\newblock {\em The Journal of Physical Chemistry A}, 101(20):3731–3740, May 1997.

\bibitem{han2017}
Han, K., Green, W.~H., and West, R.~H.
\newblock On-the-fly pruning for rate-based reaction mechanism generation.
\newblock {\em Computers \& Chemical Engineering}, 100:1–8, May 2017.

\bibitem{arspous2019}
Arús-Pous, J., Blaschke, T., Ulander, S., Reymond, J.-L., Chen, H., and Engkvist, O.
\newblock Exploring the gdb-13 chemical space using deep generative models.
\newblock {\em Journal of Cheminformatics}, 11(1), March 2019.

\bibitem{gugler2020}
Gugler, S., Janet, J.~P., and Kulik, H.~J.
\newblock Enumeration of de novo inorganic complexes for chemical discovery and machine learning.
\newblock {\em Molecular Systems Design \& Engineering}, 5(1):139–152, 2020.

\bibitem{reymond2015}
Reymond, J.-L.
\newblock The chemical space project.
\newblock {\em Accounts of Chemical Research}, 48(3):722–730, February 2015.

\bibitem{hajduk2007decade_drug_fragments}
Hajduk, P.~J. and Greer, J.
\newblock A decade of fragment-based drug design: strategic advances and lessons learned.
\newblock {\em Nature Reviews Drug Discovery}, 6(3):211--219, 2007.

\bibitem{hautier2011novel_batteries_Throughput}
Hautier, G., Jain, A., Chen, H., Moore, C., Ong, S.~P., and Ceder, G.
\newblock Novel mixed polyanions lithium-ion battery cathode materials predicted by high-throughput ab initio computations.
\newblock {\em Journal of Materials Chemistry}, 21(43):17147--17153, 2011.

\bibitem{Bhowmik2019inverse_batteries}
Bhowmik, A., Castelli, I.~E., Garcia-Lastra, J.~M., Jørgensen, P.~B., Winther, O., and Vegge, T.
\newblock A perspective on inverse design of battery interphases using multi-scale modelling, experiments and generative deep learning.
\newblock {\em Energy Storage Materials}, 21:446--456, 2019.

\bibitem{Freeze2019inverse_catalysts}
Freeze, J.~G., Kelly, H.~R., and Batista, V.~S.
\newblock Search for catalysts by inverse design: Artificial intelligence, mountain climbers, and alchemists.
\newblock {\em Chemical Reviews}, 119(11):6595--6612, 2019.
\newblock PMID: 31059236.

\bibitem{Gantzer2020inverse_drugs}
Gantzer, P., Creton, B., and Nieto-Draghi, C.
\newblock Inverse-qspr for de novo design: A review.
\newblock {\em Molecular Informatics}, 39(4):1900087, 2020.

\bibitem{von2020exploring}
von Lilienfeld, O.~A., M{\"u}ller, K.-R., and Tkatchenko, A.
\newblock Exploring chemical compound space with quantum-based machine learning.
\newblock {\em Nature Reviews Chemistry}, 4(7):347--358, 2020.

\bibitem{born1927}
Born, M. and Oppenheimer, R.
\newblock Zur quantentheorie der molekeln.
\newblock {\em Annalen der Physik}, 389(20):457–484, January 1927.

\bibitem{sutcliffe1992}
Sutcliffe, B.~T.
\newblock {\em The Born-Oppenheimer Approximation}, page 19–46.
\newblock Springer US, 1992.

\bibitem{jensen2017b}
Jensen, F.
\newblock {\em Introduction to {{Computational Chemistry}}}.
\newblock {Wiley}, 3. edition edition, 2017.

\bibitem{cramer2004}
Cramer, C.~J.
\newblock {\em Essentials of {{Computational Chemistry}}: {{Theories}} and {{Models}}}.
\newblock {Wiley}, 2. edition edition, 2004.

\bibitem{halgren1996merck}
Halgren, T.~A.
\newblock Merck molecular force field. i. basis, form, scope, parameterization, and performance of mmff94.
\newblock {\em Journal of computational chemistry}, 17(5-6):490--519, 1996.

\bibitem{rappe1992uff}
Rapp{\'e}, A.~K., Casewit, C.~J., Colwell, K., Goddard~III, W.~A., and Skiff, W.~M.
\newblock Uff, a full periodic table force field for molecular mechanics and molecular dynamics simulations.
\newblock {\em Journal of the American chemical society}, 114(25):10024--10035, 1992.

\bibitem{vanommeslaeghe2010charmm}
Vanommeslaeghe, K., Hatcher, E., Acharya, C., Kundu, S., Zhong, S., Shim, J., Darian, E., Guvench, O., Lopes, P., Vorobyov, I., et~al.
\newblock Charmm general force field: A force field for drug-like molecules compatible with the charmm all-atom additive biological force fields.
\newblock {\em Journal of computational chemistry}, 31(4):671--690, 2010.

\bibitem{bannwarth2019gfn2}
Bannwarth, C., Ehlert, S., and Grimme, S.
\newblock Gfn2-xtb—an accurate and broadly parametrized self-consistent tight-binding quantum chemical method with multipole electrostatics and density-dependent dispersion contributions.
\newblock {\em Journal of chemical theory and computation}, 15(3):1652--1671, 2019.

\bibitem{pm6}
Stewart, J. J.~P.
\newblock Optimization of parameters for semiempirical methods v: Modification of nddo approximations and application to 70 elements.
\newblock {\em Journal of Molecular Modeling}, 13(12):1173--1213, Dec 2007.

\bibitem{pm7}
Stewart, J. J.~P.
\newblock Optimization of parameters for semiempirical methods vi: more modifications to the nddo approximations and re-optimization of parameters.
\newblock {\em Journal of Molecular Modeling}, 19(1):1--32, Jan 2013.

\bibitem{om2}
Weber, W. and Thiel, W.
\newblock Orthogonalization corrections for semiempirical methods.
\newblock {\em Theoretical Chemistry Accounts}, 103(6):495--506, Apr 2000.

\bibitem{coulombmatrix}
Rupp, M., Tkatchenko, A., M\"{u}ller, K.-R., and Von~Lilienfeld, O.~A.
\newblock Fast and accurate modeling of molecular atomization energies with machine learning.
\newblock {\em Physical Review Letters}, 108(5):058301, 2012.

\bibitem{de2016comparing}
De, S., Bart{\'o}k, A.~P., Cs{\'a}nyi, G., and Ceriotti, M.
\newblock Comparing molecules and solids across structural and alchemical space.
\newblock {\em Physical Chemistry Chemical Physics}, 18(20):13754--13769, 2016.

\bibitem{behler2007generalized}
Behler, J. and Parrinello, M.
\newblock Generalized neural-network representation of high-dimensional potential-energy surfaces.
\newblock {\em Physical Review Letters}, 98(14):146401, 2007.

\bibitem{faber2018fchl}
Faber, F.~A., Christensen, A.~S., Huang, B., and Von~Lilienfeld, O.~A.
\newblock Alchemical and structural distribution based representation for universal quantum machine learning.
\newblock {\em The Journal of Chemical Physics}, 148(24), 2018.

\bibitem{dtnn}
Sch\"{u}tt, K.~T., Arbabzadah, F., Chmiela, S., M\"{u}ller, K.~R., and Tkatchenko, A.
\newblock Quantum-chemical insights from deep tensor neural networks.
\newblock {\em Nature Communications}, 8:13890, 2017.

\bibitem{smith2017ani1_potential}
Smith, J.~S., Isayev, O., and Roitberg, A.~E.
\newblock {ANI-1}: an extensible neural network potential with {DFT} accuracy at force field computational cost.
\newblock {\em Chemical Science}, 8(4):3192--3203, 2017.

\bibitem{dimenet}
Gasteiger, J., Groß, J., and G\"{u}nnemann, S.
\newblock Directional message passing for molecular graphs.
\newblock In {\em International Conference on Learning Representations}, 2020.

\bibitem{nequip}
Batzner, S., Musaelian, A., Sun, L., Geiger, M., Mailoa, J.~P., Kornbluth, M., Molinari, N., Smidt, T.~E., and Kozinsky, B.
\newblock E(3)-equivariant graph neural networks for data-efficient and accurate interatomic potentials.
\newblock {\em Nature Communications}, 13(1):2453, May 2022.

\bibitem{satorras2021n}
Satorras, V.~G., Hoogeboom, E., and Welling, M.
\newblock E(n) equivariant graph neural networks.
\newblock In Meila, M. and Zhang, T., editors, {\em Proceedings of the 38th International Conference on Machine Learning}, volume 139 of {\em Proceedings of Machine Learning Research}, pages 9323--9332. PMLR, 18--24 Jul 2021.

\bibitem{so3krates}
Frank, T., Unke, O., and M\"{u}ller, K.-R.
\newblock So3krates: Equivariant attention for interactions on arbitrary length-scales in molecular systems.
\newblock In Koyejo, S., Mohamed, S., Agarwal, A., Belgrave, D., Cho, K., and Oh, A., editors, {\em Advances in Neural Information Processing Systems}, volume~35, pages 29400--29413. Curran Associates, Inc., 2022.

\bibitem{batatia2022mace}
Batatia, I., Kovacs, D.~P., Simm, G. N.~C., Ortner, C., and Csanyi, G.
\newblock {MACE}: Higher order equivariant message passing neural networks for fast and accurate force fields.
\newblock In Oh, A.~H., Agarwal, A., Belgrave, D., and Cho, K., editors, {\em Advances in Neural Information Processing Systems}, 2022.

\bibitem{UnkeGems}
Unke, O.~T., Stöhr, M., Ganscha, S., Unterthiner, T., Maennel, H., Kashubin, S., Ahlin, D., Gastegger, M., Sandonas, L.~M., Berryman, J.~T., Tkatchenko, A., and Müller, K.-R.
\newblock Biomolecular dynamics with machine-learned quantum-mechanical force fields trained on diverse chemical fragments.
\newblock {\em Science Advances}, 10(14):eadn4397, 2024.

\bibitem{allegro}
Musaelian, A., Batzner, S., Johansson, A., Sun, L., Owen, C.~J., Kornbluth, M., and Kozinsky, B.
\newblock Learning local equivariant representations for large-scale atomistic dynamics.
\newblock {\em Nature Communications}, 14(1):579, 2023.

\bibitem{tensorfieldnet}
Thomas, N., Smidt, T., Kearnes, S., Yang, L., Li, L., Kohlhoff, K., and Riley, P.
\newblock Tensor field networks: Rotation-and translation-equivariant neural networks for 3d point clouds.
\newblock {\em arXiv preprint arXiv:1802.08219}, 2018.

\bibitem{noe2020review}
No{\'e}, F., Tkatchenko, A., M{\"u}ller, K.-R., and Clementi, C.
\newblock Machine learning for molecular simulation.
\newblock {\em Annual Review of Physical Chemistry}, 71:361--390, 2020.

\bibitem{unke2021machine_review}
Unke, O.~T., Chmiela, S., Sauceda, H.~E., Gastegger, M., Poltavsky, I., Sch\"{u}tt, K.~T., Tkatchenko, A., and M\"{u}ller, K.-R.
\newblock Machine learning force fields.
\newblock {\em Chemical Reviews}, 121(16):10142--10186, 2021.

\bibitem{gdml}
Chmiela, S., Tkatchenko, A., Sauceda, H.~E., Poltavsky, I., Sch\"{u}tt, K.~T., and M\"{u}ller, K.-R.
\newblock Machine learning of accurate energy-conserving molecular force fields.
\newblock {\em Science Advances}, 3(5):e1603015, 2017.

\bibitem{sgdml}
Chmiela, S., Sauceda, H.~E., Poltavsky, I., M\"{u}ller, K.-R., and Tkatchenko, A.
\newblock sgdml: Constructing accurate and data efficient molecular force fields using machine learning.
\newblock {\em Computer Physics Communications}, 240:38--45, 2019.

\bibitem{Chmiela2023Accurate}
Chmiela, S., Vassilev-Galindo, V., Unke, O.~T., Kabylda, A., Sauceda, H.~E., Tkatchenko, A., and M\"{u}ller, K.-R.
\newblock Accurate global machine learning force fields for molecules with hundreds of atoms.
\newblock {\em Science Advances}, 9(2):eadf0873, 2023.

\bibitem{schnet}
Sch\"{u}tt, K., Kindermans, P.-J., Sauceda~Felix, H.~E., Chmiela, S., Tkatchenko, A., and M\"{u}ller, K.-R.
\newblock {SchNet}: A continuous-filter convolutional neural network for modeling quantum interactions.
\newblock In Guyon, I., Luxburg, U.~V., Bengio, S., Wallach, H., Fergus, R., Vishwanathan, S., and Garnett, R., editors, {\em Advances in Neural Information Processing Systems}, volume~30, pages 991--1001. Curran Associates, Inc., 2017.

\bibitem{Schuett2018SchNet}
Sch{\"u}tt, K.~T., Sauceda, H.~E., Kindermans, P.-J., Tkatchenko, A., and M{\"u}ller, K.-R.
\newblock {SchNet -- A} deep learning architecture for molecules and materials.
\newblock {\em The Journal of Chemical Physics}, 148(24):241722, 2018.

\bibitem{physnet}
Unke, O.~T. and Meuwly, M.
\newblock {PhysNet}: a neural network for predicting energies, forces, dipole moments, and partial charges.
\newblock {\em Journal of Chemical Theory and Computation}, 15(6):3678--3693, 2019.

\bibitem{spookynet}
Unke, O.~T., Chmiela, S., Gastegger, M., Sch{\"u}tt, K.~T., Sauceda, H.~E., and M{\"u}ller, K.-R.
\newblock Spookynet: Learning force fields with electronic degrees of freedom and nonlocal effects.
\newblock {\em Nature communications}, 12(1):7273, 2021.

\bibitem{edm}
Hoogeboom, E., Satorras, V.~G., Vignac, C., and Welling, M.
\newblock Equivariant diffusion for molecule generation in 3{D}.
\newblock In Chaudhuri, K., Jegelka, S., Song, L., Szepesvari, C., Niu, G., and Sabato, S., editors, {\em Proceedings of the 39th International Conference on Machine Learning}, volume 162 of {\em Proceedings of Machine Learning Research}, pages 8867--8887. PMLR, 17--23 Jul 2022.

\bibitem{diffbridges}
Wu, L., Gong, C., Liu, X., Ye, M., and Liu, Q.
\newblock Diffusion-based molecule generation with informative prior bridges.
\newblock In Koyejo, S., Mohamed, S., Agarwal, A., Belgrave, D., Cho, K., and Oh, A., editors, {\em Advances in Neural Information Processing Systems}, volume~35, pages 36533--36545. Curran Associates, Inc., 2022.

\bibitem{mdm}
Huang, L., Zhang, H., Xu, T., and Wong, K.-C.
\newblock Mdm: Molecular diffusion model for 3d molecule generation.
\newblock {\em Proceedings of the AAAI Conference on Artificial Intelligence}, 37(4):5105--5112, Jun. 2023.

\bibitem{geoldm}
Xu, M., Powers, A.~S., Dror, R.~O., Ermon, S., and Leskovec, J.
\newblock Geometric latent diffusion models for 3{D} molecule generation.
\newblock In Krause, A., Brunskill, E., Cho, K., Engelhardt, B., Sabato, S., and Scarlett, J., editors, {\em Proceedings of the 40th International Conference on Machine Learning}, volume 202 of {\em Proceedings of Machine Learning Research}, pages 38592--38610. PMLR, 23--29 Jul 2023.

\bibitem{moldiff}
Peng, X., Guan, J., Liu, Q., and Ma, J.
\newblock {M}ol{D}iff: Addressing the atom-bond inconsistency problem in 3{D} molecule diffusion generation.
\newblock In Krause, A., Brunskill, E., Cho, K., Engelhardt, B., Sabato, S., and Scarlett, J., editors, {\em Proceedings of the 40th International Conference on Machine Learning}, volume 202 of {\em Proceedings of Machine Learning Research}, pages 27611--27629. PMLR, 23--29 Jul 2023.

\bibitem{geodiff}
Xu, M., Yu, L., Song, Y., Shi, C., Ermon, S., and Tang, J.
\newblock {GeoDiff}: A geometric diffusion model for molecular conformation generation.
\newblock In {\em International Conference on Learning Representations}, 2022.

\bibitem{vignac2023digress}
Vignac, C., Krawczuk, I., Siraudin, A., Wang, B., Cevher, V., and Frossard, P.
\newblock Digress: Discrete denoising diffusion for graph generation.
\newblock In {\em The Eleventh International Conference on Learning Representations}, 2023.

\bibitem{autoregressivediff}
Kong, L., Cui, J., Sun, H., Zhuang, Y., Prakash, B.~A., and Zhang, C.
\newblock Autoregressive diffusion model for graph generation.
\newblock In Krause, A., Brunskill, E., Cho, K., Engelhardt, B., Sabato, S., and Scarlett, J., editors, {\em Proceedings of the 40th International Conference on Machine Learning}, volume 202 of {\em Proceedings of Machine Learning Research}, pages 17391--17408. PMLR, 23--29 Jul 2023.

\bibitem{gschnet}
Gebauer, N., Gastegger, M., and Sch\"{u}tt, K.
\newblock Symmetry-adapted generation of 3d point sets for the targeted discovery of molecules.
\newblock In Wallach, H., Larochelle, H., Beygelzimer, A., d\textquotesingle Alch\'{e}-Buc, F., Fox, E., and Garnett, R., editors, {\em Advances in Neural Information Processing Systems}, volume~32, pages 7566--7578. Curran Associates, Inc., 2019.

\bibitem{cgschnet}
Gebauer, N. W.~A., Gastegger, M., Hessmann, S. S.~P., M{\"u}ller, K.-R., and Sch{\"u}tt, K.~T.
\newblock Inverse design of 3d molecular structures with conditional generative neural networks.
\newblock {\em Nature Communications}, 13(1):973, 2022.

\bibitem{simm2020reinforcement_3d}
Simm, G., Pinsler, R., and Hernandez-Lobato, J.~M.
\newblock Reinforcement learning for molecular design guided by quantum mechanics.
\newblock In III, H.~D. and Singh, A., editors, {\em Proceedings of the 37th International Conference on Machine Learning}, volume 119 of {\em Proceedings of Machine Learning Research}, pages 8959--8969. PMLR, 13--18 Jul 2020.

\bibitem{simm2021symmetryaware_3d}
Simm, G. N.~C., Pinsler, R., Cs{\'a}nyi, G., and Hern{\'a}ndez-Lobato, J.~M.
\newblock Symmetry-aware actor-critic for 3d molecular design.
\newblock In {\em International Conference on Learning Representations}, 2021.

\bibitem{meldgaard2021generating_3d}
Meldgaard, S.~A., Köhler, J., Mortensen, H.~L., Christiansen, M.-P.~V., No{\'{e}}, F., and Hammer, B.
\newblock Generating stable molecules using imitation and reinforcement learning.
\newblock {\em Machine Learning: Science and Technology}, 3(1):015008, 2022.

\bibitem{noe2019boltzmann}
No{\'e}, F., Olsson, S., K{\"o}hler, J., and Wu, H.
\newblock Boltzmann generators: Sampling equilibrium states of many-body systems with deep learning.
\newblock {\em Science}, 365(6457):eaaw1147, 2019.

\bibitem{kohler2020equivariant}
K{\"o}hler, J., Klein, L., and Noe, F.
\newblock Equivariant flows: Exact likelihood generative learning for symmetric densities.
\newblock In III, H.~D. and Singh, A., editors, {\em Proceedings of the 37th International Conference on Machine Learning}, volume 119 of {\em Proceedings of Machine Learning Research}, pages 5361--5370. PMLR, 13--18 Jul 2020.

\bibitem{garcia2021en_flow}
Garcia~Satorras, V., Hoogeboom, E., Fuchs, F., Posner, I., and Welling, M.
\newblock E(n) equivariant normalizing flows.
\newblock In Ranzato, M., Beygelzimer, A., Dauphin, Y., Liang, P., and Vaughan, J.~W., editors, {\em Advances in Neural Information Processing Systems}, volume~34, pages 4181--4192. Curran Associates, Inc., 2021.

\bibitem{klein2023timewarp}
Klein, L., Foong, A. Y.~K., Fjelde, T.~E., Mlodozeniec, B.~K., Brockschmidt, M., Nowozin, S., Noe, F., and Tomioka, R.
\newblock Timewarp: Transferable acceleration of molecular dynamics by learning time-coarsened dynamics.
\newblock In {\em Thirty-seventh Conference on Neural Information Processing Systems}, 2023.

\bibitem{mansimov2019molecular_graph_Translation}
Mansimov, E., Mahmood, O., Kang, S., and Cho, K.
\newblock Molecular geometry prediction using a deep generative graph neural network.
\newblock {\em Scientific Reports}, 9(1):20381, 2019.

\bibitem{simm2020generative_graph_Translation}
Simm, G. and Hernandez-Lobato, J.~M.
\newblock A generative model for molecular distance geometry.
\newblock In III, H.~D. and Singh, A., editors, {\em Proceedings of the 37th International Conference on Machine Learning}, volume 119 of {\em Proceedings of Machine Learning Research}, pages 8949--8958. PMLR, 13--18 Jul 2020.

\bibitem{gogineni2020torsionnet_graph_Translation}
Gogineni, T., Xu, Z., Punzalan, E., Jiang, R., Kammeraad, J., Tewari, A., and Zimmerman, P.
\newblock {TorsionNet}: A reinforcement learning approach to sequential conformer search.
\newblock In Larochelle, H., Ranzato, M., Hadsell, R., Balcan, M., and Lin, H., editors, {\em Advances in Neural Information Processing Systems}, volume~33, pages 20142--20153. Curran Associates, Inc., 2020.

\bibitem{ganea2021geomol}
Ganea, O., Pattanaik, L., Coley, C., Barzilay, R., Jensen, K., Green, W., and Jaakkola, T.
\newblock {GeoMol}: Torsional geometric generation of molecular 3d conformer ensembles.
\newblock In Ranzato, M., Beygelzimer, A., Dauphin, Y., Liang, P., and Vaughan, J.~W., editors, {\em Advances in Neural Information Processing Systems}, volume~34, pages 13757--13769. Curran Associates, Inc., 2021.

\bibitem{xu2021end_graph_Translation}
Xu, M., Wang, W., Luo, S., Shi, C., Bengio, Y., Gomez-Bombarelli, R., and Tang, J.
\newblock An end-to-end framework for molecular conformation generation via bilevel programming.
\newblock In Meila, M. and Zhang, T., editors, {\em Proceedings of the 38th International Conference on Machine Learning}, volume 139 of {\em Proceedings of Machine Learning Research}, pages 11537--11547. PMLR, 18--24 Jul 2021.

\bibitem{lemm2021machine_graph_Translation}
Lemm, D., von Rudorff, G.~F., and von Lilienfeld, O.~A.
\newblock Machine learning based energy-free structure predictions of molecules, transition states, and solids.
\newblock {\em Nature Communications}, 12(1):4468, 2021.

\bibitem{torsional_diff}
Jing, B., Corso, G., Chang, J., Barzilay, R., and Jaakkola, T.~S.
\newblock Torsional diffusion for molecular conformer generation.
\newblock In Oh, A.~H., Agarwal, A., Belgrave, D., and Cho, K., editors, {\em Advances in Neural Information Processing Systems}, 2022.

\bibitem{Wang_denoising}
Wang, Y., Xu, C., Li, Z., and Barati~Farimani, A.
\newblock Denoise pretraining on nonequilibrium molecules for accurate and transferable neural potentials.
\newblock {\em Journal of Chemical Theory and Computation}, 19(15):5077--5087, 2023.
\newblock PMID: 37390120.

\bibitem{pmlr-v202-feng23c}
Feng, S., Ni, Y., Lan, Y., Ma, Z.-M., and Ma, W.-Y.
\newblock Fractional denoising for 3{D} molecular pre-training.
\newblock In Krause, A., Brunskill, E., Cho, K., Engelhardt, B., Sabato, S., and Scarlett, J., editors, {\em Proceedings of the 40th International Conference on Machine Learning}, volume 202 of {\em Proceedings of Machine Learning Research}, pages 9938--9961. PMLR, 23--29 Jul 2023.

\bibitem{zaidi2023pretraining}
Zaidi, S., Schaarschmidt, M., Martens, J., Kim, H., Teh, Y.~W., Sanchez-Gonzalez, A., Battaglia, P., Pascanu, R., and Godwin, J.
\newblock Pre-training via denoising for molecular property prediction.
\newblock In {\em The Eleventh International Conference on Learning Representations}, 2023.

\bibitem{liu2023molecular}
Liu, S., Guo, H., and Tang, J.
\newblock Molecular geometry pretraining with {SE}(3)-invariant denoising distance matching.
\newblock In {\em The Eleventh International Conference on Learning Representations}, 2023.

\bibitem{godwin2022simple}
Godwin, J., Schaarschmidt, M., Gaunt, A.~L., Sanchez-Gonzalez, A., Rubanova, Y., Veli{\v{c}}kovi{\'c}, P., Kirkpatrick, J., and Battaglia, P.
\newblock Simple {GNN} regularisation for 3d molecular property prediction and beyond.
\newblock In {\em International Conference on Learning Representations}, 2022.

\bibitem{DAE}
Vincent, P., Larochelle, H., Bengio, Y., and Manzagol, P.-A.
\newblock Extracting and composing robust features with denoising autoencoders.
\newblock In {\em Proceedings of the 25th International Conference on Machine Learning}, ICML '08, page 1096–1103, New York, NY, USA, 2008. Association for Computing Machinery.

\bibitem{hsu2022score}
Hsu, T., Sadigh, B., Bertin, N., Park, C.~W., Chapman, J., Bulatov, V., and Zhou, F.
\newblock Score-based denoising for atomic structure identification.
\newblock {\em arXiv preprint arXiv:2212.02421}, 2022.

\bibitem{qm7x}
Hoja, J., Medrano~Sandonas, L., Ernst, B.~G., Vazquez-Mayagoitia, A., DiStasio~Jr., R.~A., and Tkatchenko, A.
\newblock {QM7-X}, a comprehensive dataset of quantum-mechanical properties spanning the chemical space of small organic molecules.
\newblock {\em Scientific Data}, 8(1):43, Feb 2021.

\bibitem{Mortazavi2018}
Mortazavi, M., Brandenburg, J.~G., Maurer, R.~J., and Tkatchenko, A.
\newblock Structure and stability of molecular crystals with many-body dispersion-inclusive density functional tight binding.
\newblock {\em The Journal of Physical Chemistry Letters}, 9(2):399–405, January 2018.

\bibitem{Seifert1996}
Seifert, G., Porezag, D., and Frauenheim, T.
\newblock Calculations of molecules, clusters, and solids with a simplified lcao-dft-lda scheme.
\newblock {\em International Journal of Quantum Chemistry}, 58(2):185–192, 1996.

\bibitem{Elstner1998}
Elstner, M., Porezag, D., Jungnickel, G., Elsner, J., Haugk, M., Frauenheim, T., Suhai, S., and Seifert, G.
\newblock Self-consistent-charge density-functional tight-binding method for simulations of complex materials properties.
\newblock {\em Physical Review B}, 58(11):7260–7268, September 1998.

\bibitem{Gaus2011}
Gaus, M., Cui, Q., and Elstner, M.
\newblock Dftb3: Extension of the self-consistent-charge density-functional tight-binding method (scc-dftb).
\newblock {\em Journal of Chemical Theory and Computation}, 7(4):931–948, March 2011.

\bibitem{Tkatchenko2012}
Tkatchenko, A., DiStasio, R.~A., Car, R., and Scheffler, M.
\newblock Accurate and efficient method for many-body van der waals interactions.
\newblock {\em Physical Review Letters}, 108(23), June 2012.

\bibitem{Ambrosetti2014}
Ambrosetti, A., Reilly, A.~M., DiStasio, R.~A., and Tkatchenko, A.
\newblock Long-range correlation energy calculated from coupled atomic response functions.
\newblock {\em The Journal of Chemical Physics}, 140(18), February 2014.

\bibitem{blum2009970}
Blum, L.~C. and Reymond, J.-L.
\newblock 970 million druglike small molecules for virtual screening in the chemical universe database gdb-13.
\newblock {\em Journal of the American Chemical Society}, 131(25):8732--8733, 2009.

\bibitem{Adamo1999}
Adamo, C. and Barone, V.
\newblock Toward reliable density functional methods without adjustable parameters: The pbe0 model.
\newblock {\em The Journal of Chemical Physics}, 110(13):6158–6170, April 1999.

\bibitem{Perdew1996}
Perdew, J.~P., Ernzerhof, M., and Burke, K.
\newblock Rationale for mixing exact exchange with density functional approximations.
\newblock {\em The Journal of Chemical Physics}, 105(22):9982–9985, December 1996.

\bibitem{Blum2009}
Blum, V., Gehrke, R., Hanke, F., Havu, P., Havu, V., Ren, X., Reuter, K., and Scheffler, M.
\newblock Ab initio molecular simulations with numeric atom-centered orbitals.
\newblock {\em Computer Physics Communications}, 180(11):2175–2196, November 2009.

\bibitem{Ren2012}
Ren, X., Rinke, P., Blum, V., Wieferink, J., Tkatchenko, A., Sanfilippo, A., Reuter, K., and Scheffler, M.
\newblock Resolution-of-identity approach to hartree–fock, hybrid density functionals, rpa, mp2 andgwwith numeric atom-centered orbital basis functions.
\newblock {\em New Journal of Physics}, 14(5):053020, May 2012.

\bibitem{diff_mod_sohl}
Sohl-Dickstein, J., Weiss, E., Maheswaranathan, N., and Ganguli, S.
\newblock Deep unsupervised learning using nonequilibrium thermodynamics.
\newblock In Bach, F. and Blei, D., editors, {\em Proceedings of the 32nd International Conference on Machine Learning}, volume~37 of {\em Proceedings of Machine Learning Research}, pages 2256--2265, Lille, France, 07--09 Jul 2015. PMLR.

\bibitem{DDPM_Ho}
Ho, J., Jain, A., and Abbeel, P.
\newblock Denoising diffusion probabilistic models.
\newblock In Larochelle, H., Ranzato, M., Hadsell, R., Balcan, M., and Lin, H., editors, {\em Advances in Neural Information Processing Systems}, volume~33, pages 6840--6851. Curran Associates, Inc., 2020.

\bibitem{song2021score}
Song, Y., Sohl-Dickstein, J., Kingma, D.~P., Kumar, A., Ermon, S., and Poole, B.
\newblock Score-based generative modeling through stochastic differential equations.
\newblock In {\em International Conference on Learning Representations}, 2021.

\bibitem{painn}
Sch\"{u}tt, K., Unke, O., and Gastegger, M.
\newblock Equivariant message passing for the prediction of tensorial properties and molecular spectra.
\newblock In Meila, M. and Zhang, T., editors, {\em Proceedings of the 38th International Conference on Machine Learning}, volume 139 of {\em Proceedings of Machine Learning Research}, pages 9377--9388. PMLR, 18--24 Jul 2021.

\bibitem{bishopPRML}
Bishop, C.~M.
\newblock {\em Pattern recognition and machine learning}.
\newblock Information Science and Statistics. Springer New York, 2006.

\bibitem{schuett2018schnetpack}
Schütt, K., Kessel, P., Gastegger, M., Nicoli, K., Tkatchenko, A., and Muller, K.-R.
\newblock Schnetpack: A deep learning toolbox for atomistic systems.
\newblock {\em Journal of chemical theory and computation}, 15(1):448--455, 2018.

\bibitem{schutt2023schnetpack}
Sch{\"u}tt, K.~T., Hessmann, S. S.~P., Gebauer, N. W.~A., Lederer, J., and Gastegger, M.
\newblock {SchNetPack 2.0: A neural network toolbox for atomistic machine learning}.
\newblock {\em The Journal of Chemical Physics}, 158(14):144801, 04 2023.

\bibitem{Openbabel}
O'Boyle, N.~M., Banck, M., James, C.~A., Morley, C., Vandermeersch, T., and Hutchison, G.~R.
\newblock Open babel: An open chemical toolbox.
\newblock {\em Journal of Cheminformatics}, 3(1):33, Oct 2011.

\bibitem{ase-paper}
Larsen, A.~H., Mortensen, J.~J., Blomqvist, J., Castelli, I.~E., Christensen, R., Du{\l}ak, M., Friis, J., Groves, M.~N., Hammer, B., Hargus, C., et~al.
\newblock The atomic simulation environment—a python library for working with atoms.
\newblock {\em Journal of Physics: Condensed Matter}, 29(27):273002, 2017.

\bibitem{weigend2005balanced}
Weigend, F. and Ahlrichs, R.
\newblock Balanced basis sets of split valence, triple zeta valence and quadruple zeta valence quality for h to rn: Design and assessment of accuracy.
\newblock {\em Physical Chemistry Chemical Physics}, 7(18):3297--3305, 2005.

\bibitem{Sun2015}
Sun, Q.
\newblock Libcint: An efficient general integral library for gaussian basis functions.
\newblock {\em Journal of Computational Chemistry}, 36(22):1664–1671, June 2015.

\bibitem{Sun2017}
Sun, Q., Berkelbach, T.~C., Blunt, N.~S., Booth, G.~H., Guo, S., Li, Z., Liu, J., McClain, J.~D., Sayfutyarova, E.~R., Sharma, S., Wouters, S., and Chan, G.~K.
\newblock P<scp>y</scp>scf: the python‐based simulations of chemistry framework.
\newblock {\em WIREs Computational Molecular Science}, 8(1), September 2017.

\bibitem{Sun2020}
Sun, Q., Zhang, X., Banerjee, S., Bao, P., Barbry, M., Blunt, N.~S., Bogdanov, N.~A., Booth, G.~H., Chen, J., Cui, Z.-H., Eriksen, J.~J., Gao, Y., Guo, S., Hermann, J., Hermes, M.~R., Koh, K., Koval, P., Lehtola, S., Li, Z., Liu, J., Mardirossian, N., McClain, J.~D., Motta, M., Mussard, B., Pham, H.~Q., Pulkin, A., Purwanto, W., Robinson, P.~J., Ronca, E., Sayfutyarova, E.~R., Scheurer, M., Schurkus, H.~F., Smith, J. E.~T., Sun, C., Sun, S.-N., Upadhyay, S., Wagner, L.~K., Wang, X., White, A., Whitfield, J.~D., Williamson, M.~J., Wouters, S., Yang, J., Yu, J.~M., Zhu, T., Berkelbach, T.~C., Sharma, S., Sokolov, A.~Y., and Chan, G. K.-L.
\newblock Recent developments in the p<scp>y</scp>scf program package.
\newblock {\em The Journal of Chemical Physics}, 153(2), July 2020.

\bibitem{kahouli_2024_10927872}
Kahouli, K., Hessmann, S. S.~P., Mueller, K.-R., Nakajima, S., Gugler, S., and Gebauer, N. W.~A.
\newblock Morered: Molecular relaxation by reverse diffusion with time step prediction.
\newblock Zenodo, April 2024.

\bibitem{ruddigkeit2012enumeration}
Ruddigkeit, L., Van~Deursen, R., Blum, L.~C., and Reymond, J.-L.
\newblock Enumeration of 166 billion organic small molecules in the chemical universe database {GDB-17}.
\newblock {\em Journal of Chemical Information and Modeling}, 52(11):2864--2875, 2012.

\bibitem{qm9}
Ramakrishnan, R., Dral, P.~O., Rupp, M., and von Lilienfeld, O.~A.
\newblock Quantum chemistry structures and properties of 134 kilo molecules.
\newblock {\em Scientific Data}, 1(1):140022, 2014.

\bibitem{cifar10}
Krizhevsky, A.
\newblock Learning multiple layers of features from tiny images.
\newblock 2009.

\bibitem{weininger1988smiles}
Weininger, D.
\newblock {SMILES}, a chemical language and information system. 1. {I}ntroduction to methodology and encoding rules.
\newblock {\em Journal of Chemical Information and Computer Science}, 28(1):31--36, 1988.

\bibitem{diff_edm}
Karras, T., Aittala, M., Aila, T., and Laine, S.
\newblock Elucidating the design space of diffusion-based generative models.
\newblock In Oh, A.~H., Agarwal, A., Belgrave, D., and Cho, K., editors, {\em Advances in Neural Information Processing Systems}, 2022.

\bibitem{loshchilov2018decoupled}
Loshchilov, I. and Hutter, F.
\newblock Decoupled weight decay regularization.
\newblock In {\em International Conference on Learning Representations}, 2019.

\end{thebibliography}

\end{document}